\documentclass[11pt, letterpaper]{article}
\input{epsf}
\usepackage[utf8]{inputenc}
\usepackage{amsmath}
\usepackage{amssymb}
\input{epsf}
\usepackage{natbib}
\usepackage[margin=1in]{geometry}
\usepackage[raggedright]{titlesec}
\usepackage[hang,flushmargin]{footmisc} 
\usepackage{graphicx}
\usepackage{framed}
\usepackage{color}
\usepackage{yfonts}
\usepackage[dvipsnames]{xcolor}
\usepackage{pifont}
\usepackage{amsmath,amssymb}
\usepackage{multirow}
\usepackage{amsfonts}
\usepackage{subfigure}
\usepackage{caption}
\usepackage{threeparttable}
\usepackage{esvect}
\usepackage{pgf}
\usepackage{tikz}
\usepackage{tikz-cd}
\usepackage{fancyvrb}
\usetikzlibrary{positioning}
\usepackage{bbm}
\usepackage{colortbl}
\usepackage{pdflscape}
\usepackage{booktabs}
\usepackage{amsmath}
\usepackage{mathdots}
\usepackage{yhmath}
\usepackage{cancel}
\usepackage{siunitx}
\usepackage{array}
\usepackage{multirow}
\usepackage{gensymb}
\usepackage{tabularx}
\usepackage{booktabs}
\usetikzlibrary{fadings}
\usetikzlibrary{patterns}
\usetikzlibrary{shadows.blur}
\usetikzlibrary{shapes}

\usepackage{hyperref}
\hypersetup{
    colorlinks=true,
    linkcolor=NavyBlue,
    filecolor=magenta,      
    urlcolor=cyan,
    citecolor=black
}
 
\usepackage{rotating}
\usepackage{ntheorem}
\makeatletter
\renewtheoremstyle{plain} % Adds automatic line break, if heading is too long
{\item{\theorem@headerfont ##1\ ##2\theorem@separator}~}
{\item{\theorem@headerfont ##1\ ##2\ (##3)\theorem@separator}~}
\makeatother

\theoremheaderfont{\normalfont\bfseries}

\newtheorem{theorem}{Theorem}[section]
\newtheorem{definition}{Definition}[section]
\newtheorem{lemma}{Lemma}[section]

\newtheorem{example}{Example}[section]
\newtheorem{assumption}{Assumption}

\newenvironment{proof}{\paragraph{Proof:}}{\hfill$\square$}

\newcommand{\E}{\mathbb{E}}

\newcommand{\var}{\text{var}}
\newcommand{\cov}{\text{cov}}
\newcommand{\cor}{\text{cor}}

\newcommand{\bX}{\mathbf{X}}

\newcommand{\bU}{\mathbf{U}}

\newcommand{\R}{\mathbbm{R}}

\newcommand{\Pc}{\mathcal{P}}
\newcommand{\Sc}{\mathcal{S}}

\newcommand{\cD}{\mathcal{D}}

\newcommand{\indep}{\raisebox{0.05em}{\rotatebox[origin=c]{90}{$\models$}}}%Perpendicular symbol
\definecolor{shadecolor}{gray}{0.9}

\tikzset{every picture/.style={line width=0.75pt}} %set default line width to 0.75pt        

\usepackage{enumitem}
\newlist{Step}{enumerate}{2}
\setlist[Step]{label={{Step \arabic*.}}, leftmargin=*}

\usepackage{setspace}
 \makeatletter
\newcommand\circled[1]{%
  \mathpalette\@circled{#1}%
}
\newcommand\@circled[2]{%
  \tikz[baseline=(math.base)] \node[draw,circle,inner sep=2pt] (math) {$\m@th#1#2$};%
}
\makeatother

 \makeatletter
\newcommand\circledblue[1]{%
  \mathpalette\@circledblue{#1}%
}
\newcommand\@circledblue[2]{%
  \tikz[baseline=(math.base)] \node[draw,circle, fill=blue!20, inner sep=2pt] (math) {$\m@th#1#2$};%
 }

\renewenvironment{abstract}
 {\begin{center}\normalsize\textsc{Abstract}%
 \end{center}\begin{quote}\normalsize}
 {\end{quote}}

\title{Sensitivity Analysis in the Generalization of Experimental Results\footnote{The author would like to thank Erin Hartman, Chad Hazlett, Leonard Wainstein, Eli Ben-Michael, Dan Soriano, Avi Feller, Peng Ding, Sam Pimentel, and the UCLA Causal Inference reading group. Melody Huang is supported by the National Science Foundation Graduate Research Fellowship under Grant No. 2146752. Any opinion, findings, and conclusions or recommendations expressed in this material are those of the authors(s) and do not necessarily reflect the views of the National Science Foundation.}}
\author{Melody Y. Huang\thanks{Ph.D. Candidate, Department of Statistics,
    University of California, Berkeley, CA 94720. Email:
    \href{mailto:melodyyhuang@berkeley.edu}{melodyyhuang@berkeley.edu}, URL:
    \url{https://melodyyhuang.github.io}}}
\date{}

\begin{document}
\maketitle 
\begin{abstract}
Randomized controlled trials (RCT’s) allow researchers to estimate causal effects in an experimental sample with minimal identifying assumptions. However, to generalize or transport a causal effect from an RCT to a target population, researchers must adjust for a set of treatment effect moderators. In practice, it is impossible to know whether the set of moderators has been properly accounted for.  In the following paper, I propose a three parameter sensitivity analysis for generalizing or transporting experimental results using weighted estimators, with several advantages over existing methods. First, the framework does not require assumptions on the underlying data generating process for either the experimental sample selection mechanism or treatment effect heterogeneity. Second, I show that the sensitivity parameters are guaranteed to be bounded and propose several tools researchers can use to perform sensitivity analysis: (1) graphical and numerical summaries for researchers to assess how robust a point estimate is to killer confounders; (2) an extreme scenario analysis; and (3) a formal benchmarking approach for researchers to estimate potential sensitivity parameter values using existing data. Finally, I demonstrate that the proposed framework can be easily extended to the class of doubly robust, augmented weighted estimators. The sensitivity analysis framework is applied to a set of Jobs Training Program experiments. 
\end{abstract}

\clearpage 
\doublespacing

\section{Introduction} 
Randomized controlled trials (RCT’s) provide researchers with a rich understanding of the treatment effect within an experimental sample. Because researchers have the ability to eliminate confounding by randomly assigning treatment in a controlled environment, experiments have a high degree of internal validity. However, inconveniently, a causal effect estimated from an RCT may not directly generalize to populations of interest when the experimental sample is not representative of the larger population. One prominent source of bias arises from distributional differences in treatment effect moderators---i.e., covariates that drive propensity of selection into the experimental sample, as well as treatment effect heterogeneity---between the experimental sample and the population (i.e., \cite{imai2008misunderstandings, cole2010generalizing, olsen2013external}; see \cite{egami2020elements} for discussion on alternative sources of bias). To properly generalize or transport the results from an experiment into a target population, researchers must either re-weight the experimental sample to be representative of the target population, or successfully model the treatment effect heterogeneity (\cite{stuart2011use, kern2016assessing}). 

In practice, it is impossible to know whether the set of treatment effect moderators has been correctly identified. Researchers rely on the measured variables that are available in the sample and the population, and often assume that the observed covariates sufficiently capture the confounding effect. However, when confounders are omitted from estimation, the resulting point estimates will be biased. Existing sensitivity analyses in generalizability and transportability allow researchers to assess how robust their point estimates are to omitted confounders. However, many of the existing approaches require researchers to justify sensitivity parameters that may be arbitrarily large or small, and/or invoke parametric assumptions used to model the estimated bias from confounders. For example, \cite{nguyen2017sensitivity} present a sensitivity analysis for omitted moderators, unobserved in both the experiment and the population, when modeling treatment effect heterogeneity with a linear model. However, the sensitivity analysis requires researchers to posit the degree of imbalance in the omitted confounder. Similarly, \cite{nie2021covariate} introduce a non-parametric, percentile bootstrap approach to sensitivity analysis for generalization; however, the method requires researchers to estimate a worst-case bound for the odds ratio between the misspecified and true experimental sample selection propensities. Recently, \cite{dahabreh2019generalizing} proposed an alternative approach to adjust for bias in estimation by directly modeling the bias. 

In the following paper, we introduce a sensitivity analysis framework for unobserved confounders when using a weighted estimator for generalizing or transporting a causal effect. We focus on developing a sensitivity analysis for assessing bias in the point estimate of a causal effect, with discussions for how researchers may address changes in uncertainty from omitting a confounder in Section \ref{sec:conclusion}. The proposed framework builds on the sensitivity analysis literature from observational studies (\cite{hong2021did, cinelli2020making, shen2011sensitivity}), as well as existing sensitivity analysis approaches for generalizing or transporting an estimated treatment effect (\cite{nguyen2017sensitivity} and \cite{dahabreh2019generalizing}), with several important innovations.

The paper provides three primary contributions. First, we demonstrate that the bias of a weighted estimator may be decomposed into three different components, which serve as the sensitivity parameters in the proposed framework. We show that all three parameters are guaranteed to be bounded, thereby providing researchers with a fixed set of plausible values to assess. Furthermore, the bias decomposition does not require researchers to invoke distributional or functional form assumptions for the confounder, experimental sample selection mechanism, or the individual-level treatment effect, allowing for a large degree of flexibility for when the framework can be used. 

Second, we provide several approaches to help researchers conduct their sensitivity analysis in transparent and interpretable ways. The first approach is a graphical summary of sensitivity in the form of bias contour plots. The second is a numerical summary of sensitivity, and extends the robustness value from \cite{cinelli2020making} for the weighted estimator setting. The robustness value serves as a reference for how much remaining confounding must exist in order for a point estimate to change by a given amount. Third, we introduce an extreme scenario analysis to assess a worst-case analysis of the potential bias that may arise from omitting a confounder. Finally, we propose a formal benchmarking procedure that leverages observed covariates to posit parameter values for a confounder, and allows researchers to incorporate their substantive knowledge for the relative strength of confounders. 

Third, we extend the proposed framework for the class of augmented weighted estimators. We show that the bias of an augmented weighted estimator when omitting a confounder can be similarly decomposed into three, bounded components. Furthermore, we show that the sensitivity analysis can also be adapted for cases when researchers are only modeling the individual-level treatment effect, and demonstrate that previously proposed sensitivity analyses (i.e., \cite{nguyen2017sensitivity}) are a special case of our framework.

The paper is organized as follows. Section 2 introduces the notational framework, identifying assumptions, related literature, and the running example.  Section 3 formalizes the proposed sensitivity analysis framework. In Section 4, we discuss three different tools that researchers can use to conduct the sensitivity analysis. Section 5 extends the sensitivity analysis framework to the augmented weighted estimator. Section 6 concludes. Proofs and extensions are provided in the Appendix. 

\section{Background} 
\subsection{Notation and Set-Up}
To begin, we define an infinite super-population, from which the target population and the experimental sample are drawn i.i.d. from. We define the target population as a sample of $N$ units, drawn randomly from the target population. Following \cite{buchanan2018generalizing}, we define the experimental sample of $n$ units as a potentially biased sample from the infinite super-population. Define $S_i$ as an indicator for whether the unit is in the experimental sample (i.e., $S_i = 1$ when unit $i$ is in the experiment, and $S_i = 0$ otherwise), and let $\mathcal{S}$ denote the set of indices for units included in the experimental sample. 

Let $T_i$ be a binary treatment assignment variable, where $T_i = 1$ for units assigned to treatment, and $T_i = 0$ for control. We assume full compliance, such that treatment assigned implies treatment received, and following the potential outcomes framework, define $Y_i(t)$ to be the potential outcome when unit $i$ receives treatment $T_i = t$ where $t \in \{0, 1\}$ (\cite{neyman1923, rubin1974causal}). Throughout the paper, we make the standard assumptions of no interference and that treatments are identically administered across all units (i.e., SUTVA, defined in \citet{rubin1980SUTVA}). We assume a set of pre-treatment covariates $\mathcal{X}_i$ exists across both the experimental sample and the target population. Finally, we define the individual-level treatment effect $\tau_i$ as the difference between the potential outcomes of unit $i$: 
$$\tau_i = Y_i(1) - Y_i(0)$$
Because we can never observe both potential outcomes of a specific unit, the individual-level treatment effect is unidentifiable (\cite{imbens2015causal}). To formalize, we assume that both $\{\tau_i, \mathcal{X}_i \mid S_i = 1\}_{i=1}^n$ and $\{\tau_i, \mathcal{X}_i \mid S_i = 0\}_{i=1}^N$ are drawn i.i.d. from an infinite super-population. When the experimental sample is a biased sample, the sampling distributions for the experimental sample and the target population will not be the same (i.e., $P(\tau_i, T_i, \mathcal{X}_i \mid S_i = 1) \neq P(\tau_i, T_i, \mathcal{X}_i \mid S_i = 0)$).

The sample average treatment effect (SATE) is defined as as the average treatment effect across the experimental sample (i.e., $\tau_\Sc \equiv \E\{\tau_i \mid S_i = 1\}$). When there exists randomization in an experiment with equal probability of treatment assignment, a simple difference-in-means estimator can be used to estimate the SATE:
\begin{equation}
\widehat{\tau_\Sc} \equiv \frac{1}{\sum_{i \in \Sc} T_i } \sum_{i \in \Sc} T_i Y_i - \frac{1}{\sum_{i \in \Sc} 1-T_i} \sum_{i \in \Sc} (1-T_i) Y_i,
\end{equation}
where $\Sc$ represents the set of indices that correspond to units in the experimental sample (i.e., $\Sc = \{i: \ S_i = 1\}$). The population (or target) average treatment effect (PATE) is the causal quantity of interest, formally defined as:
\begin{equation}
\tau \equiv \E\{\tau_i \mid S_i = 0\}
\end{equation}
where the expectation is taken over the realized target population.\footnote{Researchers may instead, treat the estimand of interest as the average treatment effect, across the infinite super-population, instead of the realized population. The proposed sensitivity analysis will extend for both cases. We refer readers to \cite{huang2021leveraging} for more discussion of this setting.} 

If the experimental sample is randomly drawn from the super-population, then $\widehat \tau_\Sc$ is an unbiased estimator for the PATE. However, in most settings, the experimental sample is not representative of the target population, and experimental results cannot be directly extrapolated to the population (\cite{cole2010generalizing, olsen2013external, nguyen2017sensitivity}). In these settings, an additional identifying assumption is necessary to recover the PATE from the experimental sample: 
\begin{assumption}[Conditional Ignorability of Sampling]\mbox{}\\
\label{assum-ind}
\begin{equation}
\tau_i \ \indep \ S_i \mid \mathcal{X}_i
\end{equation}
\end{assumption} 
Assumption \ref{assum-ind} states that there exists some set of pre-treatment covariates $\mathcal{X}_i$ for which, conditioned on the set $\mathcal{X}$, the distribution of the individual-level treatment effects in the sample will be equivalent to the distribution of individual-level treatment effects in the population (\cite{kern2016assessing}).\footnote{For PATE identification, Assumption \ref{assum-ind} can be relaxed for mean exchangeability. See \cite{hartman2021kpop} for more discussion.} \cite{egami2019covariate} formally define the set of covariates $\mathcal{X}_i$ that allow the sampling mechanism to be conditionally independent from the treatment effect heterogeneity as the \textit{separating set}. 

%Expand on this 
In addition to Assumption \ref{assum-ind}, we must also invoke a positivity assumption--conditional on $\mathcal{X}$, the probability of being included in the sample is non-zero (\cite{rosenbaum1983assessing}). 
\begin{assumption}[Positivity]
\begin{equation} 
0 < P(S_i = 1 \mid \mathcal{X}_i) < 1
\end{equation} 
\end{assumption} 
Violations of the positivity assumption result in attempting to generalize beyond the support of the data (see \cite{stuart2011use} and \cite{tipton2014generalizable} as two examples).

The most common approach to estimating the PATE is through a weighted estimator, where the observations in the experimental sample are re-weighted to resemble that of the target population (\cite{stuart2011use, olsen2013external}):
$$\hat \tau_{W} = \frac{1}{n_1} \sum_{i\in \Sc} w_i T_i Y_i -  \frac{1}{n_0} \sum_{i\in \Sc} w_i (1-T_i) Y_i,$$
where the weights are defined as the sampling weights (i.e., $w_i \propto P(S_i= 0 \mid \mathcal{X}_i)/P(S_i = 1 \mid \mathcal{X}_i)$), $n_1$ and $n_0$ are the number of units in the treatment and control groups, respectively. 
Weights are often estimated using logistic regression (\cite{cole2010generalizing, stuart2011use, buchanan2018generalizing}). Recently, alternative weighting methods have been proposed, including more general balancing methods, such as entropy balancing, which adjust for distributional differences between the experimental sample and population observations without explicitly modeling the underlying probability function (\cite{sarndal2003model, hainmueller2012entropy, josey2021transporting, josey2020calibration, lu2021you}; see \cite{ben2020balancing} for more discussion).

Alternative approaches to estimating the PATE include directly modeling the individual-level treatment effect using $\mathcal{X}_i$ (i.e., using a linear regression by interacting the treatment indicator with the pre-treatment covariates, or using more flexible modeling methods, such as tree-based methods--see \cite{kern2016assessing, athey2019generalized, hill2011bayesian, wager2018estimation} for more discussion and examples), or using doubly robust estimators, which allow researchers to simultaneously estimate a model for the weights and the individual-level treatment effect (\cite{kern2016assessing, dahabreh2019generalizing}). 

In practice, researchers estimate the PATE under the assumption that they have correctly identified the full separating set $\mathcal{X}_i$. When Assumption \ref{assum-ind} holds, the weighted estimators will be consistent estimators for PATE. However, violations of this assumption can result in biased estimation. The goal of this paper is to formalize a framework for assessing the sensitivity of the PATE estimates to a variable $\bU_i$ being omitted from the separating set $\mathcal{X}_i$, which we refer to as a \textit{confounder} (i.e., a variable missing from the separating set necessary for Assumption \ref{assum-ind} to hold). 

\subsection{Related Literature} 
Within the generalizability and observational causal inference literature, different sensitivity analysis frameworks have been proposed to help assess the robustness of estimators to omitted confounders. We will distinguish between sensitivity analyses developed for weighting-baesd approaches, and sensitivity analyses that have been developed for model-based approaches to estimating PATE.\\

\noindent \textit{Estimating Weights.} 
Because estimating propensity of sample selection and propensity of treatment are very similar processes, the methods developed in the observational setting with respect to propensity score weighting may be utilized in the generalization setting. \cite{rosenbaum1987sensitivity} proposed a sensitivity analysis for matched estimators that computes the largest odds ratio between the probability of treatment, conditional on $\bX_i$, and the probability of treatment, conditional on both $\bX_i$ and $\bU_i$, before the estimated treatment effect becomes zero. While the original framework inherently requires the assumption of a constant treatment effect, recent literature has allowed for treatment effect heterogeneity and other extensions (i.e.,  \cite{tan2006distributional}, \cite{shen2011sensitivity}, \cite{zhao2019sensitivity}, \cite{hong2021did}, \cite{dorn2021sharp}, \cite{nie2021covariate}). Alternative simulation-based approaches have also been developed, in which researchers must invoke a distributional assumption on the omitted confounder $\bU_i$ (i.e., \cite{ichino2008temporary, Tozzi2019CausalEO, burgette}). \\

\noindent \textit{Modeling the Individual-Level Treatment Effect.} Alternative sensitivity analysis approaches developed within the generalizability literature have focused on omitting a confounder when modeling the individual-level treatment effect. For example, \cite{nguyen2017sensitivity} proposed a sensitivity analysis for the omission of a moderator when modeling treatment effect heterogeneity. More specifically, they assume an underlying, linear potential outcomes data generating process, and show that the bias of omitting a confounder $\bU_i$ is a function of (1) the moderation effect (i.e., the coefficient in front of the omitted confounder $\bU_i$), and (2) the degree of imbalance in the omitted confounder across the population and sample.  They also show that with a weighted regression, the same bias is incurred, so long as $\bU$ is also orthogonal to any of the variables used in the estimated weights.

\cite{dahabreh2019generalizing} proposed a sensitivity analysis for bias from \textit{any} kind of violation of Assumption \ref{assum-ind} that arises from omitting a variable from the separating set. They propose estimating a bias function, conditional on the observed covariates, and use this to perform bias correction. This approach is similar to the pattern-mixture methods in the observational literature, which attempt to explicitly model specific differences between the conditional distributions of the outcomes (e.g., \cite{robins1999association}). However, in order for the bias-corrected estimators to be truly unbiased, the bias function must be correctly specified, which may be challenging. \\

In all of these existing methods, there are some common challenges. 
Many of the proposed methods have depend on either an assumed distribution of the confounder, or an assumed functional form on the bias. Neither the bias, nor the nature of the omitted confounder, are identifiable from the data. As such the validity of the sensitivity analysis rely heavily on the assumption that the functional forms practitioners have tried are sufficiently similar to the true underlying data generating process (\cite{brumback2004sensitivity}). While recent extensions (e.g., \cite{zhao2019sensitivity}) relax the distributional and parametric assumptions, these sensitivity frameworks require users to specify plausible ranges for the sensitivity parameters or bias functions. Many of these parameters are unbounded, and the justification of these plausible ranges are left to the practitioner to defend using substantive knowledge. 

The contributions of this paper are two-fold: (1) develop a sensitivity analysis framework for the generalization or transportation of experimental results without requiring distributional or functional form assumptions on the individual-level treatment effect or confounder, and (2) provide a set of tools for researchers to transparently justify plausible ranges of values for the different sensitivity parameters. The proposed sensitivity analysis extends the frameworks developed by \cite{hong2021did} and \cite{shen2011sensitivity}, and allows researchers to discuss potential substantive changes to a point estimate due to unobserved confounding; however, it does not directly address the effect of omitted confounders on uncertainty. Researchers can equivalently think of this as an asymptotic analysis, as the uncertainty associated with a point estimate will disappear as the sample size gets larger (i.e., $n \to \infty$). We discuss how researchers may consider the impact of an omitted confounder on uncertainty in Section \ref{sec:conclusion} and connections to alternative sensitivity approaches (i.e., marginal sensitivity models) in Appendix \ref{app:msm}.

\subsection{Running Example: Jobs Training Partnership Act} 
To help with our discussion of the sensitivity analysis, we will use a set of experiments conducted on the Jobs Training Partnership Act (JTPA) as a running example throughout the paper. The national JTPA study ran from 1987 to 1989, and assessed the effectiveness of the jobs training programs in helping individuals in the study find employment and increase their earnings. The original study was conducted across 16 different experimental sites. Individuals were first interviewed to determine whether or not they were eligible for JTPA services; those deemed eligible were assigned randomly to treatment and control using a 2:1 ratio. Individuals assigned to treatment were given access to JTPA services, while those assigned to control were told they were ineligible for the program. Following treatment assignment, a follow-up survey was conducted 18 months later, in which individuals were asked about their earnings (\cite{bloom1993national}). We focus our analysis on the subset of adult women, the largest target group within the JTPA study.\footnote{The estimated impacts of JTPA for the other target groups were not found to be statistically significant in the original study.}

We leverage the nature of the original multi-site experiment to perform a benchmarking exercise for the sensitivity analysis. More specifically, we pick one of the 16 experimental sites and generalize the estimated effect of JTPA access on earnings from this site to the remaining 15 sites. The benchmark PATE is defined as the average treatment effect across the units in the other 15 experimental sites. This allows us to evaluate the actual error that is incurred from generalizing. To estimate the sample selection weights, we use entropy balancing across a set of pre-treatment covariates measured in the baseline survey (\cite{hainmueller2012entropy}, \cite{josey2021transporting}). Entropy balancing directly optimizes on covariate balance (i.e., the average covariate value in the experimental sample, versus the average covariate value in the target population) to estimate the weights, instead of first estimating the probabilities of selection into sample.\footnote{The sensitivity analysis are agnostic to whether we use inverse-propensity score weights, or probability-like balancing weights. \cite{zhao2016entropy} demonstrated that entropy balancing weights are implicitly estimating propensity score weights, with a modified loss function. See \cite{wang2020minimal}, \cite{soriano2021interpretable}, and \cite{ben2020balancing} for more discussion on the connection between balancing weights and inverse-propensity score weighting.} We weight on previous earnings, age, hourly wage, years of education, whether or not the individual graduated high school (or has a GED), whether or not the individual is married, and indicators for whether the individual is black or Hispanic. 

To illustrate the sensitivity analysis, we examine the site of Omaha, Nebraska, which consists of 636 individuals, 424 of whom were assigned to treatment, with the remainder in control. The target population (i.e., the other 15 experimental sites) consists of 5,466 individuals. (See Appendix \ref{app:jtpa} for more details on the experimental site.) 

\begin{table}[!ht]
\begin{center} 
\begin{tabular}{lcc} \toprule 
               & Unweighted & Weighted\\ \midrule 
Impact of JTPA access on earnings$^*$ & 1.11 & 1.36 \\ \bottomrule 
\multicolumn{3}{l}{\small *-Estimates reported in thousands of USD}
\end{tabular} 
\end{center} 
\caption{Point estimates of impact of JTPA access on earnings, generalizing the estimated effect from the site of Omaha, Nebraska to the other 15 experimental sites.} 
\end{table} 

The within-site estimated impact of JTPA access on earnings is $\$1,100$. After weighting, the estimated impact of JTPA access earnings is $\$1,360$. In the following sections, we will introduce a sensitivity framework that allow researchers to assess how robust the estimate is to unobserved confounders. 

\section{Sensitivity Analysis for Weighted Estimators} \label{sec:weighted} 
In the following section, we will introduce a sensitivity analysis for weighted estimators when omitting a confounder from the weight estimation. In Section \ref{subsec:bias}, we show that the bias formula for the weighted estimator when omitting a confounder can be written as a function of three components: an $R^2$ parameter, a correlation value, and the variation in the individual-level treatment effect. In Section \ref{subsec:interp}, we show that all three of these parameters are guaranteed to be bounded on finite ranges, and Section \ref{subsec:summary} summarizes.

%% DEFINE THE PROBLEM 
\subsection{Bias of a Weighted Estimator when Omitting a Confounder} \label{subsec:bias} 
We consider the sensitivity of a weighted estimator to a confounder that has been omitted in the estimation of the weights. We formally define the separating set as $\mathcal{X}_i = \{ \bX_i, \bU_i\}$. In other words, for the weighted estimator to be unbiased, we should be estimating weights using both $\bX_i$ and $\bU_i$; however, we omit $\bU_i$. We write the weights estimated using just $\bX_i$ as $w_i$, and the \textit{ideal} weights that would have been estimated, had we included both $\bX_i$ and $\bU_i$, as $w_i^*$. Finally, we define $\varepsilon_i$ as the linear error in the  weights from omitting $\bU_i$ (i.e., $\varepsilon_i := w_i - w_i^*$).

In the following sections, we will assume that researchers are estimating inverse propensity score weights, and that, had they included $\bU_i$, they would be able to consistently estimate the weights.\footnote{Misspecification concerns can also be addressed with the sensitivity analysis if researchers can write the error as an omitted variable problem. For example, if a linear probability model is used, $\bU_i$ can include non-linear functions of $\bX_i$ that matter for modeling selection.} We provide extensions for balancing weights in Appendix \ref{app:balancing}. Throughout, consistent with \cite{shen2011sensitivity} and \cite{hong2021did}, we will refer to bias as the expectation of estimator minus the true value (i.e., true statistical bias). 

The error in the weights from omitting $\bU_i$ can be written as a function of (1) the estimated weights, and (2) the residual imbalance in $\bU_i$ conditional on the covariates $\bX_i$. We formalize this in the following corollary.
\begin{lemma}[Error Decomposition] \label{lem:error_decomp} \mbox{}\\
When using inverse propensity weights, the estimated weights and the ideal weights are written as: 
\begin{equation} 
w_i = \frac{P(S_i = 1)}{P(S_i = 0)} \cdot \frac{1-P(S_i = 1|\bX_i)}{P(S_i = 1| \bX_i)} \ \ \ \
w_i^* = \frac{P(S_i = 1)}{P(S_i = 0)} \cdot \frac{1-P(S_i = 1|\bX_i, \bU_i)}{P(S_i = 1| \bX_i, \bU_i)}
\end{equation} 
Then, the error in weight estimation from omitting $\bU_i$ can be decomposed in the following manner: 
\begin{align} 
\varepsilon_i &= w_i - w_i^* \nonumber \\
&= \underbrace{\frac{P(S_i = 1)}{P(S_i = 0)} \cdot \frac{P(S_i = 0 | \bX_i)}{P(S_i = 1 | \bX_i)}}_{\text{Estimated Weights } (w_i)} \cdot \underbrace{\left( \frac{P(\bU_i | \bX_i, S_i = 1) - P(\bU_i | \bX_i, S_i = 0)}{P(\bU_i | \bX_i, S_i = 1)} \right)}_{\text{Residual Imbalance in } \bU_i},
\label{eqn:eps}
\end{align} 
where $P(\bU_i \mid \bX_i, S_i = 1) - P(\bU_i \mid \bX_i, S_i = 1) $ represents the difference in the underlying probability density function of the omitted confounder $\bU_i$, conditioned on $\bX_i$, across the target population ($S_i = 0$) and the experimental sample ($S_i = 1$). 
\end{lemma} 
There are two main points to highlight from Lemma \ref{lem:error_decomp}. First, when $\bU_i$ is over-represented in the sample (represented by $P(\bU_i | \bX_i, S_i = 1) > P(\bU_i | \bX_i, S_i = 0)$), by failing to balance on $\bU_i$, the estimated weights $w_i$ will be larger than the true weights $w_i^*$. Conversely, when $\bU_i$ is under-represented in the sample, the estimated weights will be too small, relative to the ideal weights. Second, Lemma \ref{lem:error_decomp} highlights that it is the \textit{residual} imbalance in the omitted confounder, after accounting for the observed covariates $\bX_i$, that drives the error in the weight estimation. In other words, if $\bU_i$ is mostly balanced by accounting for $\bX_i$, the resulting error from omitting $\bU_i$ will be relatively low.

Consider our running example. The original study cited the latent variable motivation as a potential confounder (\cite{bloom1993national}). While we cannot include motivation directly in the weights, we have included variables such as education and previous earnings, which are likely correlated to motivation. If, by controlling for variables such as education and previous earnings, we have accounted for much of the imbalance in motivation, then we expect the error from omitting motivation from the weights to be relatively low. 

The bias of a weighted estimator from omitting a confounder $\bU_i$ is a function of $\varepsilon_i$ and the degree to which this error term is related to treatment effect heterogeneity. We formalize this in the following theorem: 
\begin{theorem}[Bias of a Weighted Estimator from Omitting a Confounder] \mbox{}\\
Assume $Y_i(1) - Y_i(0) \ \indep \ S_i \ | \ \{ \bX_i, \bU_i\}$. Let $w_i$ be the weights estimated using only $\bX_i$, and let $w_i^*$ be the (correct) weights, obtained using $\{\bX_i, \bU_i\}$. The bias of a weighted estimator from using $w_i$ instead of $w_i^*$ is given as:\footnote{The derived bias expression will be the \textit{exact} bias when researchers are using a Horvitz-Thompson style weighted estimator. In cases when researchers are using a stabilized weighted estimator, there will be finite-sample bias of order $o(1/n)$. However, the finite-sample bias will be dominated by the bias incurred from omitting a confounder from the weights (see \cite{miratrix2013adjusting}, \cite{rosenbaum2010design}, \cite{lunceford2004stratification} for more discussion).} 
\begin{equation} 
\text{Bias}(\hat \tau_W) =  \begin{cases} 
\displaystyle \rho_{\varepsilon, \tau} \sqrt{\var_\Sc(w_i) \cdot \frac{R^2_\varepsilon}{1-R^2_\varepsilon} \cdot \sigma_\tau^2} & \text{if } R^2_\varepsilon < 1\\
\rho_{\varepsilon, \tau} \sqrt{\var_\Sc(w_i^*) \cdot \sigma^2_\tau} &\text{if } R^2_\varepsilon = 1,
\end{cases} 
\label{eqn:bias_formula} 
\end{equation} 
where $\rho_{\varepsilon, \tau}$ is the correlation between $\varepsilon_i$ and $\tau_i$ (i.e., $\rho_{\varepsilon, \tau} := \cor_\Sc(\varepsilon_i, \tau_i)$), $R^2_\varepsilon$ is the ratio of variances between $\varepsilon_i$ and $w_i^*$ (i.e., $R^2_\varepsilon := \var_\Sc(\varepsilon_i)/\var_\Sc(w_i^*)$), and $\sigma_{\tau}^2$ is the variance of the individual-level treatment effect (i.e., $\sigma^2_\tau := \var_\Sc(\tau_i)$). Derivation is provided in Appendix \ref{app:proof}.
\label{thm:weight} 
\end{theorem}

Theorem \ref{thm:weight} identifies the three drivers of bias in a weighted estimator when a confounder is omitted in the weight estimation: (1) the remaining imbalance in the omitted confounder (i.e., $R^2_\varepsilon$), (2) the correlation between $\varepsilon_i$ and the individual-level treatment effect (i.e., $\rho_{\varepsilon, \tau})$, and (3) the amount of treatment effect heterogeneity (i.e., $\sigma^2_\tau$). The bias formula does not rely on any kind of distributional or parametric assumption on the confounder or the underlying data generating process. Theorem \ref{thm:weight} provides a natural foundation for a three parameter sensitivity analysis. We will discuss the interpretation of each of these parameters in the following subsection. 

%However, the crucial difference is that in the observational setting, the observations $Y_i$ are being re-weighted, while in the generalization setting, the individual-level treatment effect $\tau_i$ is being re-weighted. \\
\subsection{Interpreting the Parameters} \label{subsec:interp}
In the following subsection, we discuss the interpretation of the sensitivity parameters. Whereas previous sensitivity frameworks have relied on imposing parametric assumptions on the outcome and selection models to propose plausible ranges of values for the sensitivity parameters (i.e., \cite{shen2011sensitivity, hong2021did}), we show that all three parameters are guaranteed to be bounded across finite ranges that can be directly estimated from the observed data. 

\subsubsection{Variation in Ideal Weights Explained by $\varepsilon_i$ ($R^2_\varepsilon$)} 
The $R^2_\varepsilon$ term is defined as the ratio of variances between the error term and the ideal weights. In the following lemma, we show that the variation in the true weights can be decomposed into two components: variation explained by the estimated weights, and the variation explained by the error term $\varepsilon_i$; therefore, $R^2_\varepsilon$ is bounded on the interval of 0 and 1. As such, we can interpret $R^2_\varepsilon$ as the proportion of variation in the true weights explained by the error term $\varepsilon_i$. 

\begin{lemma}[Variance Decomposition of $w_i^*$]\mbox{}\\
For inverse propensity score weights, the variance of the true weights $w_i^*$ can be decomposed linearly into two components: 
\begin{align*} 
\var_\Sc(w_i^*) &= \var_\Sc(w_i) + \var_\Sc(\varepsilon_i)
\implies \frac{\var_\Sc(w_i)}{\var_\Sc(w_i^*)} + \underbrace{\frac{\var_\Sc(\varepsilon_i)}{\var_\Sc(w_i^*)}}_{:= R^2_\varepsilon} = 1
\end{align*} 
Therefore, $R^2_\varepsilon$ is bound between 0 and 1. 
\label{lem:var_decomp} 
\end{lemma} 
The results of Lemma \ref{lem:var_decomp} follow from the fact that for inverse propensity score weights, the covariance between the estimated weights $w_i$ and  $\varepsilon_i$ is zero. In Appendix \ref{app:balancing}, we provide an extension of this result for a class of balancing weights. 

As the amount of residual imbalance in the omitted confounder increases, $R^2_\varepsilon$ will increase.  If the residual imbalance of the omitted confounder (i.e., imbalance in $\bU_i$, conditional on $\bX_i$) is relatively small, then the estimated weights will be close to the true weights. As a result, $R^2_\varepsilon$ will be close to 0. In contrast, if the residual imbalance of the omitted confounder is large, then much of the variation in $w_i^*$ will be driven by $\varepsilon_i$, and $R^2_\varepsilon$ will be large, approaching 1. Returning to the running example, we once again consider the latent factor of motivation. If motivation is balanced after controlling for observed variables like education and previous earnings, then including motivation into the weight estimation should result in weights $w_i^*$ similar to the estimated weights $w_i$, and $R^2_\varepsilon$ will be relatively small (i.e., $R^2_\varepsilon$ is close to zero). 

In cases when researchers are unable to explain \textit{any} of the imbalance in a single moderator using the existing covariates, then $R^2_\varepsilon = 1$, which implies that all of the variation in $w_i^*$ can be explained by the error term $\varepsilon_i$. In this scenario, the bias decomposition will be undefined due to the $1-R^2_\varepsilon$ term in the denominator, and researchers must posit values for $\var(w_i^*)$ in order to perform the sensitivity analysis. However, in practice, if researchers have balanced on at least one moderator, it is unlikely that the $R_\varepsilon^2$ value will be equal to 1.

\subsubsection{Correlation between $\varepsilon_i$ and $\tau_i$ ($\rho_{\varepsilon, \tau}$)} 
The correlation between $\varepsilon_i$ and the individual-level treatment effect is a standardized measure for how much treatment effect heterogeneity $\bU_i$ explains. When $\rho_{\varepsilon, \tau}$ is very high (i.e., $\rho_{\varepsilon, \tau} \approx 1$), then units with a large $\tau_i$ are overweighted ($w_i > w_i^*$ corresponds to large $\tau_i$). Thus, in these settings, there will be positive bias. Conversely, if $\rho_{\varepsilon, \tau} \approx -1$, the opposite would be true---we underweight units with a large individual-level treatment effect, which results in a negatively biased estimated PATE. If the correlation between the error term and the individual-level treatment effect were close to zero, then the imbalance in the omitted confounder $\bU_i$ is not related to treatment effect heterogeneity, and as such, omitting $\bU_i$ would not result in much bias. 

While $\rho_{\varepsilon,\tau}$ is inherently bounded on the interval $[-1,1]$, we can decompose $\rho_{\varepsilon, \tau}$ as a function of $R^2_\varepsilon$ to restrict the set of feasible correlation values to a tighter range.
\begin{lemma}[Correlation Decomposition] \mbox{}\\
The correlation between $\varepsilon_i$ and the individual-level treatment effects can be decomposed in the following manner:
\begin{align}
\rho_{\varepsilon, \tau} &=
\begin{cases} 
\displaystyle \cor_\Sc(w_i, \tau_i) \sqrt{\frac{1-R_\varepsilon^2}{R_\varepsilon^2}} - \cor_\Sc(w^*_i, \tau_i) \cdot \sqrt{\frac{1}{R_\varepsilon^2}} & \text{when } R_\varepsilon^2 > 0 \\
0 &\text{when } R_\varepsilon^2 = 0
\end{cases} 
\label{eqn:rho_decomp}
\end{align}
 Furthermore, applying Cauchy-Schwarz bounds $\rho_{\varepsilon,\tau}$ by the following:
$$-\sqrt{1-\cor_\Sc^2(w_i, \tau_i)} \leq \rho_{\varepsilon, \tau} \leq \sqrt{1-\cor_\Sc^2(w_i, \tau_i)} $$
\label{lem:rho_bound} 
\end{lemma}
There are two primary things to highlight from the decomposition in Lemma \ref{lem:rho_bound}. The first is that there is an implicit relationship between $R^2_\varepsilon$ and $\rho_{\varepsilon, \tau}$. When $R_\varepsilon^2$ is equal to 0, this implies that there is no imbalance in the omitted confounder. In such a scenario, $\varepsilon_i$ is equal to a constant value, and the correlation between $\varepsilon_i$ and $\tau_i$ must be zero. Second, Lemma \ref{lem:rho_bound} demonstrates that $\rho_{\varepsilon, \tau}$ will be bounded between $\pm \sqrt{1-\cor_\Sc^2(w_i, \tau_i)}$. If the estimated weights $w_i$ can explain most of the variation in treatment effect heterogeneity, the additional variation that can be explained by adding in the omitted confounder must be small. 

The correlation between the estimated weights and $\tau_i$ will take on large values when (1) the covariates contained in $w_i$ explain much of the treatment effect heterogeneity, \textit{and} (2) the covariates that explain the treatment effect heterogeneity are imbalanced across the population and the experimental sample. To help provide intuition for this, consider our running example. If access to JTPA services was only effective for women who graduated high school, then if education were imbalanced across the experimental sample and the population, estimating weights on education would result in a large $|\cor_\Sc(w_i, \tau_i)|$ value. However, if education were not very imbalanced across the experimental sample and population, even though education explains much of the variation in the treatment effect heterogeneity, $\cor_\Sc(w_i, \tau_i)$ will low. In such a scenario, the true $\rho_{\varepsilon, \tau}$ value should also be small; however, this would not be reflected in the bound.

\paragraph{Remark on Estimating $\cor_\Sc(w_i, \tau_i)$:} Lemma \ref{lem:rho_bound} allows researchers to bound $\rho_{\varepsilon, \tau}$ on a more restrictive interval than $[-1,1]$. However, in practice, it is not possible to directly calculate $\cor_\Sc(w_i, \tau_i)$, since $\tau_i$ is unidentified. However, researchers may conservatively estimate the correlation of $w_i$ and $\tau_i$ by using $\cov_\Sc(w_i, Y_i(1))$ and $\cov_\Sc(w_i, Y_i(0))$, which is identified by randomization. More specifically:
\begin{align*} 
\widehat \cor_\Sc(w_i, \tau_i) &= \frac{\widehat \cov_\Sc(w_i, Y_i(1)) - \widehat \cov_\Sc(w_i, Y_i(0))}{\sqrt{\sigma^2_\tau \cdot \widehat \var_\Sc(w_i)}}
\end{align*} 
Because $\widehat \cor_\Sc(w_i, \tau_i)$ is a function of the variation in the individual-level treatment effect (i.e., $\sigma^2_{\tau}$), if researchers use a more conservative estimate of $\sigma^2_{\tau}$, this will subsequently lead to a more conservative estimate on $\cor_\Sc(w_i, \tau_i)$, and by extension, a more conservative estimate for the bounds on $\rho_{\varepsilon, \tau}$. See Section \ref{subsec:TE_het} for details on specifying $\sigma^2_\tau$.\\

To help illustrate the relationship between $R_\varepsilon^2$, $\rho_{\varepsilon, \tau}$, and bias, consider the following example. 
\begin{example}[Difference-in-Means]
Consider the case in which we use the difference-in-means estimator in the experimental sample as an estimator for PATE. In that situation, $\varepsilon_i = \frac{1}{n} - w_i^*$, as there is equal weights (i.e., $w_i = 1/n$) for all units in the sample. The bias of a difference-in-means estimator (i.e., $\hat \tau_\Sc$) for the PATE is: 
$$\text{Bias}(\hat \tau_\Sc) = \cor_\Sc(w^*_i, \tau_i) \sqrt{\var_\Sc(w_i^*) \cdot \sigma^2_{\tau}}.$$
The relative reduction in bias from weighting is as follows:
\begin{align*} 
\text{Relative Reduction} = \left| \frac{\text{Bias}(\hat \tau_W)}{\text{Bias}(\hat \tau_\Sc)} \right|
&= \left| 1-\sqrt{(1-R_\varepsilon^2)} \cdot \frac{\cor_\Sc(w_i, \tau_i)}{\cor_\Sc(w^*_i, \tau_i)} \right|.
\end{align*} 
See Appendix \ref{app:additional_deriv} for derivation.  
 \label{ex:dim} 
\end{example}

Example \ref{ex:dim} illustrates the fact that even when a confounder is omitted, weighting can help reduce the bias. The closer $w_i$ is to $w_i^*$, the greater the relative reduction in bias.  This underscores a point that has been made in the sub-classification literature: while it is generally impossible to perfectly adjust for confounding effects, adjusting for \textit{some} confounding is usually better than none at all (\cite{o2014generalizing, tipton2013improving}).

\subsubsection{Treatment Effect Heterogeneity ($\sigma^2_\tau$)} \label{subsec:TE_het} 

The last term in the bias formula is the variance in the individual-level treatment effect. This term is independent of the estimation process and is intrinsic to the analysis at hand. The magnitude of treatment effect heterogeneity ($\sigma^2_{\tau}$) acts as a scaling factor. When there exists a large degree of treatment effect heterogeneity, the task of recovering the PATE becomes harder, and even small imbalances in the moderators can result in a large degree of bias. When there is less treatment effect heterogeneity, we have more leeway in mis-specifying the weights without incurring large amounts of bias. In the most extreme case of no treatment effect heterogeneity, we need not adjust for any confounders to have unbiased estimation. Because treatment effect heterogeneity is inherent to the underlying data generating process, regardless of what variables are included in the weights, $\sigma^2_\tau$ is fixed. In the following subsection, we apply the results from \cite{ding2019decomposing} to show that $\sigma^2_\tau$, while unidentifiable, can be bounded. We suggest that researchers set $\sigma^2_\tau$ to a conservative upper bound, and vary the other sensitivity parameters, which correspond to parameters for which they \textit{do} have control over in the weighting process. 

To begin, decompose $\sigma^2_{\tau}$ as: 
$$\sigma^2_{\tau} = \var_\Sc(Y_i(1)) + \var_\Sc(Y_i(0)) - 2 \cov_\Sc(Y_i(1), Y_i(0))$$ 
The decomposition illustrates that the magnitude of treatment effect heterogeneity will be driven by two factors: (1) the total variation in the outcomes (i.e., $\var_\Sc(Y_i(1)) + \var_\Sc(Y_i(0))$), and (2) how correlated the potential outcomes are. Because we cannot estimate the covariance between the potential outcomes, $\sigma^2_\tau$ can never be identified. However, \cite{ding2019decomposing} showed that sharp bounds for $\sigma^2_\tau$ can be obtained by applying Fréchet-Hoeffding bounds \citep{hoeffding1941masstabinvariante, frechet1951tableaux}:
\begin{equation}
\int_0^1 \Big\{ F_{Y_1}^{-1}(u) - F_{Y_0}^{-1}(u) \Big\} du \leq \sigma^2_\tau \leq 
\int_0^1 \Big\{ F_{Y_1}^{-1}(u) - F_{Y_0}^{-1}(1-u) \Big\} du,
\label{eqn:vartau_bound1}
\end{equation} 
where $F_{Y_1}$ and $F_{Y_0}$ represent the empirical cumulative distribution functions of the treatment and control potential outcomes, respectively. Intuitively, the lower bound of $\sigma^2_\tau$ is reached when the potential outcomes are perfectly correlated (i.e., $\cor_\Sc(Y_i(1), Y_i(0)) = 1$). The upper bound of $\sigma^2_\tau$ is reached when the potential outcomes are perfectly anti-correlated (i.e., $\cor_\Sc(Y_i(1), Y_i(0)) = -1$). As such, researchers may use the upper bound detailed in Equation \ref{eqn:vartau_bound1} as a conservative estimate for $\sigma^2_\tau$. The bound in Equation \eqref{eqn:vartau_bound1} will \textit{always} hold. However, it can span a large range of values. If researchers are willing to impose additional assumptions, a tighter bound  on $\sigma^2_{\tau}$ can be obtained. See Appendix \ref{app:treat_het} for more details. 

\subsection{Summary of the Sensitivity Framework} \label{subsec:summary}
To summarize the sensitivity analysis framework thus far, we have parameterized the bias of a weighted estimator when omitting a confounder in the estimation of the weights with the following components: (1) an $R^2$ measure that is bounded between 0 and 1 (i.e., $R^2_\varepsilon$), (2) the correlation between the error term $\varepsilon_i$ and the individual-level treatment effect (i.e., $\rho_{\varepsilon, \tau}$), and (3) variation in the individual-level treatment effect (i.e., $\sigma^2_\tau$). We summarize this below. \\

\noindent\fbox{%
\vspace{2mm}
\parbox{0.98\textwidth}{%
\vspace{2mm}
\noindent \underline{\textbf{Summary of Sensitivity Framework for Weighted Estimators}}
\begin{Step} 
\item Estimate an upper bound for $\sigma^2_{\tau}$ (i.e., $\sigma^2_{\tau, \max}$).
\item Using $\sigma^2_{\tau, \max}$, estimate $\widehat{\cor_\Sc}^2(w_i, \tau_i)$ as a bound for $\cor_\Sc^2(w_i, \tau_i)$.
\item Vary $\rho_{\varepsilon, \tau}$ from $-\sqrt{1-\widehat{\cor_\Sc}^2(w_i, \tau_i)}$ to $\sqrt{1-\widehat{\cor_\Sc}^2(w_i, \tau_i)}$. 
\item Vary $R_\varepsilon^2$ from the range of $[0,1)$. 
\item Evaluate the bias. 
\end{Step} 
} 
} \\
\vspace{2mm}

\section{Tools for Sensitivity Analysis} \label{sec:tools} 
In the following section, we provide different tools that researchers can use to help understand the degree of sensitivity associated with a point estimate. First, we introduce two summary measures: (1) a graphical representation of sensitivity, in the form of bias contour plots, and (2) a numerical measure, referred to as a robustness value, which summarizes how much confounding must be present for an omitted confounder to result in change in the estimated effect. Second, we propose an extreme scenario analysis that evaluates an upper bound for the bias that would occur in the extreme case that the error term $\varepsilon_i$ is maximally correlated with the individual-level treatment effect. Last, we introduce a formal benchmarking approach that allows researchers to use observed covariates to calibrate their understanding of plausible parameter values. 

\subsection{Summary Measures of Sensitivity} 
We provide two approaches for researchers to summarize the sensitivity in their point estimates. The first approach is a graphical summary, while the second is a numerical measure.

\subsubsection{Graphical Summary: Contour Plots} 
A simple way to summarize and visualize the sensitivity of the point estimates is through bias contour plots (see Figure \ref{fig:jtpa_contour}). To generate the plots, the $y$-axis represents values that the correlation term can take on (i.e., the estimated range from Lemma \ref{lem:rho_bound}), and the $x$-axis represents values of $R_\varepsilon^2$ across the interval of $[0,1)$. 
 
Furthermore, we recommend researchers shade in the ``killer confounder'' region. The killer confounder region represents the set of $\{R_\varepsilon^2, \rho_{\varepsilon, \tau}\}$ values for which we expect, given an omitted confounder in this set, the bias is large enough to substantively alter the estimated effect (i.e., either changing the directional sign, or rendering the treatment effect equal to zero). If the killer confounder region is large, then there exists a greater degree of sensitivity to violations to the conditional ignorability assumption. If the region is small, there is less sensitivity.

\subsubsection{Numerical Summary: Robustness Value} 
In practice, justifying whether the killer confounder region is large or small can be challenging. As such, we propose the robustness value as a standardized, numerical summary of how sensitive a point estimate is to confounders that may change the substantive interpretation of an estimated treatment effect. This is an extension of the robustness value proposed by \cite{cinelli2020making}.  

The robustness value measures how strong a confounder must be in order for the bias to equal $100 \times q\%$ of the estimated effect:
\begin{equation} 
RV_q = \frac{1}{2} \left( \sqrt{a_q^2 + 4a_q} - a_q \right), \ \ \ \text{where } a_q = \frac{q^2 \cdot \hat \tau_W^2}{\sigma^2_{\tau} \cdot \var_\Sc(w_i)}
\label{eqn:RV} 
\end{equation} 
Evaluating the robustness value at $q=1$ provides a measure for minimum confounding strength in order for the bias to equal the point estimate, which would result in the point estimate being equal to zero. $RV_q$ is interpreted as the minimum amount of variation in treatment effect heterogeneity \textit{and} the true sample selection weights $w_i^*$, that the error term $\varepsilon_i$ must explain (i.e., $\rho^2_{\varepsilon, \tau} = R^2_\varepsilon \geq RV_q$) for the bias to be $q \times 100$\% that of the point estimate. More details and derivations are provided in Appendix \ref{app:additional_deriv}. 

A key property of the robustness value is that it exists on a scale from 0 to 1. When $RV_1$ is close to 1, then this implies that $\varepsilon_i$ must explain close to 100\% of the variation in both $\tau_i$ and $w_i^*$ for the bias to result in a zero point estimate. In contrast, if $RV_q$ is close to zero, then if $\varepsilon_i$ is able to explain a small amount of variation in both $\tau_i$ and $w_i^*$, the bias will be large enough to bring the point estimate down to zero. While the robustness value cannot rule out the possibility of a killer confounder, it can help researchers discuss the plausibility of such a confounder. 

\paragraph{Geometric Connection to Bias Contour Plots.} The robustness value is connected to the boundary of the killer confounder region. More specifically, the point of the boundary of the killer confounder region for which $\rho_{\varepsilon, \tau}^2 = R_\varepsilon^2$ will be representative of the robustness value when $q=1$ (i.e., $RV_1$). The boundary represents the set of \textit{all} potential values for which we would expect a reduction in the point estimate to 0. As such, we recommend researchers report both the robustness value and the bias contour plots when performing sensitivity analysis.

\subsubsection{Example: Sensitivity Summary Measures in JTPA}
We illustrate the proposed sensitivity summary measures in our running example. To conduct the sensitivity analysis, we use an estimated bound of 8.4 for $\sigma^2_\tau$. (Details on how $\hat \sigma^2_{\tau, \max}$ was chosen is provided in Appendix \ref{app:jtpa_sigma_tau}.) Table \ref{tbl:sensitivity_summary} provides the different sensitivity statistics:
\begin{table}[!ht] 
\centering 
\begin{tabular}{lccc} \toprule 
& Unweighted & Weighted &  $RV_{q=1}$  \\ \midrule
Impact of JTPA access on earnings$^*$ & 1.11 & 1.36 & 0.41 \\ \bottomrule
\multicolumn{4}{l}{\small $\hat \sigma^2_{\tau, \max} = 8.4$; $\widehat{\cor}_\Sc(w_i, \tau_i) = 0.07$, *-Estimates reported in thousands of USD}
\end{tabular} 
\caption{Summary of point estimates and sensitivity statistics. } 
\label{tbl:sensitivity_summary} 
\end{table} 

\noindent We see that the estimated robustness value is 0.41, which implies that the error in the weights for omitting a confounder (i.e., $\varepsilon_i$) must explain 41\% of the variation in the individual-level treatment effect, as well as 41\% of the variation in the ideal weights in order for the treatment effect to be brought down to 0. Whether or not the robustness value is large or small depends on whether researchers believe that it is plausible for the error in omitting a confounder to explain 41\% of the variation in both the ideal weights and the treatment effect heterogeneity.

\begin{figure}[!ht]
\centering
\includegraphics[width=0.55\textwidth]{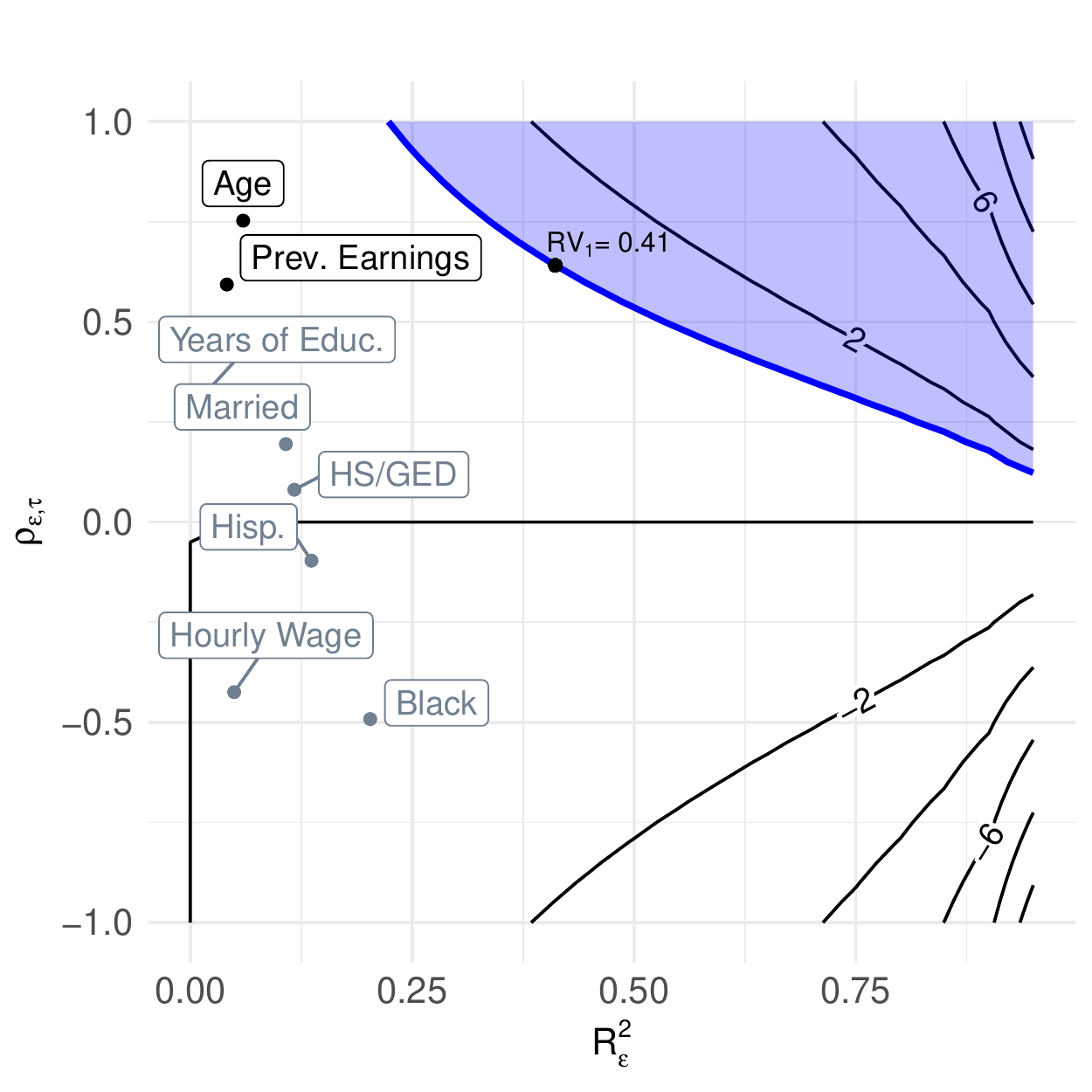}
\caption{Bias Contour Plot for Omaha, Nebraska. The blue region represents the killer confounder region. Whether or not the killer confounder region is large or small (and by extension, the robustness value) can be difficult to justify. To aid in our discussion, we use formal benchmarking (introduced in Section \ref{sec:calib}) to estimate the parameter values for an omitted confounder with similar confounding strength as an observed covariate. We see that the benchmarked points are far away from the killer confounder region, which implies that an omitted confounder would have to be much stronger than an observed covariate to change the direction of the point estimate.} 
\label{fig:jtpa_contour} 
\end{figure}

We also examine a bias contour plot, in which we shade in the killer confounder region (see Figure \ref{fig:jtpa_contour}). In the context of the example, the killer confounder region represents the part of the plot in which the bias is large enough to reduce the estimated impact of JTPA access on earnings to zero. The boundary of the killer confounder region visualizes the full set of $\{R^2_\varepsilon, \rho_{\varepsilon, \tau} \}$ that corresponds to a confounder strong enough to reduce the estimated treatment effect to zero. For example, an omitted confounder that results in an error term that explains very little variation in the ideal weights (i.e., $R^2_\varepsilon = 0.25$), but explains a large amount of variation in the individual-level treatment effect (i.e., $\rho_{\varepsilon, \tau} = 0.93$) would be a killer confounder. Similarly, a confounder that results in an error term that explains a large amount of the variation in the ideal weights (i.e., $R^2_\varepsilon = 0.75$), but a small portion of the variation in the individual-level treatment effect (i.e., $\rho_{\varepsilon, \tau} = 0.31$) would also be a killer a confounder. 

\subsection{Extreme Scenario Analysis} \label{subsubsec:extreme} 
We propose an extreme scenario analysis for researchers to evaluate the bias when the error term $\varepsilon_i$ is maximally correlated to the individual-level treatment effect. Under this scenario, the maximum values that $\rho_{\varepsilon, \tau}$ and $R^2_\varepsilon$ (referred to as $\rho_{max}$ and $R^2_{max}$, respectively) can take on will be a function of $1-\cor_\Sc(w_i, \tau_i)^2$: 
$$\rho^2_{max} = R^2_{max} = 1- \cor_\Sc(w_i, \tau_i)^2$$
As such, evaluating the bias at $\big(\rho_{max}, R^2_{max} \big)$ will result in an upper bound on the bias. 

In practice, this will be an extremely conservative estimate of the bias. More specifically, \cite{miratrix_2018} demonstrated that the sample selection weights estimated in their survey data were weakly correlated with the treatment effect heterogeneity. For example, in the JTPA application, calculating the extreme scenario bound using our conservative estimate of $\cor_\Sc(w_i, \tau_i)$ results in a bound of 0.99 for the maximum value $\rho_{\varepsilon, \tau}$ and $R^2_{\varepsilon}$. The extreme scenario would arise if the error term explained 99\% of the variation in the individual-level treatment effect and the ideal weights. Researchers can choose to evaluate less conservative estimates of $\big(\rho_{max}, R^2_{max} \big)$ by relaxing how much variation in the treatment effect they believe the true weights $w_i^*$ can explain. More detail is provided in Appendix \ref{app:extreme_scenario}. 

\subsection{Formal Benchmarking to Infer Reasonable Parameters} \label{sec:calib}
A challenge in sensitivity analysis is positing reasonable values for the sensitivity parameters to take on. In the previous section, we showed that all three sensitivity parameters are bounded across finite ranges. This restricts the set of potential parameter values, but the question of what parameter values \textit{within} those ranges are plausible remains open. Furthermore, justifying whether the killer confounder region of a bias contour plot, or the robustness value, is large or small can be challenging in practice. In the following subsection, we introduce a formal benchmarking approach for researchers to use observed covariates to calibrate their understanding of plausible parameter values using relative strength. 

To begin, let $\bX^{(j)}$ be an observed covariate (i.e., $\bX^{(j)} \in \{\bX\}$, and $\bX \in \R^{n\times p}$, $j \in \{1, ..., p\}$). Define $\varepsilon_i^{-(j)}$ as the error term that compares the weights estimated using all covariates $\bX_i$ with the weights estimated using all the covariates, except for $\bX_i^{(j)}$: 
\begin{equation} 
\varepsilon^{-(j)}_{i} := w^{-(j)}_{i} - w_i,
\label{eqn:informal_eps} 
\end{equation} 
where $w_{i}^{-(j)}$ is the set of weights estimated using all the covariates $\bX_i$, except for $\bX_i^{(j)}$, and $w_i$ is the set of weights estimated using all available covariates $\bX_i$. 

We define the amount of confounding strength an omitted confounder has by how much variation $\varepsilon_i$ explains in the ideal weights $w_i^*$ and the individual-level treatment effect $\tau_i$. Thus, to obtain formal benchmarks, we posit the amount of variation explained in $w_i^*$ and $\tau_i$ by $\varepsilon_i$, in comparison to $\varepsilon_i^{-(j)}$. More formally, define: 
\begin{equation} 
k_\sigma = \frac{\var_\Sc(\varepsilon_i)/\var_\Sc(w_i^*)}{\var_\Sc(\varepsilon_i^{-(j)})/\var_\Sc(w_i^*)}, \ \ \ \ \ \ k_\rho = \frac{\cor_\Sc(\varepsilon_i, \tau_i)}{\cor_\Sc(\varepsilon_i^{-(j)}, \tau_i)}, 
\label{eqn:k} 
\end{equation}
where the numerators (i.e., $\var_\Sc(\varepsilon_i)/\var_\Sc(w_i^*)$ and $\cor_\Sc(\varepsilon_i, \tau_i)$) correspond to the sensitivity parameters introduced in Section \ref{sec:weighted}. $k_\sigma$ represents how much relative variation in the true sample selection weights $w_i^*$ the error term $\varepsilon_i$ explains, relative to the observed $\varepsilon_i^{-(j)}$ term. If the residual imbalance in the omitted confounder $\bU_i$ is greater than the observed residual imbalance in the covariate $\bX_i^{(j)}$, then we expect $k_\sigma > 1$. $k_\rho$ represents correlated the individual-level treatment effect and the error term $\varepsilon_i$ are, relative to $\varepsilon_i^{-(j)}$. $k_\sigma$ and $k_\rho$ intuitively represent the relative confounding strength of an observed covariate. When $k_\sigma = k_\rho = 1$, then we say that an omitted confounder has equivalent confounding strength to an observed covariate. 

With a researcher-specified $k_\sigma$ and $k_\rho$, we obtain the formally benchmarked sensitivity parameters. Theorem \ref{thm:benchmark} formalizes this. 
\begin{theorem}[Formal Benchmarking for Sensitivity Parameters]\label{thm:benchmark} \mbox{}\\
Let $k_\sigma$ and $k_\rho$ be defined as in Equation \eqref{eqn:k}. Let $R^{2-(j)}_\varepsilon := \var_\Sc(\varepsilon_i^{-(j)})/\var_\Sc(w_i)$, and $\rho_{\varepsilon, \tau}^{-(j)} := \cor_\Sc(\varepsilon_i^{-(j)}, \tau_i)$. The sensitivity parameters $R^2_\varepsilon$ and $\rho_{\varepsilon, \tau}$ can be written as a function of $k_\sigma$ and $k_\rho$: 
$$R^2_\varepsilon = \frac{k_\sigma \cdot R^{2-(j)}_\varepsilon}{1+ k_\sigma \cdot R^{2-(j)}_\varepsilon}, \ \ \ \ \ \  \rho_{\varepsilon, \tau} = k_\rho \cdot \rho_{\varepsilon, \tau}^{-(j)} $$ 
\end{theorem} 
\paragraph{Remark on Informal Benchmarking:} There are connections between the proposed approach and informal benchmarking approaches (e.g., see \cite{imbens2003sensitivity, carnegie2016assessing, hong2021did}).  Informal benchmarking directly uses $R^{2-(j)}_\varepsilon$ and $\rho_{\varepsilon, \tau}^{-(j)}$ to calibrate the sensitivity parameters. From Theorem \ref{thm:benchmark}, we see that to recover $R^2_\varepsilon$, we have to adjust for the change in baseline variation by scaling $R^{2-(j)}_\varepsilon$ by $1+k_\sigma \cdot R^{2-(j)}_\varepsilon$. As such, using just $R^{2-(j)}$ as an estimate for $R^2_\varepsilon$ could result in misleading estimates. \\

Theorem \ref{thm:benchmark} provides a way for researchers to estimate parameter values for an omitted confounder, after specifiying the confounding strength, relative to an observed covariate. There are several key takeaways to highlight. First,  Theorem \ref{thm:benchmark} can be extended for a subset of covariates. This is helpful if researchers believe that a subset of observed covariates (or interactions) is particularly important to explaining the sample selection process or treatment effect heterogeneity, and wish to assess the effect of omitting a confounder with similar strength to the entire group of covariates. Secondly, because both $R^2_\varepsilon$ and $\rho_{\varepsilon, \tau}$ are inherently bounded, $k_\sigma$ and $k_\rho$ will also be bounded. As such, researchers can estimate the maximum confounding strength of an omitted confounder, relative to an observed covariate. Finally, in addition to providing a better understanding of potential parameter values, formal benchmarking can be used to assess the plausibility of killer confounders. We elaborate on this final point in the following subsection. 

\subsubsection{Using Benchmarking to Understand Killer Confounders}
\paragraph{Comparing benchmarked bias with point estimate: } Benchmarking the sensitivity parameters allows researchers to estimate the resulting bias a confounder with fixed relative confounding strength as a covariate. A natural measure for how much relative confounding strength an omitted variable must have to result in a killer confounder is found by dividing the point estimate with the estimated bias when $k_\rho = k_\sigma = 1$. We refer to this as the \textit{minimum relative confounding strength} (MRCS): 
\begin{equation} 
\text{MRCS}(\bX_i^{-(j)}) = \frac{\hat \tau_W}{\widehat{\text{Bias}}(\varepsilon_i^{-(j)}, k_\rho = 1, k_\sigma = 1)}.
\label{eqn:rel_bias} 
\end{equation} 
If the estimated MRCS is small (i.e., MRCS $< 1$), then this implies that an omitted confounder, with weak confounding strength, relative to the covariate $\bX_i^{-(j)}$, could lead to the point estimate being reduced to zero. On the other hand, if MRCS is large (i.e., MRCS $> 1$), then this indicates that an omitted confounder must be stronger than the observed covariate to result in a killer confounder. MRCS is an especially helpful measure when researchers have strong substantive priors for what may be important covariates.

\paragraph{Comparing benchmarking results with $RV_q$:} From benchmarking, researchers can estimate the necessary $k_\rho$ and $k_\sigma$ values in order for $R^2_\varepsilon = \rho_{\varepsilon, \tau}^2 = RV_q$. We denote these values as $\{k_\rho^{min}, k_\sigma^{min}\}$. If the estimated $k_\rho$ or $k_\sigma$ values are both greater than 1, then this implies that the error from omitting the confounder would have to explain more of the variation in the sample selection process and the treatment effect heterogeneity than the observed covariate, in order to be a killer confounder. The interpretation of $k_\rho^{min}$ and $k_\sigma^{min}$ is similar to that of the MRCS; however, researchers can now look at the drivers of bias with respect to the confounder's relationship to the sample selection process and treatment effect heterogeneity separately. 

\subsubsection{Example: Applying Formal Benchmarking in JTPA} 
To help assess plausible sensitivity parameters in the JTPA application, we perform formal benchmarking. Table \ref{tbl:benchmarking} presents the results. For each of the covariates included in the weights, we estimate $R^2_\varepsilon$ and $\rho_{\varepsilon, \tau}$, the MRCS, and $\{k_\sigma^{min}, k_\rho^{min}\}$. We see that omitting a confounder with equivalent confounding strength to the covariates age, whether or not the individual is black, or previous earnings will result in the largest amount of bias. This is consistent with the substantive findings from the original study, which reported strong subgroup effects when looking at race, age, and previous earnings (\cite{bloom1993national}). 
\begin{table}[ht]
\centering
\begin{tabular}{lcccccc}
  \toprule
Covariate & $R^2_\varepsilon$ & $\rho_{\varepsilon, \tau}$ & Est. Bias & MRCS & $k_\sigma^{min}$ & $k_\rho^{min}$ \\ 
  \midrule
Prev. Earnings & 0.04 & 0.59 & 0.31 & 4.4 & 10.0 & 1.1 \\ 
  Age & 0.06 & 0.75 & 0.48 & 2.83 & 7.0 & 0.9 \\ 
  Married & 0.11 & 0.19 & 0.17 & 7.9 & 3.8 & 3.3 \\ 
  Hourly Wage & 0.05 & -0.42 & -0.25 & -5.5 & 8.3 & -1.5 \\ 
  Black & 0.20 & -0.49 & -0.63 & -2.2 & 2.0 & -1.3 \\ 
  Hisp. & 0.14 & -0.10 & -0.10 & -14.0 & 3.0 & -6.7 \\ 
  HS/GED & 0.12 & 0.08 & 0.07 & 18.2 & 3.5 & 7.9 \\ 
  Years of Educ. & 0.00 & 0.28 & 0.02 & 59.7 & 409.0 & 2.3 \\ 
   \bottomrule
   \multicolumn{6}{l}{\small Point Estimate ($\hat \tau_W$): $1.36$; $\hat \sigma^2_{\tau, \max} = 8.4$; $RV_1 = 0.41$}
\end{tabular}
\caption{Formal benchmarking results for Omaha, Nebraska. The estimated bias is reported in thousands of USD.}
\label{tbl:benchmarking}
\end{table}

Using the results from formal benchmarking, we estimate the MRCS for each covariate. We see that a confounder would have to be over twice as strong as the covariate for whether or not the individual is black, or $3-4$ times as strong as either age or previous earnings in order to be a killer confounder. Similarly, we compare the robustness value to the benchmarked sensitivity parameters. Most of the $k_\sigma^{min}$ and $k_\rho^{min}$ values are greater than 1. In particular, all of the $k_\sigma^{min}$ values are larger than 2, which implies that the error from omitting an unobserved confounder would have to explain at least twice as much variation in the ideal weights than the error from omitting an observed covariate to be a killer confounder. As such, we conclude that in order for an omitted confounder to be a killer confounder, it would have to be substantially stronger than any of the observed covariates. This can be visualized on the bias contour plots by plotting each of the benchmarked parameter values (see Figure \ref{fig:jtpa_contour}). 

\paragraph{Comparison with Actual Error.} From the sensitivity analysis, we conclude that the estimated treatment effects are quite robust to confounders. Because of the nature of our empirical application, we can compare the point estimate with the ``true'' benchmark PATE (i.e., the estimated average treatment effect across the other 15 sites). The true benchmark PATE is
\$1,250, which is similar to the weighted estimate of \$1,360. As such, we see that in alignment with the results from the sensitivity analysis, there was minimal bias in the weighted estimator. 

The experimental site used in this running example was shown to be relatively robust to confounders. An example of a sensitivity analysis for a similarly sized experimental site with a greater degree of sensitivity to confounders is provided in Appendix \ref{app:jtpa}.

%%%%%%%%%%%%%%%%%%%%%%%%%%%%%%%%%%%%%%%%%%%%%%%%%%%%%%%%%%%%%%%%%%%%%%%%%%%%%%%%%%%%%%%%%%%%%%%%%%%%%%%%%%%%%%%%%%%%%%%%
\section{Extension for Augmented Weighted Estimators} \label{sec:aug} 
In the following section, we extend the proposed sensitivity analysis for the class of augmented weighted, doubly robust estimators. Doubly robust estimators are a popular approach used to help improve the robustness of estimators to potential misspecfications (\cite{dahabreh2019generalizing, tan2007comment, robins1994estimation, bang2005doubly}). There are many different doubly robust estimators (\cite{kang2007demystifying}), but we will focus on the augmented weighted estimator:
\begin{definition}[Augmented Weighted Estimator] 
\begin{align*} 
\hat \tau^{Aug}_W = \hat \tau_W - \underbrace{\frac{1}{n} \sum_{i \in \Sc} w_i \hat \tau(\bX_i) + \frac{1}{N} \sum_{i \in \mathcal{P}} \hat \tau(\bX_i)}_{\text{Augmented Component}}
\end{align*} 

\end{definition} 
where $\mathcal{P}$ represents the set of indices of the units in the target population, $\hat \tau(\bX_i)$ is the estimated individual-level treatment effect, and the weights are defined in the same manner as before. Doubly robust estimators, like the augmented weighted estimator, allow practitioners to model both the probability of sample selection and the treatment effect heterogeneity simultaneously. When one of these processes is specified correctly, then the estimator will be unbiased and asymptotically consistent. 

In the following section, we introduce a sensitivity analysis for the augmented weighted estimator when omitting a confounder from the minimum separating set. We show that there are strong parallels between the sensitivity analysis for the augmented weighted estimator and the sensitivity analysis for the weighted estimator.

\subsection{Bias Formula}
To begin, we show that the bias of an augmented weighted estimator when omitting a variable from the minimum separating set can be written as a function of three components: (1) the correlation between the error in the weights and the error in modeling the treatment effect heterogeneity, (2) residual imbalance in the omitted confounder, and (3) the variation in the residual individual-level treatment effect, not accounted for by $\hat \tau(\bX_i)$. We formalize this in the following theorem. 

\begin{theorem}[Bias of Augmented Weighted Estimator] \mbox{}\\
The bias of an augmented weighted estimator when a variable has been omitted from the minimum separating set (resulting in both the weight and treatment effect heterogeneity being misspecified): 
\begin{align} 
\text{Bias}(\hat \tau_{W}^{Aug})
&= \rho_{\varepsilon, \xi} \cdot \sqrt{\var(w_i) \frac{R^2_\varepsilon}{1-R^2_\varepsilon}\cdot \sigma^2_\xi}
\label{eqn:bias_dr} 
\end{align} 
where $\xi_i$ represents the difference between the true individual-level treatment effect and estimated treatment effect (i.e., $\xi_i = \tau_i - \hat \tau(\bX_i)$). 
\label{thm:dr} 
\end{theorem} 
There are several key takeaways from Theorem \ref{thm:dr}. First, the double robustness of the augmented weighted estimator is apparent from Theorem \ref{thm:dr} by noting that if there is no error in the estimated weights (i.e., $\varepsilon_i = 0$), or there is no error in estimating the treatment effect heterogeneity (i.e., $\hat \tau(\bX_i)$ is a consistent model for $\tau_i$), then $\xi_i$ will be made up of random noise, and the correlation between $\xi_i$ and $\varepsilon_i$ will be zero (i.e., $\rho_{\varepsilon, \xi} =0$). Second, Theorem \ref{thm:dr} highlights that the bias of an augmented weighted estimator from omitting a confounder is very similar to the bias of a weighted estimator (i.e., Equation \eqref{eqn:bias_formula}). The primary difference is that instead of the individual-level treatment effect $\tau_i$, we are interested in $\xi_i$, which is the residual component of $\tau_i$ that cannot be explained by $\hat \tau(\bX_i)$. 

Theorem \ref{thm:dr} provides a very flexible formulation of the bias that can be adapted to different frameworks. For example, if we are not modeling the individual-level treatment effect and are focused only on estimating weights, then $\hat \tau(\bX_i)$ is equal to zero for all $i$, and $\var(\xi_i) = \var(\tau_i) \equiv \sigma^2_{\tau}$. Thus, the resulting bias formula is equal to the bias formula from Theorem \ref{thm:weight}. Similarly, researchers can adapt Theorem \ref{thm:dr} to the case where they are not re-weighting the data at hand, and are focused solely on modeling the individual-level treatment effect $\hat \tau(\bX_i)$. If we assume that the individual-level treatment effect follows a linear model, then we can recover the results from \cite{nguyen2017sensitivity} (see Appendix \ref{app:aug_tools} for more details). In other words, previously proposed sensitivity analysis frameworks that rely on parametric assumptions are special cases of our proposed bias decomposition. In cases when researchers do not wish to impose parametric assumptions, Theorem \ref{thm:dr} provides a flexible approach for sensitivity analysis. 

\subsection{Sensitivity Analysis for Augmented Weighted Estimators} 
In the previous subsection, we showed that the primary differentiation between the bias formula for the augmented weighted estimator and the weighted estimator is $\xi_i$ (i.e., the residuals in the treatment effect model). This results in two new parameters in the augmented weighted estimator setting: $\rho_{\varepsilon, \xi}$ and $\sigma^2_\xi$. The third parameter in the bias decomposition is $R^2_\varepsilon$, which is identical in both the weighted and augmented weighted estimator setting. We show in Appendix \ref{app:aug_tools} that similar bounds to the ones derived in Section \ref{sec:weighted} apply to this setting. As such, after estimating an adequate upper bound for $\sigma^2_\xi$, researchers may vary both $R^2_\varepsilon$ and $\rho_{\varepsilon, \xi}$ across bounded ranges to assess the sensitivity of an augmented weighted estimator to omitted confounders. Similarly, the sensitivity tools in Section \ref{sec:tools} can also be extended for the augmented weighted estimator case. Details are provided in Appendix \ref{app:aug_tools}, and Appendix \ref{app:jtpa_aug} illustrates the sensitivity analysis using JTPA. 

\section{Conclusion} \label{sec:conclusion} 
Generalizing or transporting causal effects from an experiment to a different, or larger, population requires researchers to correctly identify a separating set of pre-treatment covariates that allow the confounding effect of sample selection to be conditionally ignorable. When this separating set is not correctly identified, PATE estimation will be biased.

In this paper, we formalize a sensitivity analysis framework for weighted estimators in the generalization or transportability setting, with extensions for augmented weighted estimators. We demonstrate that the proposed framework is a more general version of previously proposed sensitivity analysis frameworks. Furthermore, we introduce a set of tools for both the weighted and augmented weighted estimators that allow researchers to quantitatively and graphically summarize how much sensitivity there is in their estimation.

The proposed framework has several advantages to existing approaches. First, it is agnostic to the underlying data generating process, and can be used for any set of weighted or augmented weighted estimators, regardless of how researchers choose to estimate the weights or the individual-level treatment effect model. Second, the sensitivity parameters in the framework are all bounded on a finite scale, which removes the onus on researchers to justify plausible ranges for the parameters. Third, we propose a set of sensitivity analysis tools to help researchers understand and summarize the degree of sensitivity that is present in their estimation. We introduce two summary measures, and demonstrate that the proposed sensitivity parameters can be bounded in an extreme scenario analysis, allowing researchers to quantify worst-case scenarios for their estimates. Furthermore, we introduce a formal benchmarking approach for researchers to use observed covariates to posit plausible sensitivity parameter values. While the focus of this paper is on the bias that arises from an omitted confounder, we note that researchers can fix the different sensitivity parameters, and then perform bootstrapping to measure uncertainty associated with point estimates.

Finally, in concluding this paper, it is important to emphasize the limits of the sensitivity tools. The proposed sensitivity framework provides researchers with different quantitative and graphical measures to assess the degree of robustness that is present in their point estimate. However, these tools cannot be used to \textit{eliminate} the possibility of killer confounders, and akin to \cite{cinelli2020making}, we do not provide cutoff measures for measures such as the robustness value or the minimum relative confounding strength. We caution researchers from using these tools without also considering substantive judgment. 
The sensitivity framework provides a strong foundation for researchers to discuss the plausibility of killer confounders, but should not be used in lieu of substantive understanding of the underlying covariates and context.

\clearpage
%BIBLIOGRAPHY: 
\bibliographystyle{chicago} 
\bibliography{bibliography}

\begin{thebibliography}{}

\bibitem[\protect\citeauthoryear{Athey, Tibshirani, Wager, et~al.}{Athey
  et~al.}{2019}]{athey2019generalized}
Athey, S., J.~Tibshirani, S.~Wager, et~al. (2019).
\newblock Generalized random forests.
\newblock {\em The Annals of Statistics\/}~{\em 47\/}(2), 1148--1178.

\bibitem[\protect\citeauthoryear{Bang and Robins}{Bang and
  Robins}{2005}]{bang2005doubly}
Bang, H. and J.~M. Robins (2005).
\newblock Doubly robust estimation in missing data and causal inference models.
\newblock {\em Biometrics\/}~{\em 61\/}(4), 962--973.

\bibitem[\protect\citeauthoryear{Ben-Michael, Hirschberg, Feller, and
  Zubizarreta}{Ben-Michael et~al.}{2020}]{ben2020balancing}
Ben-Michael, E., D.~Hirschberg, A.~Feller, and J.~Zubizarreta (2020).
\newblock The balancing act for causal inference.

\bibitem[\protect\citeauthoryear{Bloom et~al.}{Bloom
  et~al.}{1993}]{bloom1993national}
Bloom, H.~S. et~al. (1993).
\newblock The national jtpa study. title ii-a impacts on earnings and
  employment at 18 months.

\bibitem[\protect\citeauthoryear{Brumback, Hern{\'a}n, Haneuse, and
  Robins}{Brumback et~al.}{2004}]{brumback2004sensitivity}
Brumback, B.~A., M.~A. Hern{\'a}n, S.~J. Haneuse, and J.~M. Robins (2004).
\newblock Sensitivity analyses for unmeasured confounding assuming a marginal
  structural model for repeated measures.
\newblock {\em Statistics in medicine\/}~{\em 23\/}(5), 749--767.

\bibitem[\protect\citeauthoryear{Buchanan, Hudgens, Cole, Mollan, Sax, Daar,
  Adimora, Eron, and Mugavero}{Buchanan
  et~al.}{2018}]{buchanan2018generalizing}
Buchanan, A.~L., M.~G. Hudgens, S.~R. Cole, K.~R. Mollan, P.~E. Sax, E.~S.
  Daar, A.~A. Adimora, J.~J. Eron, and M.~J. Mugavero (2018).
\newblock Generalizing evidence from randomized trials using inverse
  probability of sampling weights.
\newblock {\em Journal of the Royal Statistical Society: Series A (Statistics
  in Society)\/}~{\em 181\/}(4), 1193--1209.

\bibitem[\protect\citeauthoryear{Burgette, Griffin, Pane, and
  McCaffrey}{Burgette et~al.}{2020}]{burgette}
Burgette, L., B.~A. Griffin, J.~Pane, and D.~F. McCaffrey (2020).
\newblock Efficient sensitivity analysis for propensity score weighting.
\newblock Joint Statistical Meetings.

\bibitem[\protect\citeauthoryear{Carnegie, Harada, and Hill}{Carnegie
  et~al.}{2016}]{carnegie2016assessing}
Carnegie, N.~B., M.~Harada, and J.~L. Hill (2016).
\newblock Assessing sensitivity to unmeasured confounding using a simulated
  potential confounder.
\newblock {\em Journal of Research on Educational Effectiveness\/}~{\em
  9\/}(3), 395--420.

\bibitem[\protect\citeauthoryear{Cinelli and Hazlett}{Cinelli and
  Hazlett}{2020}]{cinelli2020making}
Cinelli, C. and C.~Hazlett (2020).
\newblock Making sense of sensitivity: Extending omitted variable bias.
\newblock {\em Journal of the Royal Statistical Society: Series B (Statistical
  Methodology)\/}~{\em 82\/}(1), 39--67.

\bibitem[\protect\citeauthoryear{Cole and Stuart}{Cole and
  Stuart}{2010}]{cole2010generalizing}
Cole, S.~R. and E.~A. Stuart (2010).
\newblock Generalizing evidence from randomized clinical trials to target
  populations: The actg 320 trial.
\newblock {\em American journal of epidemiology\/}~{\em 172\/}(1), 107--115.

\bibitem[\protect\citeauthoryear{Dahabreh, Robertson, Tchetgen, Stuart, and
  Hern{\'a}n}{Dahabreh et~al.}{2019}]{dahabreh2019generalizing}
Dahabreh, I.~J., S.~E. Robertson, E.~J. Tchetgen, E.~A. Stuart, and M.~A.
  Hern{\'a}n (2019).
\newblock Generalizing causal inferences from individuals in randomized trials
  to all trial-eligible individuals.
\newblock {\em Biometrics\/}~{\em 75\/}(2), 685--694.

\bibitem[\protect\citeauthoryear{Ding, Feller, and Miratrix}{Ding
  et~al.}{2019}]{ding2019decomposing}
Ding, P., A.~Feller, and L.~Miratrix (2019).
\newblock Decomposing treatment effect variation.
\newblock {\em Journal of the American Statistical Association\/}~{\em
  114\/}(525), 304--317.

\bibitem[\protect\citeauthoryear{Djebbari and Smith}{Djebbari and
  Smith}{2008}]{djebbari2008heterogeneous}
Djebbari, H. and J.~Smith (2008).
\newblock Heterogeneous impacts in progresa.
\newblock {\em Journal of Econometrics\/}~{\em 145\/}(1-2), 64--80.

\bibitem[\protect\citeauthoryear{Dorn and Guo}{Dorn and
  Guo}{2021}]{dorn2021sharp}
Dorn, J. and K.~Guo (2021).
\newblock Sharp sensitivity analysis for inverse propensity weighting via
  quantile balancing.
\newblock {\em arXiv preprint arXiv:2102.04543\/}.

\bibitem[\protect\citeauthoryear{Egami and Hartman}{Egami and
  Hartman}{2019}]{egami2019covariate}
Egami, N. and E.~Hartman (2019).
\newblock Covariate selection for generalizing experimental results.
\newblock {\em arXiv preprint arXiv:1909.02669\/}.

\bibitem[\protect\citeauthoryear{Egami and Hartman}{Egami and
  Hartman}{2020}]{egami2020elements}
Egami, N. and E.~Hartman (2020).
\newblock Elements of external validity: Framework, design, and analysis.
\newblock {\em Design, and Analysis (June 30, 2020)\/}.

\bibitem[\protect\citeauthoryear{Fogarty and Hasegawa}{Fogarty and
  Hasegawa}{2019}]{fogarty2019extended}
Fogarty, C.~B. and R.~B. Hasegawa (2019).
\newblock Extended sensitivity analysis for heterogeneous unmeasured
  confounding with an application to sibling studies of returns to education.
\newblock {\em The Annals of Applied Statistics\/}~{\em 13\/}(2), 767--796.

\bibitem[\protect\citeauthoryear{Fr{\'e}chet}{Fr{\'e}chet}{1951}]{frechet1951tableaux}
Fr{\'e}chet, M. (1951).
\newblock Sur les tableaux de corr{\'e}lation dont les marges sont donn{\'e}es.
\newblock {\em Ann. Univ. Lyon, 3\^{} e serie, Sciences, Sect. A\/}~{\em 14},
  53--77.

\bibitem[\protect\citeauthoryear{Hainmueller}{Hainmueller}{2012}]{hainmueller2012entropy}
Hainmueller, J. (2012).
\newblock Entropy balancing for causal effects: A multivariate reweighting
  method to produce balanced samples in observational studies.
\newblock {\em Political analysis\/}~{\em 20\/}(1), 25--46.

\bibitem[\protect\citeauthoryear{Hartman, Hazlett, and Sterbenz}{Hartman
  et~al.}{2021}]{hartman2021kpop}
Hartman, E., C.~Hazlett, and C.~Sterbenz (2021).
\newblock Kpop: A kernel balancing approach for reducing specification
  assumptions in survey weighting.
\newblock {\em arXiv preprint arXiv:2107.08075\/}.

\bibitem[\protect\citeauthoryear{Heckman, Smith, and Clements}{Heckman
  et~al.}{1997}]{heckman1997making}
Heckman, J.~J., J.~Smith, and N.~Clements (1997).
\newblock Making the most out of programme evaluations and social experiments:
  Accounting for heterogeneity in programme impacts.
\newblock {\em The Review of Economic Studies\/}~{\em 64\/}(4), 487--535.

\bibitem[\protect\citeauthoryear{Hill}{Hill}{2011}]{hill2011bayesian}
Hill, J.~L. (2011).
\newblock Bayesian nonparametric modeling for causal inference.
\newblock {\em Journal of Computational and Graphical Statistics\/}~{\em
  20\/}(1), 217--240.

\bibitem[\protect\citeauthoryear{Hoeffding}{Hoeffding}{1941}]{hoeffding1941masstabinvariante}
Hoeffding, W. (1941).
\newblock Masstabinvariante korrelationsmasse f{\"u}r diskontinuierliche
  verteilungen.
\newblock {\em Archiv f{\"u}r mathematische Wirtschafts-und
  Sozialforschung\/}~{\em 7}, 49--70.

\bibitem[\protect\citeauthoryear{Hong, Yang, and Qin}{Hong
  et~al.}{2021}]{hong2021did}
Hong, G., F.~Yang, and X.~Qin (2021).
\newblock Did you conduct a sensitivity analysis? a new weighting-based
  approach for evaluations of the average treatment effect for the treated.
\newblock {\em Journal of the Royal Statistical Society: Series A (Statistics
  in Society)\/}~{\em 184\/}(1), 227--254.

\bibitem[\protect\citeauthoryear{Huang, Egami, Hartman, and Miratrix}{Huang
  et~al.}{2021}]{huang2021leveraging}
Huang, M., N.~Egami, E.~Hartman, and L.~Miratrix (2021).
\newblock Leveraging population outcomes to improve the generalization of
  experimental results.
\newblock {\em arXiv preprint arXiv:2111.01357\/}.

\bibitem[\protect\citeauthoryear{Ichino, Mealli, and Nannicini}{Ichino
  et~al.}{2008}]{ichino2008temporary}
Ichino, A., F.~Mealli, and T.~Nannicini (2008).
\newblock From temporary help jobs to permanent employment: what can we learn
  from matching estimators and their sensitivity?
\newblock {\em Journal of applied econometrics\/}~{\em 23\/}(3), 305--327.

\bibitem[\protect\citeauthoryear{Imai, King, and Stuart}{Imai
  et~al.}{2008}]{imai2008misunderstandings}
Imai, K., G.~King, and E.~A. Stuart (2008).
\newblock Misunderstandings between experimentalists and observationalists
  about causal inference.
\newblock {\em Journal of the royal statistical society: series A (statistics
  in society)\/}~{\em 171\/}(2), 481--502.

\bibitem[\protect\citeauthoryear{Imbens}{Imbens}{2003}]{imbens2003sensitivity}
Imbens, G.~W. (2003).
\newblock Sensitivity to exogeneity assumptions in program evaluation.
\newblock {\em American Economic Review\/}~{\em 93\/}(2), 126--132.

\bibitem[\protect\citeauthoryear{Imbens and Rubin}{Imbens and
  Rubin}{2015}]{imbens2015causal}
Imbens, G.~W. and D.~B. Rubin (2015).
\newblock {\em Causal inference in statistics, social, and biomedical
  sciences}.
\newblock Cambridge University Press.

\bibitem[\protect\citeauthoryear{Josey, Berkowitz, Ghosh, and Raghavan}{Josey
  et~al.}{2021}]{josey2021transporting}
Josey, K.~P., S.~A. Berkowitz, D.~Ghosh, and S.~Raghavan (2021).
\newblock Transporting experimental results with entropy balancing.
\newblock {\em Statistics in Medicine\/}.

\bibitem[\protect\citeauthoryear{Josey, Yang, Ghosh, and Raghavan}{Josey
  et~al.}{2020}]{josey2020calibration}
Josey, K.~P., F.~Yang, D.~Ghosh, and S.~Raghavan (2020).
\newblock A calibration approach to transportability with observational data.
\newblock {\em arXiv preprint arXiv:2008.06615\/}.

\bibitem[\protect\citeauthoryear{Kang, Schafer, et~al.}{Kang
  et~al.}{2007}]{kang2007demystifying}
Kang, J.~D., J.~L. Schafer, et~al. (2007).
\newblock Demystifying double robustness: A comparison of alternative
  strategies for estimating a population mean from incomplete data.
\newblock {\em Statistical science\/}~{\em 22\/}(4), 523--539.

\bibitem[\protect\citeauthoryear{Kent, Paulus, Van~Klaveren, D'Agostino,
  Goodman, Hayward, Ioannidis, Patrick-Lake, Morton, Pencina, et~al.}{Kent
  et~al.}{2020}]{kent2020predictive}
Kent, D.~M., J.~K. Paulus, D.~Van~Klaveren, R.~D'Agostino, S.~Goodman,
  R.~Hayward, J.~P. Ioannidis, B.~Patrick-Lake, S.~Morton, M.~Pencina, et~al.
  (2020).
\newblock The predictive approaches to treatment effect heterogeneity (path)
  statement.
\newblock {\em Annals of internal medicine\/}~{\em 172\/}(1), 35--45.

\bibitem[\protect\citeauthoryear{Kern, Stuart, Hill, and Green}{Kern
  et~al.}{2016}]{kern2016assessing}
Kern, H.~L., E.~A. Stuart, J.~Hill, and D.~P. Green (2016).
\newblock Assessing methods for generalizing experimental impact estimates to
  target populations.
\newblock {\em Journal of research on educational effectiveness\/}~{\em
  9\/}(1), 103--127.

\bibitem[\protect\citeauthoryear{Lu, Ben-Michael, Feller, and Miratrix}{Lu
  et~al.}{2021}]{lu2021you}
Lu, B., E.~Ben-Michael, A.~Feller, and L.~Miratrix (2021).
\newblock Is it who you are or where you are? accounting for compositional
  differences in cross-site treatment variation.
\newblock {\em arXiv preprint arXiv:2103.14765\/}.

\bibitem[\protect\citeauthoryear{Lunceford and Davidian}{Lunceford and
  Davidian}{2004}]{lunceford2004stratification}
Lunceford, J.~K. and M.~Davidian (2004).
\newblock Stratification and weighting via the propensity score in estimation
  of causal treatment effects: a comparative study.
\newblock {\em Statistics in medicine\/}~{\em 23\/}(19), 2937--2960.

\bibitem[\protect\citeauthoryear{Miratrix, Sekhon, Theodoridis, and
  Campos}{Miratrix et~al.}{2018}]{miratrix_2018}
Miratrix, L.~W., J.~S. Sekhon, A.~G. Theodoridis, and L.~F. Campos (2018).
\newblock Worth weighting? how to think about and use weights in survey
  experiments.
\newblock {\em Political Analysis\/}~{\em 26\/}(3), 275–291.

\bibitem[\protect\citeauthoryear{Miratrix, Sekhon, and Yu}{Miratrix
  et~al.}{2013}]{miratrix2013adjusting}
Miratrix, L.~W., J.~S. Sekhon, and B.~Yu (2013).
\newblock Adjusting treatment effect estimates by post-stratification in
  randomized experiments.
\newblock {\em Journal of the Royal Statistical Society: Series B (Statistical
  Methodology)\/}~{\em 75\/}(2), 369--396.

\bibitem[\protect\citeauthoryear{Neyman}{Neyman}{1923}]{neyman1923}
Neyman, J. (1923).
\newblock {On the Application of Probability Theory to Agricultural
  Experiments. Essay on Principles (with discussion). Section 9 (translated)}.
\newblock {\em Statistical Science\/}~{\em 5\/}(4), 465--472.

\bibitem[\protect\citeauthoryear{Nguyen, Ebnesajjad, Cole, Stuart,
  et~al.}{Nguyen et~al.}{2017}]{nguyen2017sensitivity}
Nguyen, T.~Q., C.~Ebnesajjad, S.~R. Cole, E.~A. Stuart, et~al. (2017).
\newblock Sensitivity analysis for an unobserved moderator in
  rct-to-target-population generalization of treatment effects.
\newblock {\em The Annals of Applied Statistics\/}~{\em 11\/}(1), 225--247.

\bibitem[\protect\citeauthoryear{Nie, Imbens, and Wager}{Nie
  et~al.}{2021}]{nie2021covariate}
Nie, X., G.~Imbens, and S.~Wager (2021).
\newblock Covariate balancing sensitivity analysis for extrapolating randomized
  trials across locations.
\newblock {\em arXiv preprint arXiv:2112.04723\/}.

\bibitem[\protect\citeauthoryear{Olsen, Orr, Bell, and Stuart}{Olsen
  et~al.}{2013}]{olsen2013external}
Olsen, R.~B., L.~L. Orr, S.~H. Bell, and E.~A. Stuart (2013).
\newblock External validity in policy evaluations that choose sites
  purposively.
\newblock {\em Journal of Policy Analysis and Management\/}~{\em 32\/}(1),
  107--121.

\bibitem[\protect\citeauthoryear{O'Muircheartaigh and Hedges}{O'Muircheartaigh
  and Hedges}{2014}]{o2014generalizing}
O'Muircheartaigh, C. and L.~V. Hedges (2014).
\newblock Generalizing from unrepresentative experiments: a stratified
  propensity score approach.
\newblock {\em Journal of the Royal Statistical Society: Series C (Applied
  Statistics)\/}~{\em 63\/}(2), 195--210.

\bibitem[\protect\citeauthoryear{Raudenbush and Bloom}{Raudenbush and
  Bloom}{2015}]{raudenbush2015learning}
Raudenbush, S.~W. and H.~S. Bloom (2015).
\newblock Learning about and from a distribution of program impacts using
  multisite trials.
\newblock {\em American Journal of Evaluation\/}~{\em 36\/}(4), 475--499.

\bibitem[\protect\citeauthoryear{Robins}{Robins}{1999}]{robins1999association}
Robins, J.~M. (1999).
\newblock Association, causation, and marginal structural models.
\newblock {\em Synthese\/}, 151--179.

\bibitem[\protect\citeauthoryear{Robins, Rotnitzky, and Zhao}{Robins
  et~al.}{1994}]{robins1994estimation}
Robins, J.~M., A.~Rotnitzky, and L.~P. Zhao (1994).
\newblock Estimation of regression coefficients when some regressors are not
  always observed.
\newblock {\em Journal of the American statistical Association\/}~{\em
  89\/}(427), 846--866.

\bibitem[\protect\citeauthoryear{Rosenbaum}{Rosenbaum}{1987}]{rosenbaum1987sensitivity}
Rosenbaum, P.~R. (1987).
\newblock Sensitivity analysis for certain permutation inferences in matched
  observational studies.
\newblock {\em Biometrika\/}~{\em 74\/}(1), 13--26.

\bibitem[\protect\citeauthoryear{Rosenbaum}{Rosenbaum}{2010}]{rosenbaum2010design}
Rosenbaum, P.~R. (2010).
\newblock {\em Design of observational studies}, Volume~10.
\newblock Springer.

\bibitem[\protect\citeauthoryear{Rosenbaum and Rubin}{Rosenbaum and
  Rubin}{1983}]{rosenbaum1983assessing}
Rosenbaum, P.~R. and D.~B. Rubin (1983).
\newblock Assessing sensitivity to an unobserved binary covariate in an
  observational study with binary outcome.
\newblock {\em Journal of the Royal Statistical Society: Series B
  (Methodological)\/}~{\em 45\/}(2), 212--218.

\bibitem[\protect\citeauthoryear{Rubin}{Rubin}{1974}]{rubin1974causal}
Rubin, D.~B. (1974).
\newblock {Estimating Causal Effects of Treatments in Randomized and
  Nonrandomized Studies}.
\newblock {\em Journal of Educational Psychology\/}~{\em 66\/}(5), 688.

\bibitem[\protect\citeauthoryear{Rubin}{Rubin}{1980}]{rubin1980SUTVA}
Rubin, D.~B. (1980).
\newblock {Discussion of `Randomization analysis of experimental data: The
  Fisher randomization test comment' by Basu}.
\newblock {\em Journal of the American Statistical Association\/}~{\em
  75\/}(371), 591--593.

\bibitem[\protect\citeauthoryear{S{\"a}rndal, Swensson, and
  Wretman}{S{\"a}rndal et~al.}{2003}]{sarndal2003model}
S{\"a}rndal, C.-E., B.~Swensson, and J.~Wretman (2003).
\newblock {\em Model Assisted Survey Sampling}.
\newblock Springer Science \& Business Media.

\bibitem[\protect\citeauthoryear{Shen, Li, Li, and Were}{Shen
  et~al.}{2011}]{shen2011sensitivity}
Shen, C., X.~Li, L.~Li, and M.~C. Were (2011).
\newblock Sensitivity analysis for causal inference using inverse probability
  weighting.
\newblock {\em Biometrical Journal\/}~{\em 53\/}(5), 822--837.

\bibitem[\protect\citeauthoryear{Soriano, Ben-Michael, Bickel, Feller, and
  Pimentel}{Soriano et~al.}{2021}]{soriano2021interpretable}
Soriano, D., E.~Ben-Michael, P.~J. Bickel, A.~Feller, and S.~D. Pimentel
  (2021).
\newblock Interpretable sensitivity analysis for balancing weights.
\newblock {\em arXiv preprint arXiv:2102.13218\/}.

\bibitem[\protect\citeauthoryear{Stuart, Cole, Bradshaw, and Leaf}{Stuart
  et~al.}{2011}]{stuart2011use}
Stuart, E.~A., S.~R. Cole, C.~P. Bradshaw, and P.~J. Leaf (2011).
\newblock The use of propensity scores to assess the generalizability of
  results from randomized trials.
\newblock {\em Journal of the Royal Statistical Society: Series A (Statistics
  in Society)\/}~{\em 174\/}(2), 369--386.

\bibitem[\protect\citeauthoryear{Tan}{Tan}{2006}]{tan2006distributional}
Tan, Z. (2006).
\newblock A distributional approach for causal inference using propensity
  scores.
\newblock {\em Journal of the American Statistical Association\/}~{\em
  101\/}(476), 1619--1637.

\bibitem[\protect\citeauthoryear{Tan}{Tan}{2007}]{tan2007comment}
Tan, Z. (2007).
\newblock Comment: Understanding or, ps and dr.
\newblock {\em Statistical Science\/}~{\em 22\/}(4), 560--568.

\bibitem[\protect\citeauthoryear{Tipton}{Tipton}{2013}]{tipton2013improving}
Tipton, E. (2013).
\newblock Improving generalizations from experiments using propensity score
  subclassification: Assumptions, properties, and contexts.
\newblock {\em Journal of Educational and Behavioral Statistics\/}~{\em
  38\/}(3), 239--266.

\bibitem[\protect\citeauthoryear{Tipton}{Tipton}{2014}]{tipton2014generalizable}
Tipton, E. (2014).
\newblock How generalizable is your experiment? an index for comparing
  experimental samples and populations.
\newblock {\em Journal of Educational and Behavioral Statistics\/}~{\em
  39\/}(6), 478--501.

\bibitem[\protect\citeauthoryear{Tozzi, Lertxundi, Ibarluzea, and
  Baccini}{Tozzi et~al.}{2019}]{Tozzi2019CausalEO}
Tozzi, V., A.~Lertxundi, J.~Ibarluzea, and M.~Baccini (2019).
\newblock Causal effects of prenatal exposure to pm2.5 on child development and
  the role of unobserved confounding.
\newblock {\em International Journal of Environmental Research and Public
  Health\/}~{\em 16}.

\bibitem[\protect\citeauthoryear{Wager and Athey}{Wager and
  Athey}{2018}]{wager2018estimation}
Wager, S. and S.~Athey (2018).
\newblock Estimation and inference of heterogeneous treatment effects using
  random forests.
\newblock {\em Journal of the American Statistical Association\/}~{\em
  113\/}(523), 1228--1242.

\bibitem[\protect\citeauthoryear{Wang and Zubizarreta}{Wang and
  Zubizarreta}{2020}]{wang2020minimal}
Wang, Y. and J.~R. Zubizarreta (2020).
\newblock Minimal dispersion approximately balancing weights: asymptotic
  properties and practical considerations.
\newblock {\em Biometrika\/}~{\em 107\/}(1), 93--105.

\bibitem[\protect\citeauthoryear{Zhao}{Zhao}{2019}]{zhao2019covariate}
Zhao, Q. (2019).
\newblock Covariate balancing propensity score by tailored loss functions.
\newblock {\em The Annals of Statistics\/}~{\em 47\/}(2), 965--993.

\bibitem[\protect\citeauthoryear{Zhao and Percival}{Zhao and
  Percival}{2016}]{zhao2016entropy}
Zhao, Q. and D.~Percival (2016).
\newblock Entropy balancing is doubly robust.
\newblock {\em Journal of Causal Inference\/}~{\em 5\/}(1).

\bibitem[\protect\citeauthoryear{Zhao, Small, and Bhattacharya}{Zhao
  et~al.}{2019}]{zhao2019sensitivity}
Zhao, Q., D.~S. Small, and B.~B. Bhattacharya (2019).
\newblock Sensitivity analysis for inverse probability weighting estimators via
  the percentile bootstrap.
\newblock {\em Journal of the Royal Statistical Society: Series B (Statistical
  Methodology)\/}~{\em 81\/}(4), 735--761.

\end{thebibliography}

\clearpage
%%%%%%%%%%%%%%%%%%%%%%%%%%%%%%%%%%%%%%%%%%%%%%%%%%%%%%%%%%%%%%%%%%%%%%%%%%%%%%%%%%%%%%%%%%%%%%%%%%
\appendix
\setcounter{page}{1}
\begin{center}
    \singlespacing
    \Large
    \textbf{Supplementary Materials:} \\ Sensitivity Analysis for Generalizing Experimental Results 
\end{center}

\begingroup \allowdisplaybreaks
\section{Extensions and Additional Discussion} 

\subsection{Extension of Sensitivity Framework for Balancing Weights} \label{app:balancing} 
The proposed sensitivity framework can be extended for balancing weights. Balancing weights directly optimize for covariate balance (i.e., \cite{hainmueller2012entropy, ben2020balancing, wang2020minimal}, to name a few). There is a connection between balancing weights and propensity scores; \cite{zhao2019covariate} and \cite{wang2020minimal} show that balancing weights are a more general formulation of regularized propensity scores. 

We argue that for the class of balancing weights that meet the following conditions, the sensitivity framework can be directly applied:

\noindent \textbf{Condition 1.} $\E_\Sc(w_i)/\E_\Sc(w_i^*) = 1$

\noindent \textbf{Condition 2.} $\E_\Sc(w_i^* \mid \bX_i) = w_i$

When Condition 1 is met, this implies that the bias decomposition introduced in Theorem \ref{thm:weight} will hold. When Condition 2 is met, this implies that the bounds derived for $R^2_\varepsilon$ and $\rho_{\varepsilon, \tau}$ will apply. Condition 1 states that the estimated weights and the ideal weights must be centered at the same value. This is not a very stringent condition, as most weights (by definition) will be centered at mean 1. Condition 2 states that by conditioning on the observed covariates $\bX_i$, the ideal weights must be centered at the estimated weights $w_i$. The sensitivity analysis can still be applied to balancing weights that meet Condition 1, but not 2; however, the estimated bounds on the parameters may not necessarily hold. 

\subsection{Relationship to Marginal Sensitivity Models} \label{app:msm} 
Several recent papers have proposed using an alternative approach to performing sensitivity analysis in the form of marginal sensitivity models (\cite{zhao2019sensitivity}, \cite{soriano2021interpretable}). Marginal sensitivity models define a class of sensitivity models that bound the underlying error in the selection probabilities. Under the assumption that researchers have specified the correct bound, a Monte Carlo approach can be used to estimate asymptotic confidence intervals that provide nominal coverage. The approach tends to be conservative, and recent extensions of this approach allow researchers to obtain sharper or tighter bounds (i.e., \cite{dorn2021sharp}, \cite{nie2021covariate}).

The proposed sensitivity analysis framework is distinct from the marginal sensitivity models in how the bias is parameterized. In particular, marginal sensitivity models use the worst-case multiplicative error in the underlying selection probabilities to bound the potential bias from an omitted confounder, whereas Theorem \ref{thm:weight} parameterizes the bias in terms of characteristics related to the linear error in the estimated weights. There are two potential benefits from using our proposed approach over the marginal sensitivity approach. The first benefit is our ability to construct bounded, interpretable sensitivity parameters, whereas using the worst-case multiplicative error as the sensitivity parameter can potentially be more challenging in practice. The second is that the marginal sensitivity models assess a \textit{worst-case} error. Especially in generalizability settings, where weights tend to be extreme, this may result in very extreme parameter values. (See \cite{fogarty2019extended} for more discussion.) The proposed framework allows for a less conservative assessment of sensitivity, and allows researchers to transparently include their substantive priors into the sensitivity analysis. We leave exploring the more explicit relationship between the proposed bias decomposition approach and marginal sensitivity models for future work.

\subsection{Relaxing Bounds on $\sigma^2_\tau$} \label{app:treat_het}
We will discuss two examples of assumptions that researchers may wish to impose. Figure \ref{fig:treatment_het} provides a summary.

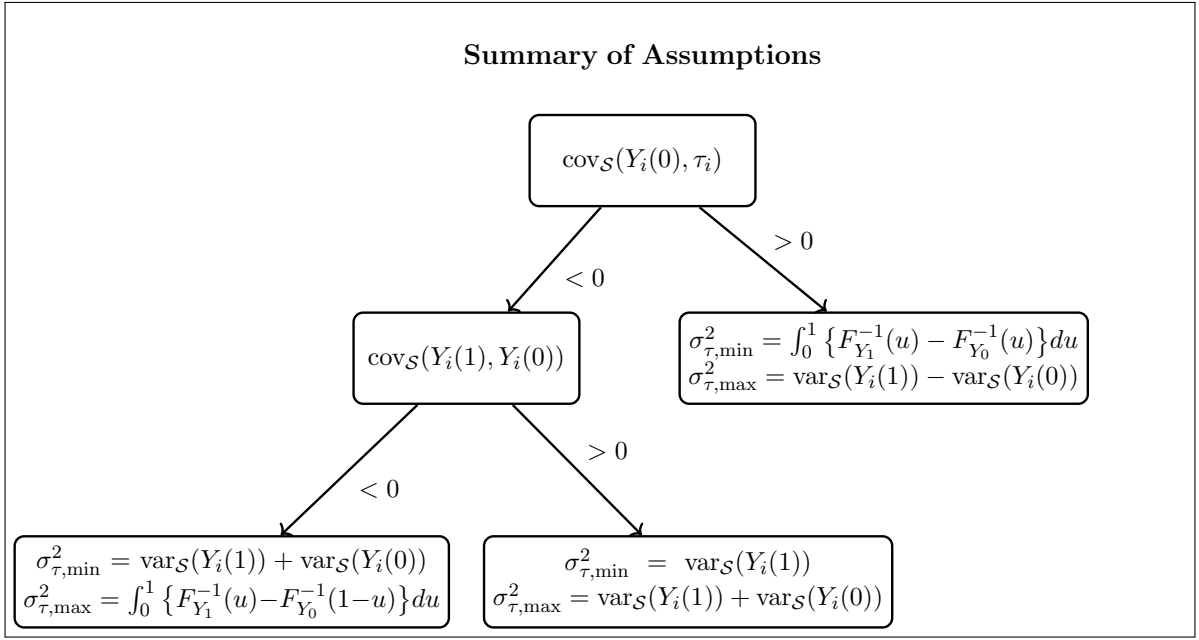
\begin{figure}[!ht]
\begin{center} 
\fbox{
\scalebox{0.9}{
    \tikzstyle{block} = [rectangle, draw, line width = 1pt, 
    text width=8em, text centered, rounded corners, minimum height=3.5em]
\tikzstyle{line} = [draw, ->, line width = 1pt]

\hspace{-4cm}
\begin{tikzpicture}[node distance = 12em, auto]
    % Place nodes
    \node [block, draw = none, text width = 65em] at (0em, 8em) (title) {\mbox{\large\textbf{Summary of Assumptions}}};
    
    %% Design
    \node [block] at (0, 4em) (covariance) {$\cov_\Sc(Y_i(0), \tau_i)$};
    \node [block, below right=4em and -3em of covariance, text width = 15em] (cov_pos) { $\sigma^2_{\tau, \min}=\int_0^1 \big\{ F_{Y_1}^{-1}(u) - F_{Y_0}^{-1}(u) \big\} du$ \\ $\sigma^2_{\tau, \max}=\var_\Sc(Y_i(1))-\var_\Sc(Y_i(0))$};  
    \node [block, below left=4em and -2em of covariance] (cov_neg) {$\cov_\Sc(Y_i(1), Y_i(0))$};  
    
    %second layer 
    \node [block, below left=5em and -3.75em of cov_neg, text width = 16em] (cov_neg_2) { $\sigma^2_{\tau, \min}=\var_\Sc(Y_i(1))+\var_\Sc(Y_i(0))$\\ $\sigma^2_{\tau, \max}=\int_0^1 \big\{ F_{Y_1}^{-1}(u) - F_{Y_0}^{-1}(1-u) \big\} du$};  
    \node [block, below right=5em and -3.75em of cov_neg, text width = 15em] (cov_pos_2) { $\sigma^2_{\tau, \min}=\var_\Sc(Y_i(1))$ \\ $\sigma^2_{\tau, \max}=\var_\Sc(Y_i(1))+\var_\Sc(Y_i(0))$};

    \path [line] (covariance) -- node {$<0$}(cov_neg);
    \path [line] (covariance) -- node {$>0$}(cov_pos); 
    \path [line] (cov_neg) -- node {$<0$}(cov_neg_2); 
    \path [line] (cov_neg) -- node {$>0$}(cov_pos_2); 
\end{tikzpicture}
\hspace{-5cm} 
}
}
\end{center} 
\caption{Summary of assumptions that researchers may invoke to help tighten the bound on $\sigma^2_\tau$. The above diagram provides the tightened minimum and maximum values for $\sigma^2_\tau$, depending on the assumption researchers wish to invoke. Researchers can estimate $m$ to check if $m < 1$. If $m < 1$, then it is guaranteed that $\cov_\Sc(Y_i(0), \tau_i) < 0$. Substantive knowledge can be used to help justify the different assumptions.} 
\label{fig:treatment_het} 
\end{figure} 

\paragraph{Directional sign of the correlation between $\tau_i$ and $Y_i(0)$:} \cite{ding2019decomposing} and \cite{raudenbush2015learning} show that information about the correlation between the individual-level treatment effect and the control potential outcomes could be inferred from $m$ (i.e., the ratio of variance between the treatment and control outcomes). More specifically, when $m < 1$ (i.e., the variance of the control outcomes is greater than the variance of the treatment outcomes), then the correlation between the individual treatment effect and the control potential outcome is negative, and the lower bound on $\sigma^2_\tau$ can be tightened:\footnote{This follows simply from the fact that we may rewrite the decomposed treatment effect heterogeneity as: 
\begin{align*} 
\sigma^2_{\tau} &= \var_\Sc(Y_i(1))+ \var_\Sc(Y_i(0)) - 2 \cov(Y_i(1), Y_i(0)) \\
&=\var_\Sc(Y_i(1)) - \var_\Sc(Y_i(0)) - 2 \cov_\Sc(\tau_i, Y_i(0))
\end{align*} 
Because $\sigma^2_{\tau} \geq 0$, then $2 \cov_\Sc(\tau_i, Y_i(0)) \leq \underbrace{\var_\Sc(Y_i(1))-\var_\Sc(Y_i(0))}_{(*)}$. When $m <1$, this implies that the term in $(*)$ is going to be negative, which in turn, implies $\cov_\Sc(\tau_i, Y_i(0)) \leq 0$.}
\begin{equation} 
\var_\Sc(Y_i(1)) \leq \sigma^2_\tau 
\label{eqn:vartau_bound2} 
\end{equation} 

Unfortunately, the converse cannot be shown to be true (i.e., $m > 1$ does not necessarily imply a positive relationship). However, researchers may have substantive knowledge to justify a positive relationship. For example, in the original JTPA study, researchers compared the estimated impact of jobs training programs across women by previous earnings and employment history (\cite{bloom1993national}). They found that women who had a higher hourly wage in their work history had a higher estimated impact from accessibility to jobs training programs. Similarly, women who came from families with greater household income also saw a greater impact from jobs training programs. As such, we assume there exists a non-negative association between the individual-level treatment effect and the outcomes under control (i.e., $\cov(Y_i(0), \tau_i) > 0$)

In cases where researchers are willing to assume that $\cov_\Sc(\tau_i, Y_i(0)) \geq 0$, the upper bound of $\sigma^2_\tau$ becomes: 
\begin{equation} 
\sigma^2_{\tau} \leq \var_\Sc(Y_i(1)) - \var_\Sc(Y_i(0))
\label{eqn:vartau_bound_pos_cov} 
\end{equation} 

\paragraph{Directional sign of the correlation between potential outcomes:} Alternatively, researchers may assume information about the relationship between $Y_i(1)$ and $Y_i(0)$. In particular, if researchers believe that the correlation between $Y_i(1)$ and $Y_i(0)$ is non-negative, then a tighter upper bound may be obtained on $\sigma^2_{\tau}$: 
\begin{equation} 
\sigma^2_{\tau} \leq \var_\Sc(Y_i(1)) + \var_\Sc(Y_i(0))
\end{equation} 
\label{eqn:vartau_bound2} 
Alternatively, if researchers assume the correlation between $Y_i(1)$ and $Y_i(0)$ is negative, then a tighter lower bound is obtained: 
\begin{equation} 
\var_\Sc(Y_i(1)) + \var_\Sc(Y_i(0)) \leq \sigma^2_{\tau} 
\end{equation} 
We note that in order for the correlation between $Y_i(1)$ and $Y_i(0)$ to be negative, $\cov(Y_i(0), \tau_i)$ must be negative.\footnote{This follows from the following:
\begin{align*} 
\cov(Y_i(0), Y_i(1)) < 0 &\implies \cov(Y_i(0), Y_i(0)) + \cov(Y_i(0), \tau_i) < 0 \\
&\implies \cov(Y_i(0), \tau_i) < - \var(Y_i(0)) < 0
\end{align*} 
}  (The converse is not true--i.e., $\cov(\tau_i, Y_i(0))$ may be negative, without $\cov(Y_i(1), Y_i(0)) < 0$.) \\

These two assumptions can be combined in conjunction to help tighten the bound on $\sigma^2_{\tau}$. We recommend researchers first estimate $m$ to determine whether or not $\cov(Y_i(0), \tau_i)$ must be negative, which can help narrow down the plausible assumptions that can be used. A summary is provided in Figure \ref{fig:treatment_het}.

Another approach to tighten the bound on plausible $\sigma^2_\tau$ values is to directly model the individual-level treatment effect (e.g., see \cite{kent2020predictive, athey2019generalized, wager2018estimation, hill2011bayesian}) Many existing approaches leverage flexible, machine learning methods to estimate $\hat \tau_i$ without relying heavily on parametric assumptions, such as linearity. Therefore, researchers can model $\hat \tau_i$, and then directly estimate $\var_\Sc(\hat \tau_i)$ to understand what may be plausible values for $\sigma^2_\tau$. Because we are only concerned about treatment effect heterogeneity across the experimental sample, this can be especially advantageous in settings where researchers have a richer set of covariates within the experimental sample that may not be measured across the population.\footnote{We note that in cases when researchers have access to a rich set of covariates across both the population and the experimental sample and strongly believe that they can accurately parametrically model $\tau_i$, it may be advantageous to use a doubly robust estimator, instead of just the weighted estimator. (See Section \ref{sec:aug} for discussion.)}

We highlight two ways researchers can leverage parametrically modeling $\tau_i$ to help bound $\sigma^2_\tau$. First, similar to \cite{ding2019decomposing}, researchers may use the estimated $\var_\Sc(\hat \tau_i)$ and posit how many times larger the actual variation in individual-level treatment effect is. In general, we caution researchers from directly using $\var_\Sc(\hat \tau_i)$ as the estimate for $\sigma^2_\tau$. Even in the scenario that the true conditional expectation of the individual-level treatment effect is used to estimate $\tau_i$, $\var_\Sc(\hat \tau_i)$ will be an underestimation of the true variation in the individual-level treatment effect.\footnote{This can be formalized in the following. Assume $g(\bX_i) = \E(\tau_i | \bX_i)$. Thus, we may decompose $\tau_i$ into the component that can be explained by $g(\bX_i)$, and the component that cannot:
$$\tau_i = \E(\tau_i | \bX_i) + u_i$$
Because the covariance between $u_i$ and the estimated $\hat \tau_i$ values must be 0: $\sigma^2_{\tau} = \var_\Sc(\hat \tau_i) + \var_\Sc(u_i)$, and $\sigma^2_{\tau} \geq \var_\Sc(\hat \tau_i)$.} The second way researchers can benefit from parametrically modeling $\tau_i$ is by using $\hat \tau_i$ to help aid the substantive justification of one of the assumptions used to tighten the bounds on $\sigma^2_\tau$. For example, if researchers wish to assume that $\cov_\Sc(Y_i(0), \tau_i) > 0$, they can use the estimated $\hat \tau_i$ to check if $\cov_\Sc(Y_i(0), \hat \tau_i)$ is positive. 

\subsection{Less Conservative Extreme Scenario Bounds} \label{app:extreme_scenario} 
The extreme scenario bound proposed in Section \ref{subsubsec:extreme} allows researchers to evaluate the bias when the error term $\varepsilon_i$ explains all residual variation in the individual-level treatment effect. When $\rho_{\varepsilon, \tau} = \rho_{max}$, the maximum value of $R^2_\varepsilon$ is equal to $1-\cor_\Sc(w_i, \tau_i)^2$. However, we can use Lemma \ref{lem:rho_bound} to show that when $\rho_{\varepsilon, \tau}$ is equal to the upper bound of $1-\cor_\Sc(w_i, \tau_i)^2$, then $R_\varepsilon^2$ can actually take on a range of values, defined by the following: 
\begin{equation} 
R^2_{max} =1-\left( \cor_\Sc(w_i, \tau_i) \cdot \cor_\Sc(w^*_i, \tau_i) \pm \sqrt{(1-\cor_\Sc(w^*_i, \tau_i)^2) \cdot (1-\cor_\Sc(w_i, \tau_i)^2)} \right)^2 
\label{eqn:r2_bound} 
\end{equation} 
Equation \eqref{eqn:r2_bound} represents the degree of imbalance that must be present in order for $\rho_{\varepsilon, \tau}$ to equal to the upper bound of $\rho_{max}$. However, Equation \eqref{eqn:r2_bound} depends on $\cor_\Sc(w^*_i, \tau_i)$ (i.e., the relationship between the true selection weights $w_i^*$ and the individual-level treatment effect), which cannot be estimated. Evaluating Equation \eqref{eqn:r2_bound} for the extreme case that $|\cor(w^*_i, \tau_i) = 1|$ removes the dependency on $\cor(w^*_i, \tau_i)$ and results in the upper bound proposed in Section \ref{subsubsec:extreme}: 
$$R^2_{max} \leq 1-\cor_\Sc(w_i, \tau_i)^2$$

Researchers may not wish to assume that $\cor(w^*_i, \tau_i)$ is at $-1$ or $1$. As such, evaluating Equation \eqref{eqn:r2_bound} at lower values of $\cor(w_i^*, \tau_i)$ will result in a less conservative bound on $R^2_\varepsilon$. One approach researchers can take to posit plausible values for $\cor(w^*_i, \tau_i)$ is by using the estimated $\cor(w_i, \tau_i)$ and specifying how much \textit{additional} variation in $\tau_i$ they believe $w_i^*$ is able to explain. For example, if $\widehat{\cor}(w_i, \tau_i)$ is very low (i.e., $\approx 0.1$), it may be unlikely that the true weights $w_i^*$ would be $10 \times$ more correlated with the individual-level treatment effect, such that $\cor(w^*_i, \tau_i) \approx 1$. This allows researchers to obtain less conservative estimates of an extreme scenario bound, depending on what they deem is a ``reasonable'' choice for $\cor_\Sc(w^*_i, \tau_i)$.

\subsection{Details for Sensitivity Analysis for the Augmented Weighted Estimator} \label{app:aug_tools} 
\subsubsection{Interpreting the Parameters} 
\paragraph{Correlation between $\varepsilon_i$ and $\xi_i$ (i.e., $\rho_{\varepsilon, \xi}$)} The correlation term between $\varepsilon_i$ and $\xi_i$ represents the relationship between the error in the weight estimation, and the error in the treatment effect modeling. In other words, $\rho_{\varepsilon, \xi}$ is a measure for how related the residual imbalance in the omitted confounder $\bU_i$ is to the residuals in the individual-level treatment effect model. In general, we expect $|\rho_{\varepsilon, \xi}|$ to be less than $|\rho_{\varepsilon, \tau}|$ because the residual imbalance in the omitted confounder is likely to be less correlated to the residuals $\xi_i$ than the overall individual-level treatment effect $\tau_i$. 

We can extend Lemma \ref{lem:rho_bound} to bound $\rho_{\varepsilon, \xi}$ on the range $\left[- \sqrt{1-\cor_\Sc(w_i, \xi_i)}, \sqrt{1-\cor_\Sc(w_i, \xi_i)} \right]$. If the estimated weights $w_i$ are highly correlated with the residuals $\xi_i$, then the range of values that $\rho_{\varepsilon, \xi}$ may take on will be more restricted. 

\paragraph{Variation in $\xi_i$ (i.e., $\sigma^2_\xi$)} 
$\sigma^2_\xi$ is the total variation leftover in the treatment effect heterogeneity that is not explained by the estimated treatment effect model. $\sigma^2_\xi$ is often referred to in the literature as the \textit{idiosyncratic treatment effect variation} (\cite{ding2019decomposing, djebbari2008heterogeneous, heckman1997making}). $\sigma^2_\xi$ can be written as a function of $\sigma^2_\tau$: 
$$\sigma^2_\xi = \sigma^2_\tau - \var_\Sc(\hat \tau_i) - 2 \cov_\Sc(\hat \tau_i, \xi_i),$$
where both $\var(\hat \tau_i)$ and $\cov(\hat \tau_i, \xi_i)$ can be estimated from observed data. Thus, researchers can use the same bounds derived in Section \ref{subsec:TE_het} to estimate an upper bound for $\sigma^2_\tau$ (denoted as $\sigma^2_{\tau,max}$, and bound $\sigma^2_\xi$ in the following manner: 
\begin{equation} 
\sigma^2_\xi \leq \sigma^2_{\tau,max} - \var_\Sc(\hat \tau_i) - 2 \cov_\Sc(\hat \tau_i, \xi_i)
\label{eqn:var_xi_bound} 
\end{equation} 

Alternatively, researchers can choose to bound $\sigma^2_\xi$ directly. For example, the bound from Equation \eqref{eqn:vartau_bound1} can be extended for the residuals across the potential outcomes \citep{ding2019decomposing}. The derived bounds can be sharpened by invoking additional assumptions on the residuals between the potential outcomes.

\subsubsection{Summary of Sensitivity Framework} We summarize the sensitivity analysis framework for augmented weighted estimators below. \\

\noindent\fbox{%
\vspace{2mm}
\parbox{0.98\textwidth}{%
\vspace{2mm}
\noindent \underline{\textbf{Summary of Sensitivity Framework for Augmented Weighted Estimators}}
\begin{Step} 
\item Estimate a conservative upper bound for $\sigma^2_{\xi}$ (i.e., $\sigma^2_{\xi, \max}$). 
\item Using $\sigma^2_{\xi,\max}$, estimate $\widehat{\cor_\Sc}^2(w_i, \xi_i)$ (as a conservative bound for $\cor_\Sc^2(w_i, \xi_i)$).
\item Vary $\rho_{\varepsilon, \xi}$ from $-\sqrt{1-\widehat{\cor_\Sc}^2(w_i, \xi_i)}$ to $\sqrt{1-\widehat{\cor_\Sc}^2(w_i, \xi_i)}$. 
\item Vary $R_\varepsilon^2$ from the range of $[0,1)$. 
\item Evaluate the bias. 
\end{Step} 
} 
} \\
\vspace{2mm}
\subsubsection{Relationship with Sensitivity Analysis from \cite{nguyen2017sensitivity}}\label{proof_ex:nguyen} 
Consider the case in which only the treatment effect heterogeneity is modeled, using $\hat \tau(\bX_i)$. Denote this estimator as $\hat \tau_{model}$. The bias formula for failing to account for $\bU_i$ in the individual-level treatment model is: 
\begin{equation} 
\text{Bias}(\hat \tau_{model}) = \rho_{w^*, \xi} \cdot \sqrt{\var_\Sc(w_i^*) \cdot \sigma^2_\xi},
\label{eqn:TE_only} 
\end{equation} 
where $\rho_{w^*, \xi} := \cor_\Sc(w_i^*, \xi_i)$. If we assume the following linear model:
$$\E(\tau_i) = \tau + \beta_X \bX_i + \beta_U \bU_i,$$
where $\bX_i \indep \bU_i \mid S_i$, then Equation \eqref{eqn:TE_only} is equivalent to the bias formula from \cite{nguyen2017sensitivity}:
$$\beta_U \cdot \big( \E( \bU_i | S_i = 0) - \E(\bU_i | S_i=1) \big)$$
\begin{proof} 
Assume we estimate the following $\hat \tau(\bX_i)$ model: 
$$\hat \tau(\bX_i) := \beta_X \bX_i$$
This is equivalent to fitting two linear regressions to the control and treatment potential outcomes, using only $\bX_i$. As such, 
$$\xi_i = \tau_i - \hat \tau(\bX_i) = \beta_U \bU_i$$ 
Therefore, using the bias formula: 
\begin{align*} 
cor_\Sc(w_i^*, \xi_i) &\cdot \sqrt{\var_\Sc(w_i^*) \cdot \sigma^2_\xi}\\ &\equiv \cov_\Sc( \xi_i, w_i^*) \\
&= \E \left( \beta_U \bU_i \cdot w_i^* | S_i = 1 \right) - \E(\beta_U \bU_i| S_i = 1 ) \cdot \E(w_i^*| S_i = 1 ) 
\intertext{Using the decomposition of $w_i^* = w_i \cdot P(\bU_i | S_i = 0)/P(\bU_i | S_i = 1)$ from Lemma \ref{lem:error_decomp}:} 
&= \E \left( \left. \beta_U \bU_i \cdot w_i \cdot \frac{P(\bU_i | S_i = 0)}{P(\bU_i | S_i = 1)}\right| S_i = 1 \right) - \E(\beta_U \bU_i| S_i = 1 ) \cdot \E(w_i^*| S_i = 1 ) \\
&= \E(w_i| S_i = 1 ) \cdot \E \left( \left.\beta_U \bU_i \cdot \frac{P(\bU_i | S_i = 0)}{P(\bU_i | S_i = 1)} \right| S_i = 1  \right) - \E(\beta_U \bU_i| S_i = 1 ) \cdot \E(w_i^*| S_i = 1 ) \\
&= \E(w_i| S_i = 1 ) \cdot \beta_U \cdot \underbrace{\E \left( \left.\bU_i \cdot \frac{P(\bU_i | S_i = 0)}{P(\bU_i | S_i = 1)} \right| S_i = 1\right)}_{:= \E(\bU_i | S_i = 0)} - \beta_U \cdot \E( \bU_i | S_i = 1) \cdot \E(w_i^* | S_i = 1) 
\intertext{By definition of balancing weights:}
&= \E(w_i| S_i = 1 ) \cdot \beta_U \cdot \E \left(  \bU_i \mid S_i = 0 \right) - \beta_U \cdot \E(\bU_i \mid S_i = 1) \cdot \E(w_i^*| S_i = 1 ) \\
&= \beta_U \cdot \bigg( \E \left(  \bU_i \mid S_i = 0 \right) - \E(\bU_i \mid S_i = 1)\bigg),
\end{align*} 
which is equivalent to the expression from \cite{nguyen2017sensitivity}. 
\end{proof} 

\subsubsection{Tools for Sensitivity Analysis for the Augmented Weighted Estimator} 
\paragraph{Robustness Value} An analogous robustness value to the one introduced in Section \ref{subsec:summary} can be derived for the augmented weighted estimator. In particular: 
$$RV_q^{Aug} = \frac{1}{2} \left( \sqrt{b_q^2 + 4b_q} - b_q \right), \ \ \ \text{where } \ \ b_q = \frac{q^2 \cdot  \left(\hat\tau_W^{Aug}\right)^2}{\sigma^2_\xi \cdot \var(w_i)}$$
The primary difference between $RV_q^{Aug}$ and the previously proposed $RV_q$ is that the robustness value for the augmented weighted estimator is a function of $\sigma^2_\xi$, instead of $\sigma^2_{\tau}$. This highlights the fact that the relative robustness of the augmented weighted estimator, compared to the weighted estimator, depends directly on how much variation is explained by the individual-level treatment effect model $\hat \tau_i$. 

\paragraph{Extreme Scenario Analysis} In the augmented weighted estimator setting, the extreme scenario analysis represents the case in which the error term $\varepsilon_i$ is able to explain all residual variation in the idiosyncratic treatment effect (i.e., $\rho^2_{\varepsilon, \xi} = 1 - \cor(w_i, \xi_i)^2$). Thus, the maximum parameter values may be evaluated at $\rho^2_{\varepsilon, \xi} = R^2_\varepsilon = 1- \cor(w_i, \xi_i)^2$. In practice, the correlation between the estimated weights and the residual component of the treatment effect heterogeneity, unexplained by the observed covariates, is likely to be relatively low. As such, we expect the extreme scenario analysis to be conservative in nature. Researchers can employ similar methods to the weighted estimator case to evaluate less conservative scenarios (see Section \ref{subsubsec:extreme}). 

\paragraph{Formal Benchmarking} 
To formally benchmark the sensitivity parameters in the augmented weighted estimator framework, we must also account for the error from misspecifying the treatment effect heterogeneity model. More specifically, let $\hat \tau(\bX_i^{-(j)})$ be the estimated individual-level treatment effect, omitting covariates $\bX_i^{-(j)}$. Then, define the following error term: 
$$\xi_i^{-(j)} := \hat \tau(\bX_i) - \hat \tau(\bX_i^{-(j)})$$
$\xi_i^{-(j)}$ represents the error incurred from omitting $\bX_i^{(j)}$ from estimating $\tau_i$. 

We define the following: 
$$k_{\rho}^{\xi} := \frac{\cor_\Sc(\varepsilon_i, \xi_i)}{\cor_\Sc(\varepsilon_i^{-(j)}, \xi_i^{-(j)})},$$
where $k_\rho^{\xi}$ compares the amount of variation that $\varepsilon_i$ can explain in $\xi_i$, relative to the amount of variation that $\varepsilon_i^{-(j)}$ can explain in $\xi_i^{-(j)}$. To calibrate $\rho_{\varepsilon, \xi}$, researchers can estimate $\cor_\Sc(\varepsilon_i^{-(j)}, \xi_i^{-(j)})$ and scale by the inputted $k_{\rho}^{\xi}$ value. At $k_\rho^{\xi} = 1$, this implies that the correlation between $\varepsilon_i$ and $\xi_i$ is equivalent to the correlation between $\varepsilon_i^{-(j)}$ and $\xi_i^{-(j)}$. It is worth noting that researchers can choose to additionally benchmark $\sigma^2_\xi$. However, if researchers are bounding $\sigma^2_\xi$ using Equation \eqref{eqn:var_xi_bound}, there is no need to calibrate $\sigma^2_\xi$ because we will have bounded it using $\sigma^2_{\tau, max}$ (which is not dependent on any covariates) and two estimable quantities.

\clearpage 
\section{Proofs for Theorems and Lemmas} \label{app:proof} 
\subsection{Proof of Lemma \ref{lem:error_decomp} (Error Decomposition)} 
\noindent\fbox{%
\vspace{2mm}
\parbox{\textwidth}{%
\vspace{2mm}
When using inverse propensity weights, the estimated weights and the ideal weights are written as: 
\begin{equation*} 
w_i = \frac{P(S_i = 1)}{P(S_i = 0)} \cdot \frac{1-P(S_i = 1|\bX_i)}{P(S_i = 1| \bX_i)} \ \ \ \
w_i^* = \frac{P(S_i = 1)}{P(S_i = 0)} \cdot \frac{1-P(S_i = 1|\bX_i, \bU_i)}{P(S_i = 1| \bX_i, \bU_i)}
\end{equation*} 
Then, the error in weight estimation from omitting $\bU_i$ can be decomposed in the following manner: 
\begin{align*} 
\varepsilon_i &= w_i - w_i^* \nonumber \\
&= \underbrace{\frac{P(S_i = 1)}{P(S_i = 0)} \cdot \frac{P(S_i = 0 | \bX_i)}{P(S_i = 1 | \bX_i)}}_{\text{Estimated Weights } (w_i)} \cdot \underbrace{\left( \frac{P(\bU_i | \bX_i, S_i = 1) - P(\bU_i | \bX_i, S_i = 0)}{P(\bU_i | \bX_i, S_i = 1)} \right)}_{\text{Residual Imbalance in } \bU_i},
\end{align*} 
where $P(\bU_i \mid \bX_i, S_i = 1) - P(\bU_i \mid \bX_i, S_i = 1) $ represents the difference in the underlying probability density function of the omitted confounder $\bU_i$, conditioned on $\bX_i$, across the target population ($S_i = 0$) and the experimental sample ($S_i = 1$). 
}
}
\begin{proof} 
We will substitute in the IPW forms for both $w_i$ and $w_i^*$ and then apply Baye's rule: 
\begin{align*} 
\varepsilon_i = w_i - w_i^* =&  \frac{P(S_i = 1)}{P(S_i = 0)} \cdot \frac{P(S_i = 0 | \bX_i)}{P(S_i = 1 | \bX_i)} -  \frac{P(S_i = 1)}{P(S_i = 0)} \cdot\frac{P(S_i = 0 | \bX_i, U_i)}{P(S_i = 1 | \bX_i, \bU_i)} \\
=&  \frac{P(S_i = 1)}{P(S_i = 0)} \cdot\left(\frac{1}{P(S_i = 1 | \bX_i)} - 1 - \frac{1}{P(S_i = 1 | \bX_i, \bU_i)} + 1 \right)\\
=&  \frac{P(S_i = 1)}{P(S_i = 0)} \cdot \underbrace{\left(\frac{1}{P(S_i = 1 | \bX_i)} - \frac{1}{P(S_i = 1 | \bX_i, \bU_i)}\right)}_{(*)}
\end{align*} 

Using Baye's Rule, we can show that $\varepsilon_i$ is proportional to the imbalance in the omitted confounder $U_i$, conditional on $X_i$. This is done by re-writing the term $(*)$:  
\begin{align*} 
&\frac{1}{P(S_i = 1 | \bX_i)} - \frac{1}{P(S_i = 1 | \bX_i, \bU_i)} \\
&= \frac{1}{P(S_i = 1 | \bX_i)} - \frac{P(\bU_i | \bX_i)}{P(\bU_i | S_i = 1, \bX_i) \cdot P(S_i = 1 | \bX_i)} \\
&= \frac{1}{P(S_i = 1 | \bX_i)} \cdot\left(1 - \frac{P(\bU_i | \bX_i)}{P(\bU_i | S_i = 1, \bX_i)} \right)\\
&= \frac{1}{P(S_i = 1 | \bX_i)} \cdot \left( \frac{P(\bU_i | \bX_i, S_i = 1)}{P(\bU_i | \bX_i, S_i = 1)} - \frac{P(\bU_i | \bX_i, S_i =1) P(S_i = 1 | \bX_i) + P(\bU_i | \bX_i, S_i = 0)P(S_i = 0|\bX_i)}{P(\bU_i | \bX_i, S_i=1)} \right) \\
&= \frac{1}{P(S_i = 1 | \bX_i)} \cdot \left( \frac{P(\bU_i | \bX_i, S_i = 1)(1-P(S_i = 1 | \bX_i)) - P(\bU_i | \bX_i, S_i = 0) P(S_i = 0 |\bX_i)}{P(\bU_i | S_i = 1, \bX_i)} \right) \\
&= \frac{P(S_i = 0 | \bX_i)}{P(S_i=1|\bX_i)} \left( \frac{P(\bU_i | S_i = 1, \bX_i) - P(\bU_i | S_i = 0, \bX_i)}{P(\bU_i | S_i = 1, \bX_i)} \right)
\end{align*} 
\end{proof}

\subsection{Proof of Lemma \ref{lem:var_decomp} (Variance Decomposition of $w_i^*$)} 
\noindent\fbox{%
\vspace{2mm}
\parbox{\textwidth}{%
\vspace{2mm}
For inverse propensity score weights, the variance of the true weights $w_i^*$ can be decomposed linearly into two components: 
$$\var_\Sc(w_i^*) = \var_\Sc(w_i) + \var_\Sc(\varepsilon_i)$$
Therefore, $R^2_\varepsilon$ is bounded between 0 and 1. 
}
}
\begin{proof} 

The proof of Lemma \ref{lem:var_decomp} will proceed in two parts. To begin, we will first show that for inverse propensity score weights, $\E_\Sc(w_i^* | \bX_i) = w_i$. Then, we will show that $\var(w_i^*)$ can be written as the sum of the variance of the estimated weights $w_i$ and the error term $\varepsilon_i$. 

Recall from Lemma \ref{lem:error_decomp}, we showed that $w_i^*$ could be decomposed in the following terms: 
$$w_i^* = \frac{P(S_i = 0)}{P(S_i = 1)} \frac{P(S_i = 0 \mid \bX_i)}{P(S_i = 1 \mid \bX_i)} \frac{P(\bU_i \mid \bX_i, S_i = 0)}{P(\bU_i \mid \bX_i, S_i = 1)} \equiv w_i \cdot \frac{P(\bU_i \mid \bX_i, S_i = 0)}{P(\bU_i \mid \bX_i, S_i = 1)}$$
We can then show that the expectation of $w_i^*$, conditioned on $\bX_i$, will be equal to $w_i$: 
\begin{align*} 
\E_\Sc(w_i^* \mid \bX_i) &= \E_\Sc \left( \left.w_i \cdot \frac{P(\bU_i \mid \bX_i, S_i = 0)}{P(\bU_i \mid \bX_i, S_i = 1)}\right| \bX_i \right) \\
&=  w_i \cdot \E_\Sc \left( \left. \frac{P\bU_i \mid \bX_i, S_i = 0)}{P(\bU_i  \mid \bX_i, S_i = 1)} \right| \bX_i \right) \\
&= w_i \cdot \E \left( \left. \frac{P(\bU_i \mid \bX_i, S_i = 0)}{P(\bU_i  \mid \bX_i, S_i = 1)} \right| \bX_i, S_i = 1\right)\\
&=w_i \cdot \left( \sum_{u \in \mathcal{U}} \frac{P(\bU_i =u \mid \bX_i, S_i = 0)}{P(\bU_i =u \mid \bX_i, S_i = 1)} P(\bU_i=u \mid \bX_i, S_i = 1) \right) \\
&=w_i \cdot \underbrace{\left( \sum_{u \in \mathcal{U}} P(\bU_i = u \mid \bX_i, S_i = 0) \right)}_{=1} \\
&= w_i
\end{align*} 
Now we will show that the variance of $\varepsilon_i$ can be written as the difference between the variance of $w_i$ and the variance of $w_i^*$: 
\begin{align*} 
\var_\Sc(\varepsilon_i) &= \var_\Sc(w_i - w_i^*) \\
&= \var_\Sc(w_i) + \var_\Sc(w_i^*) - 2 \cov_\Sc(w_i, w_i^*) \\
&= \var_\Sc(w_i) + \var_\Sc(w_i^*) - 2 \left( \E_\Sc(w_i \cdot w_i^*) - \E_\Sc(w_i) \E_\Sc(w_i^*) \right)
\intertext{Making use of the fact that $\E_\Sc(w_i) = \E_\Sc(w_i^*)$ and by Law of Iterated Expectation:} 
&= \var_\Sc(w_i) + \var_\Sc(w_i^*) - 2 \left( \E_\Sc(\E_\Sc(w_i \cdot w_i^* | \bX_i =x )) - \E_\Sc(w_i)^2 \right) 
\intertext{From above, we have shown that $\E_\Sc(w_i^* | \bX_i =x ) = w_i$:}
&= \var_\Sc(w_i) + \var_\Sc(w_i^*) - 2 \left( \E_\Sc(w_i^2) - \E_\Sc(w_i)^2 \right) \\
&= \var_\Sc(w_i) + \var_\Sc(w_i^*) - 2 \var_\Sc(w_i) \\
&= \var_\Sc(w_i^*) - \var_\Sc(w_i)
\end{align*} 
Thus, we have shown that $\var_\Sc(w_i^*)$ can be decomposed into the sum of $\var_\Sc(w_i)$ and $\var_\Sc(\varepsilon_i)$. It naturally follows that $R^2_\varepsilon := \var_\Sc(\varepsilon_i)/\var_\Sc(w_i^*)$ is bounded on the interval $[0,1]$. 

\paragraph{Remark.} The extension for balancing weights in Section \ref{app:balancing} states that for Lemma \ref{lem:error_decomp} to hold for a set of balancing weights, the condition of $\E_\Sc(w_i^* \mid \bX_i) = w_i$ must hold. \\
\end{proof}

\subsection{Proof of Lemma \ref{lem:rho_bound} (Correlation Decomposition)} 
\noindent\fbox{%
\vspace{2mm}
\parbox{\textwidth}{%
\vspace{2mm}
The correlation between $\varepsilon_i$ and the individual-level treatment effects can be decomposed in the following manner:
\begin{align*}
\rho_{\varepsilon, \tau} &=
\begin{cases} 
\displaystyle \cor_\Sc(w_i, \tau_i) \sqrt{\frac{1-R_\varepsilon^2}{R_\varepsilon^2}} - \cor_\Sc(w^*_i, \tau_i) \cdot \frac{1}{\sqrt{R_\varepsilon^2}} R_\varepsilon^2 > 0 \\
0 &\text{when } R_\varepsilon^2 = 0 
\end{cases} 
\end{align*}
Furthermore, $\rho_{\varepsilon,\tau}$ is bounded by the following range: 
$$-\sqrt{1-\cor^2(w_i, \tau_i)} \leq \rho_{\varepsilon, \tau} \leq \sqrt{1-\cor^2(w_i, \tau_i)} $$
}
}
\begin{proof} 
To begin, we can rewrite $\rho_{\varepsilon, \tau}$ as follows:
\begin{align} 
\rho_{\varepsilon, \tau} &= \frac{\cov_\Sc(w_i, \tau_i) - \cov(w_i^*, \tau_i)}{\sqrt{\var_\Sc(\varepsilon_i) \cdot \var_\Sc(\tau_i)}} \nonumber \\
&= \frac{\cor_\Sc(w_i, \tau_i) \cdot \sqrt{\var_\Sc(w_i) \cdot \var_\Sc(\tau_i)} - \cor_\Sc(w^*_i, \tau_i) \cdot \sqrt{\var_\Sc(w_i^*) \cdot \var_\Sc(\tau_i)}}{\sqrt{\var_\Sc(\varepsilon_i) \cdot \var_\Sc(\tau_i)}}\nonumber \\
&= \frac{\cor_\Sc(w_i, \tau_i) \cdot \sqrt{\var_\Sc(w_i)} - \cor_\Sc(w^*_i, \tau_i) \cdot \sqrt{\var_\Sc(w_i^*)}}{\sqrt{\var_\Sc(w_i^*) \cdot  R^2_\varepsilon }}  \nonumber \\
&= \cor_\Sc(w_i, \tau_i)\sqrt{\frac{\var_\Sc(w_i)}{\var_\Sc(w_i^*)}} \cdot \frac{1}{\sqrt{R^2_\varepsilon}} - \cor_\Sc(w^*_i, \tau_i) \cdot \frac{1}{\sqrt{R^2_\varepsilon}} \nonumber \\
&= \cor_\Sc(w_i, \tau_i) \sqrt{\frac{1-R_\varepsilon^2}{R_\varepsilon^2}} - \cor_\Sc(w^*_i, \tau_i) \cdot \frac{1}{\sqrt{R_\varepsilon^2}} %\\
%&\in \pm \sqrt{1- \cor^2(w_i, \tau_i)}
\label{eqn:rho_decomp2} 
\end{align} 
Now, note that $\cor(w^*_i, \tau_i)$ can be bounded using the recursive formula of partial correlation:\footnote{This follows from applying the recursive formula of partial correlation for a single variable, and applying the fact that the partial correlation must be bounded by 1 and -1. Alternatively, we may note that for any three random vectors $a$, $b$, and $c$, the correlation of $a$ and $c$ can be bounded by: 
$$\cor(a,b) \cor(c, b) -\sqrt{1-\cor^2(a, b)}\cdot \sqrt{1-\cor^2(b,c)} \leq \cor(a,c) \leq \cor(a,b) \cor(c, b) + \sqrt{1-\cor^2(a, b)}\cdot \sqrt{1-\cor^2(b,c)}.$$ This can be shown by orthogonally decomposing $a$ and $c$ into the components that can be explained by $b$ and the components that cannot be explained (i.e., the orthogonal component).} 
$$\cor_\Sc(w^*_i, \tau_i) \in \cor_\Sc(w_i, \tau_i) \cdot \cor_\Sc(w_i, w_i^*) \pm \sqrt{1-\cor_\Sc^2(w_i, \tau_i)} \sqrt{1-\cor_\Sc^2(w_i, w_i^*)}$$
Because $\cor_\Sc(w_i, w_i^*) = \sqrt{\frac{\var_\Sc(w_i)}{\var_\Sc(w_i^*)}}$, the above simplifies to the following: 
$$\cor_\Sc(w_i, \tau_i) \sqrt{1-R_\varepsilon^2} - \sqrt{1-\cor_\Sc^2(w_i, \tau_i)} \sqrt{R_\varepsilon^2} \leq \cor_\Sc(w^*_i, \tau_i) \leq \cor_\Sc(w_i, \tau_i) \sqrt{ 1-R_\varepsilon^2} + \sqrt{1-\cor^2(w_i, \tau_i)} \sqrt{R_\varepsilon^2}$$ 

Thus, substituting in the bounds for $\cor(w^*_i, \tau_i)$ into Equation \eqref{eqn:rho_decomp2}, we obtain the bound: 
$$- \sqrt{1-\cor_\Sc^2(w_i, \tau_i)} \leq \rho_{\varepsilon, \tau} \leq \sqrt{1-\cor_\Sc^2(w_i, \tau_i)}$$

\end{proof} 

\subsection{Proof of Theorem \ref{thm:weight} (Bias of Weighted Estimator)} 
\noindent\fbox{%
\vspace{2mm}
\parbox{\textwidth}{%
\vspace{2mm}
Assume $Y_i(1) - Y_i(0) \ \indep \ S_i \ | \ \{ \bX_i, \bU_i\}$. Let $w_i$ be the weights estimated using only $\bX_i$, and let $w_i^*$ be the (correct) weights, obtained using $\{\bX_i, \bU_i\}$. The bias of a weighted estimator from using $w_i$ instead of $w_i^*$ is given as: 
$$\text{Bias}(\hat \tau_W) = \rho_{\varepsilon, \tau} \cdot \sqrt{\var_\Sc(w_i) \cdot \frac{R^2_\varepsilon}{1-R^2_\varepsilon} \cdot \sigma^2_\tau},$$
where $\varepsilon_i$ is defined as the difference between the estimated weights and the correct weights (i.e., $\varepsilon_i = w_i - w_i^*$), and $\tau_i$ is the individual-level treatment effect. 
}
}
\begin{proof} 
I will first show the proof for a Horvitz-Thompson style weighted estimator. The proof for a Hajek style weighted estimator (with stabilized weights) follows similarly, but with the addition of a finite-sample bias term. 
A Horvitz-Thompson style weighted estimator is defined as: 
$$\hat \tau_W = \frac{1}{n_1} \sum_{i \in \Sc} w_i T_i Y_i - \frac{1}{n_0} \sum_{i \in \Sc} w_i(1-T_i) Y_i$$

We begin by showing that if we were to have estimated weights with the full separating set $\mathcal{X}_i$, the weighted estimator will be an unbiased estimator for PATE. We will denote the weighted estimator using $w_i^*$ as $\hat \tau_W^*$. We will denote expectations with a subscript $\Sc$ as the expectation over the experimental sample (i.e., $\E_\Sc(\cdot) = \E(\cdot \mid S_i = 1)$), and expectations with a subscript $\Pc$ as the expectation over the target population. Expectations with no subscripts will represent the expectation over both the experimental sample and the target population. We define $\mathcal{D}$ as the set of all indices corresponding to units in the experimental sample and the target population. 
\begin{align} 
\E(\hat \tau_W^*) =& \E \left(\frac{1}{n_1} \sum_{i \in \Sc} w_i^* T_i Y_i - \frac{1}{n_0} \sum_{i \in \Sc} w_i^* T_i Y_i \right) \nonumber \\
 =& \E \left(\frac{1}{n_1} \sum_{i \in \cD} w_i^* T_i Y_i S_i - \frac{1}{n_0} \sum_{i \in \cD} w_i^* (1-T_i) Y_i S_i \right) \nonumber \\ 
=& \frac{1}{n_1} \E \left(\sum_{i \in \cD} w_i^* S_i T_i Y_i(1) \right)- \frac{1}{n_0} \E \left(\sum_{i \in \cD} w_i^* (1-T_i) Y_i(0) \right) \nonumber 
\intertext{By Linearity of Expectation:} 
=& \frac{1}{n_1}\sum_{i \in \cD} \E \left( w_i^* S_i T_i Y_i(1) \right)- \frac{1}{n_0}  \sum_{i \in \cD} \E \left( w_i^* (1-T_i) Y_i(0) \right) \nonumber 
\intertext{By Law of Total Expectation:} 
=&  \frac{1}{n_1}\sum_{i \in \cD} \E \left( w_i^* S_i T_i Y_i(1) | S_i = 1, T_i = 1 \right) P(S_i = 1 \text{ and } T_i = 1) +\nonumber \\
 &\frac{1}{n_0}\sum_{i \in \cD}  \E \left( w_i^* S_i (1-T_i) Y_i(0) | S_i = 1, T_i = 0 \right) \nonumber \\
 =&  \frac{1}{n_1}\sum_{i \in \cD} \frac{n}{n+N} \cdot \frac{n_1}{n} \E \left( w_i^* S_i T_i Y_i(1) | S_i = 1, T_i = 1 \right) +\nonumber \\
 &\frac{1}{n_0}\sum_{i \in \cD} \frac{n}{n+N} \cdot \frac{n_0}{n} \E \left( w_i^* S_i (1-T_i) Y_i(0) | S_i = 1, T_i = 0 \right) \nonumber \\
=& \E(w_i^* S_i Y_i(1) | S_i = 1, T_i = 1) - 
\E(w_i^* S_i Y_i(0) | S_i = 1, T_i = 0)  \nonumber 
\intertext{From random treatment assignment:} 
=& \E(w_i^* S_i Y_i(1) | S_i = 1) - 
\E(w_i^* S_i Y_i(0) | S_i = 1) \nonumber \\
=& \E(w_i^* S_i (Y_i(1) - Y_i(0)) | S_i = 1) \nonumber \\
\equiv& \E_\Sc(w_i^* \tau_i) 
\label{eqn:exp_weight} 
\end{align} 
To show that $\E_\Sc(w_i^* \tau_i) = \tau$, we first apply Baye's Rule: 
\begin{align*} 
\E_\Sc(w_i^* \tau_i) &=
\sum_{\tau_i, \mathcal{X}_i} w_i^* \tau_i \cdot P(\mathcal{X}_i, \tau_i | S_i = 1) \\
&=
\sum_{\tau, \mathcal{X}_i} w_i^* \tau_i \cdot \frac{P(S_i = 1 | \mathcal{X}_i, \tau_i) \cdot P(\mathcal{X}_i, \tau_i)}{P(S_i = 1)}
\intertext{By the conditional ignorability assumption that $\tau_i \indep S_i \mid \mathcal{X}_i$:} 
&= \sum_{\tau, \mathcal{X}_i} w_i^* \tau_i \cdot \frac{P(S_i = 1 | \mathcal{X}_i) \cdot P(\mathcal{X}_i, \tau_i)}{P(S_i = 1)}\\
&=
\sum_{\tau, \mathcal{X}_i} \frac{P(S_i = 1)}{P(S_i = 0)} \frac{P(S_i = 0 | \mathcal{X}_i)}{P(S_i = 1 | \mathcal{X}_i)} \tau_i \cdot \frac{P(S_i = 1 | \mathcal{X}_i) \cdot P(\mathcal{X}_i, \tau_i)}{P(S_i = 1)}\\
&=\sum_{\tau, \mathcal{X}_i} \tau_i \cdot \frac{P(S_i = 0 | \mathcal{X}_i) \cdot P(\mathcal{X}_i, \tau_i)}{P(S_i = 0)} \\
&=\sum_{\tau, \mathcal{X}_i} \tau_i \cdot P(\mathcal{X}_i, \tau_i | S_i = 0)\\
&= \E_\mathcal{P}(\tau_i) \\
&= \E_\mathcal{P}(\tau_i) \equiv \tau 
\end{align*} 

As such, the bias of a weighted estimator when omitting a confounder is: 
\begin{align*} 
\text{Bias}(\hat \tau_W) &= \E_\Sc(\hat \tau_W) - \tau 
\intertext{Using the result from Equation \eqref{eqn:exp_weight} with $w_i$ and the fact that $\E_\Sc(w_i^* \tau_i) = \tau$:} 
&= \E_\Sc(w_i \tau_i) - \E_\Sc( w_i^* \tau_i ) \\
&= \E_\Sc(\underbrace{(w_i - w_i^*)}_{:= \varepsilon_i} \tau_i) \\
&= \E_\Sc(\varepsilon_i \tau_i) \\
\intertext{By construction, $\E_\Sc(w_i) = \E_\Sc(w_i^*) = 1$, which implies that $\E_\Sc(\varepsilon_i) = 0$:} 
&= \E_\Sc(\varepsilon_i \tau_i ) - \E_\Sc(\varepsilon_i) \cdot \E_\Sc(\tau_i)\\
&= \cov_\Sc(\varepsilon_i, \tau_i) \\
&= \cor_\Sc(\varepsilon_i, \tau_i) \cdot \sqrt{\var_\Sc(\varepsilon_i) \cdot \var_\Sc(\tau_i)} 
\intertext{Define $R^2_\varepsilon := \var_\Sc(\varepsilon_i)/\var_\Sc(w_i^*)$ and making use of Lemma \ref{lem:var_decomp}:} 
&= \cor_\Sc(\varepsilon_i, \tau_i) \cdot \sqrt{\var_\Sc(w_i) \cdot \frac{R^2_\varepsilon}{1-R^2_\varepsilon} \cdot \var_\Sc(\tau_i)} \\
&\equiv \rho_{\varepsilon, \tau} \cdot \sqrt{\var_\Sc(w_i) \cdot \frac{R^2_\varepsilon}{1-R^2_\varepsilon} \cdot \sigma^2_\tau}
\end{align*} 
\end{proof}

%%%%%%%%%%%%%%%%%%%%%%%%%%%%%%%%%%%%%%%%%%%%%%%%%%%%%%%%%%%%%%%%%%%
\subsection{Proof of Theorem \ref{thm:benchmark}}
\noindent\fbox{%
\vspace{2mm}
\parbox{\textwidth}{%
\vspace{2mm}
Let $k_\sigma$ and $k_\rho$ be defined as in Equation \eqref{eqn:k}. Let $R^{2-(j)}_\varepsilon := \var_\Sc(\varepsilon_i^{-(j)})/\var_\Sc(w_i)$, and $\rho_{\varepsilon, \tau}^{-(j)} := \cor_\Sc(\varepsilon_i^{-(j)}, \tau_i)$. The sensitivity parameters $R^2_\varepsilon$ and $\rho_{\varepsilon, \tau}$ can be written as a function of $k_\sigma$ and $k_\rho$: 
$$R^2_\varepsilon = \frac{k_\sigma \cdot R^{2-(j)}_\varepsilon}{1+ k_\sigma \cdot R^{2-(j)}_\varepsilon}, \ \ \ \ \ \  \rho_{\varepsilon, \tau} = k_\rho \cdot \rho_{\varepsilon, \tau}^{-(j)} $$ 
}
}

\begin{proof} 
It follows immediately from Equation \eqref{eqn:k} that $\rho_{\varepsilon, \tau} = k_\rho \cdot \rho_{\varepsilon, \tau}^{-(j)}$. Therefore, we just need to show that $R^2_\varepsilon$ can be written as a function of $R^{2-(j)}_\varepsilon$.
\begin{align*} 
R^2_\varepsilon &= \frac{\var_\Sc(\varepsilon_i)}{\var_\Sc(w_i^*)}\\
\intertext{By Equation \eqref{eqn:k}:} 
&= k_\sigma \cdot \frac{\var_\Sc(\varepsilon^{-(j)}_i)}{\var_\Sc(w_i^*)} \\
&= k_\sigma \cdot \frac{\var_\Sc(\varepsilon^{-(j)}_i)}{\var_\Sc(w_i)+ \var_\Sc(\varepsilon_i)} \\
&= \frac{k_\sigma \cdot \var_\Sc(\varepsilon_i^{-(j)})/\var_\Sc(w_i)}{1+k_\sigma \var_\Sc(\varepsilon_i^{-(j)})/\var_\Sc(w_i)} \\
&= \frac{k_\sigma \cdot R^{2-(j)}_{\varepsilon}}{1+k_\sigma \cdot R^{2-(j)}_{\varepsilon}}
\end{align*} 
\end{proof} 
 %%%%%%%%%%%%%%%%%%%%%%%%%%%%%%%%%%%%%%%%%%%%%%%%%%%%%%%%%
\subsection{Proof of Theorem \ref{thm:dr}}
\noindent\fbox{%
\vspace{2mm}
\parbox{\textwidth}{%
\vspace{2mm}
The bias of an augmented weighted estimator when both the weight and outcome model are mis-specified is given as: 
$$\text{Bias}(\hat \tau_{W}^{Aug}) =\rho_{\varepsilon, \xi} \cdot \sqrt{\var_\Sc(w_i) \cdot \frac{R^2_\varepsilon}{1-R^2_\varepsilon} \cdot \var_\Sc(\xi_i)}$$
where $\varepsilon_i$ is defined consistent with before, and $\xi_i$ represents the difference between the true individual-level treatment effect and estimated treatment effect (i.e., $\xi_i = \tau_i - \hat \tau_i$). 
}
}
\begin{proof} 
\begin{align*} 
\hat \tau^{Aug}_W = \hat \tau_W - \underbrace{\frac{1}{n} \sum_{i \in \Sc} w_i \hat \tau(\bX_i) + \frac{1}{N} \sum_{i \in \mathcal{P}} \hat \tau(\bX_i)}_{\text{Augmented Component}}
\end{align*} 
From Theorem \ref{thm:weight}, we showed that $\E(\hat \tau_W) = \E_\Sc(w_i \tau_i)$. We will now derive the expectation of the augmented component. To begin, we take the expectation of the $1/n \sum_{i \in \Sc} w_i \hat \tau(\bX_i)$ component: 
\begin{align*} 
\E \left( \frac{1}{n} \sum_{i \in \Sc} w_i \hat \tau(\bX_i) \right) &= \frac{1}{n} \E_\Sc\left( \sum_{i \in \Sc} w_i \hat \tau(\bX_i) \right) \\
&=  \frac{1}{n} \E_\Sc\left( \sum_{i \in \mathcal{D}} S_i w_i \hat \tau(\bX_i) \right) \\
&= \frac{1}{n} \sum_{i \in \mathcal{D}} \E(S_i w_i \hat \tau(\bX_i)) \\
&= \frac{1}{n} \sum_{i \in \mathcal{D}} \E(S_i w_i \hat \tau(\bX_i) \mid S_i = 1) P(S_i = 1) \\
&= \E(w_i \hat \tau(\bX_i) \mid S_i = 1) \\
&= \E_\Sc(w_i \hat \tau(\bX_i)
\end{align*} 
For the $1/N \sum_{i \in \mathcal{P}} \hat \tau(\bX_i)$ component: 
\begin{align*} 
\E \left(\frac{1}{N} \sum_{i \in \mathcal{P}} \hat \tau(\bX_i) \right) &= \E \left(\frac{1}{N} \sum_{i \in \mathcal{D}} (1-S_i) \cdot \hat \tau(\bX_i) \right) \\
&= \frac{1}{N} \sum_{i \in \mathcal{D}} \E ((1-S_i) \hat \tau(\bX_i)) \\
&= \frac{1}{N} \sum_{i \in \mathcal{D}} \E ((1-S_i) \hat \tau(\bX_i) \mid S_i= 0) P(S_i = 0) \\
&= \E( \hat \tau(\bX_i) \mid S_i = 0) \\
&= \E_\mathcal{P}( \hat \tau(\bX_i))
\end{align*} 
As such, the bias of the augmented weighted estimator can be written as follows: 
\begin{align*} 
\text{Bias}(\hat \tau_W^{Aug}) &= \E( \hat \tau_W^{Aug}) - \tau \\
&= \E_\Sc(w_i (\tau_i - \hat \tau(\bX_i)) + \E_\mathcal{P}(\hat \tau(\bX_i)) - \E_\mathcal{P}(\tau_i) \\
&= \E_\Sc(w_i(\tau_i - \hat \tau(\bX_i)) - \E_\Pc(\tau_i - \hat \tau_i) 
\intertext{By definition, $\varepsilon_i = w_i - w_i^*$:} 
&= \E_\Sc(\varepsilon_i(\tau_i - \hat \tau_i)) + \E_\Sc(w_i^* (\tau_i - \hat \tau_i)) - \E_\Pc(\tau_i - \hat \tau_i) \\
&= \E_\Sc(\varepsilon_i(\tau_i - \hat \tau_i)) + \E_\Pc(\tau_i - \hat \tau_i) - \E_\Pc(\tau_i - \hat \tau_i)  \\
&= \E_\Sc(\varepsilon_i(\tau_i - \hat \tau_i))
\intertext{Defining $\xi_i := \tau_i - \hat \tau_i$:} 
&= \E_\Sc(\varepsilon_i \cdot \xi_i) \\
&= \cov_\Sc(\varepsilon_i, \xi_i) \\
&= \cor_\Sc(\varepsilon_i, \xi_i) \cdot \sqrt{\var_\Sc(\varepsilon_i) \cdot \var_\Sc(\xi_i)}\\
&= \rho_{\varepsilon, \xi} \cdot \sqrt{\var_\Sc(w_i) \cdot \frac{R^2_\varepsilon}{1-R^2_\varepsilon} \cdot \var_\Sc(\xi_i)}
\end{align*} 
\end{proof}
%%%%%%%%%%%%%%%%%%%%%%%%%%%%%%%%%%%%%%%%%%%%%%%%%%%%%%%%%%%%%%%%%%%%%%%%

\section{Additional Derivations} \label{app:additional_deriv} 
\subsection{Proof of Example \ref{ex:dim}}
Consider the case in which no weighting adjustment has been made, and we use the difference-in-means estimate in the experimental sample as an estimator for PATE. In that situation, $\varepsilon_i = \frac{1}{n} - w_i^*$, as there is equal weighting (i.e., $w_i = 1/n$) for all units in the sample. As a result, the bias of a difference-in-means estimator in estimating the PATE will be: 
$$\text{Bias}(\hat \tau_\Sc) = \cor_\Sc(w^*_i, \tau_i) \sqrt{\var_\Sc(w_i^*) \cdot \sigma^2_{\tau}},$$
which corresponds to the bias expression for using SATE to estimate PATE derived in \cite{miratrix_2018}. The relative reduction in bias from weighting is as follows:
\begin{align*} 
\text{Relative Reduction} = \left| \frac{\text{Bias}(\hat \tau_W)}{\text{Bias}(\hat \tau_\Sc)} \right|
&= \left| 1-\sqrt{(1-R_\varepsilon^2)} \cdot \frac{\cor_\Sc(w_i, \tau_i)}{\cor_\Sc(w^*_i, \tau_i)} \right|.
\end{align*} 
\begin{proof} 
We will begin by substituting in the correlation decomposition from Lemma \ref{lem:rho_bound} to our bias decomposition of a weighted estimator: 
\begin{align*} 
\text{Bias}(\hat \tau_W) &= \rho_{\varepsilon, \tau} \cdot \sqrt{\var_\Sc(w_i) \cdot \frac{R^2_\varepsilon}{1-R^2_\varepsilon}  \cdot \sigma^2_\tau} 
\intertext{Substitute in the correlation decomposition from Lemma \ref{lem:rho_bound}:} 
&= \left( \cor_\Sc(w_i, \tau_i) \cdot \sqrt{\frac{1-R^2_\varepsilon}{R^2_\varepsilon}} - \cor_\Sc(w_i^*, \tau_i) \cdot \sqrt{\frac{1}{R^2_\varepsilon}} \right) \cdot \sqrt{\var_\Sc(w_i) \cdot \frac{R^2_\varepsilon}{1-R^2_\varepsilon}  \cdot \sigma^2_\tau} \\
&= \left(\cor_\Sc(w_i, \tau_i) - \cor_\Sc(w_i^*, \tau_i) \cdot \frac{1}{\sqrt{1-R^2_\varepsilon}}\right) \cdot \sqrt{\var_\Sc(w_i) \cdot \sigma^2_\tau}
\end{align*} 
Now, calculating the relative reduction: 
\begin{align*} 
\text{Relative Reduction} &= \left| \frac{\text{Bias}(\hat \tau_W)}{\text{Bias}(\hat \tau_\Sc)} \right|\\
&= \left| \frac{\left(\cor_\Sc(w_i, \tau_i) - \cor_\Sc(w_i^*, \tau_i) \cdot \frac{1}{\sqrt{1-R^2_\varepsilon}}\right) \cdot \sqrt{\var_\Sc(w_i) \cdot \sigma^2_\tau}}{\cor_\Sc(w_i^*, \tau_i) \cdot \sqrt{\var_\Sc(w_i^*) \cdot \sigma^2_\tau}}\right|\\
&=\left| \frac{\left(\cor_\Sc(w_i, \tau_i) - \cor_\Sc(w_i^*, \tau_i) \cdot \frac{1}{\sqrt{1-R^2_\varepsilon}}\right) \cdot \sqrt{\var_\Sc(w_i)}}{\cor_\Sc(w_i^*, \tau_i) \cdot \sqrt{\var_\Sc(w_i^*)}} \right|
\intertext{Note that $\var_\Sc(w_i)/\var_\Sc(w_i^*) = 1- R^2_\varepsilon$:}
&= \left|\frac{\left(\cor_\Sc(w_i, \tau_i) - \cor_\Sc(w_i^*, \tau_i) \cdot \frac{1}{\sqrt{1-R^2_\varepsilon}}\right)}{\cor(w_i^*, \tau_i)} \cdot \sqrt{1-R^2_\varepsilon} \right|\\
&= \left| 1-\sqrt{(1-R_\varepsilon^2)} \cdot \frac{\cor_\Sc(w_i, \tau_i)}{\cor_\Sc(w^*_i, \tau_i)} \right|
\end{align*} 
\end{proof} 
\subsection{Robustness Value} 
To derive the robustness value, recall that we are interested in the bias that arises from a confounder with equal impact on the overall imbalance and the individual-level treatment effect. In particular, we define the robustness value such that $RV = \rho_{\varepsilon, \tau}^2 = R_\varepsilon^2$. Thus, for the bias to equal $q \times 100\%$ of a given point estimate:
$$\text{Bias}(\hat \tau_W) = q \cdot \hat \tau_W $$
From Theorem \ref{thm:weight}:
 \begin{align} 
 \implies \rho_{\varepsilon, \tau} \sqrt{\var_\Sc(w_i) \cdot \frac{R_\varepsilon^2}{1-R_\varepsilon^2} \cdot \sigma^2_\tau} &= q \cdot \hat \tau_W  \nonumber \\ 
 \rho_{\varepsilon, \tau}^2 \cdot \var_\Sc(w_i) \cdot \frac{R_\varepsilon^2}{1-R_\varepsilon^2} \cdot \sigma^2_\tau &= q^2 \cdot \hat \tau_W^2 \nonumber \\
 \rho_{\varepsilon, \tau}^2 \cdot \frac{R_\varepsilon^2}{1-R_\varepsilon^2} &= \underbrace{\frac{q^2 \cdot \hat \tau_W^2}{\var_\Sc(w_i) \cdot \sigma^2_\tau}}_{:=a_q} \nonumber 
 \intertext{Defining $RV = \rho^2_{\varepsilon, \tau} = R^2_\varepsilon$:}
 RV \cdot \frac{RV}{1-RV} &= a_q
 \label{eqn:rv_final} 
 \end{align} 
Let $RV_q$ be the value of $RV$ for a given $q$. Thus, solving Equation \eqref{eqn:rv_final} for $RV_q$:
$$RV_q = \frac{1}{2} \left(\sqrt{a_q^2 + 4a_q} - a_q \right)$$ 
A similar derivation can be applied for the augmented weighted estimator, with the primary difference being that instead of $a_q$, the robustness value is a function of $b_q$, where: 
$$b_q = \frac{q^2 \cdot \left(\hat \tau_W^{Aug} \right)^2}{\sigma^2_\xi \cdot \var(w_i)}$$
\subsection{Extreme Bounds} 
To derive $R^2_{max}$, we set $\rho_{\varepsilon, \tau}$ to be at the extreme bounds of $\pm \sqrt{1-\cor_\Sc^2(w_i, \tau_i)}$, and solve for $R_\varepsilon^2$ using the correlation decomposition from Lemma \ref{lem:rho_bound}. 
\begin{align*} 
\cor_\Sc(w_i, \tau_i) \cdot \sqrt{\frac{1-R_\varepsilon^2}{R_\varepsilon^2}} - \cor_\Sc(w^*_i, \tau_i) \cdot \frac{1}{\sqrt{R_\varepsilon^2}} \leq \sqrt{1-\cor_\Sc^2(w_i, \tau_i)} \\
\cor_\Sc(w_i, \tau_i) \sqrt{1-R_\varepsilon^2} - \cor_\Sc(w^*_i, \tau_i) \leq \sqrt{1-\cor_\Sc^2(w_i, \tau_i)} \cdot \sqrt{1-\cor_\Sc^2(w_i^*, \tau_i)}
\end{align*} 
Using the quadratic formula, we solve for $\sqrt{R_\varepsilon^2}$: 
\begin{align*} 
\sqrt{R_\varepsilon^2} = 1-\cor_\Sc(w_i, \tau_i) \cdot \cor_\Sc(w^*_i, \tau_i) \pm \sqrt{(1-\cor_\Sc^2(w_i^*, \tau_i)} \cdot \sqrt{(1-\cor_\Sc^2(w_i, \tau_i)} 
\end{align*} 
Setting $\cor_\Sc(w^*_i, \tau_i) = \pm 1$, we find that 
$$\underbrace{R_\varepsilon^2}_{:=R^2_{max}} = 1- \cor_\Sc(w_i, \tau_i)^2$$

\clearpage 

\section{Extended Results for Empirical Application: JTPA}\label{app:jtpa}

\subsection{Bounding $\sigma^2_\tau$} \label{app:jtpa_sigma_tau} 
To estimate a bound for $\sigma^2_\tau$, we follow Figure \ref{fig:treatment_het} and use the upper bound estimated in Equation \eqref{eqn:vartau_bound_pos_cov} (i.e., for cases when we assume $\cov_\Sc(\tau_i, Y_i(0)) \geq 0$). This is in line with the substantive findings from the original JTPA study. To reiterate the example from Section \ref{app:treat_het}, researchers found that women with the greatest estimated impact from JTPA services also had higher hourly wages in their work history, and or came from families with greater household income. (See \cite{bloom1993national} for more discussion.) 

Therefore, we bound $\sigma^2_\tau$ by taking the difference between the estimated variance in the treated outcomes and the estimated variance in the control outcomes and obtain a bound of $8.4$. 

\subsection{Applying the Augmented Weighted Sensitivity Analysis} \label{app:jtpa_aug} 
To illustrate the sensitivity analysis for augmented weighted estimators, we return to our JTPA application. To estimate the individual-level treatment effect model, we use a causal random forest, estimated on the same set of covariates included in the weights (\cite{athey2019generalized}). We then estimate the individual-level treatment effect for all units across both the experimental sample and the target population. Using the bound from Equation \eqref{eqn:var_xi_bound}, we estimate an upper bound for $\sigma^2_\xi$ to be 6.79. After obtaining the upper bound for $\sigma^2_\xi$, we proceed with the sensitivity analysis. 

\paragraph{Summarizing Sensitivity.} To begin, we visualize the bias contour plot, as well as estimate the robustness value and the extreme scenario bound. See Figure \ref{fig:jtpa_contour_aug} for the bias contour plot. 

\begin{table}[!ht] 
\centering 
\begin{tabular}{lccccc} \toprule 
& Unweighted & Aug-Weighted &  $RV^{Aug}_{q=1}$  \\ \midrule
Impact of JTPA access on earnings & 1.11 & 1.36 & 0.44 \\ \bottomrule
\multicolumn{3}{l}{\small $\hat \sigma^2_{\xi, \max} = 6.79$; $\widehat{\cor}_\Sc(w_i, \xi_i) = 0.002$ (Extreme scenario bound: $0.99$)}
\end{tabular} 
\caption{Summary of point estimates and sensitivity statistics.} 
\label{tbl:sensitivity_summary_aug} 
\end{table} 

The robustness value for the augmented weighted estimator is 0.44, which implies that if the error from omitting the confounder can explain 44\% of the variation in the idiosyncratic treatment effect (i.e., $\xi_i$), as well as 44\% of the variation in the ideal weights, then the bias will be large enough to reduce the point estimate to 0. We see that the robustness value for the augmented weighted estimator is slightly higher than the robustness value for the weighted estimator. This is likely due to the fact that we have modeled some of the variation in $\tau_i$ with our estimated treatment effect heterogeneity model. 

The extreme scenario bound is extremely conservative, as the estimated correlation between $w_i$ and $\xi_i$ is very low. This is likely due to the fact that the component of $\tau_i$ that is related to the weights $w_i$ are also related to the covariates $\bX_i$. Because $\xi_i$ represents the component of $\tau_i$ that cannot be explained by $\bX_i$, the correlation between $w_i$ and $\xi_i$ is relatively low. 

\paragraph{Formal Benchmarking Results.} We now perform formal benchmarking across the observed covariates for the augmented weighted estimator. The formally benchmarked parameter values for the $R^2_\varepsilon$ parameter will be identical to the formally benchmarked $R^2_\varepsilon$ values in the weighted estimator setting. In general, we that the estimated bias values from formal benchmarking in the augmented weighted estimator case is lower than the estimated bias values for the weighted estimator case; this is likely due to the fact that the bound on the idiosyncratic treatment effect variation is lower than the bound on the overall treatment effect heterogeneity (i.e., $\hat \sigma^2_{\xi, \max} \leq \hat \sigma^2_{\tau, \max}$). 
\begin{table}[ht]
\centering
\begin{tabular}{lcccccc}
  \toprule
Covariate & $R^2_\varepsilon$ & $\rho_{\varepsilon, \xi}$ & Est. Bias & MRCS & $k_\sigma^{min}$ & $k_\rho^{min}$ \\ 
  \midrule
  Prev. Earnings & 0.04 & 0.10 & 0.05 & 28.65 & 10.81 & 6.63 \\ 
  Age & 0.06 & 0.38 & 0.22 & 6.23 & 7.47 & 1.75 \\ 
  Married & 0.11 & 0.00 & 0.00 & 689.99 & 4.13 & 267.86 \\ 
  Hourly Wage & 0.05 & -0.10 & -0.05 & -25.13 & 8.99 & -6.41 \\ 
  Black & 0.20 & -0.18 & -0.20 & -6.65 & 2.19 & -3.75 \\ 
  Hisp. & 0.14 & -0.06 & -0.05 & -25.56 & 3.26 & -11.36 \\ 
  HS/GED & 0.12 & 0.05 & 0.04 & 32.03 & 3.79 & 13.04 \\ 
  Years of Educ. & 0.00 & 0.24 & 0.02 & 79.71 & 442.21 & 2.83 \\ 
   \bottomrule
\end{tabular}
\caption{Formal benchmarking for Omaha, Nebraska, for an augmented weighted estimator. We see a greater degree of robustness in omitting a confounder with equivalent confounding strength to the observed covariates for the augmented weighted estimator, relative to the weighted estimator. This is reflected in the larger MRCS, $k_\sigma^{min}$ and $k_\rho^{min}$ values.} 
\end{table}

\begin{figure}[!ht]
\centering
\includegraphics[width=0.5\textwidth]{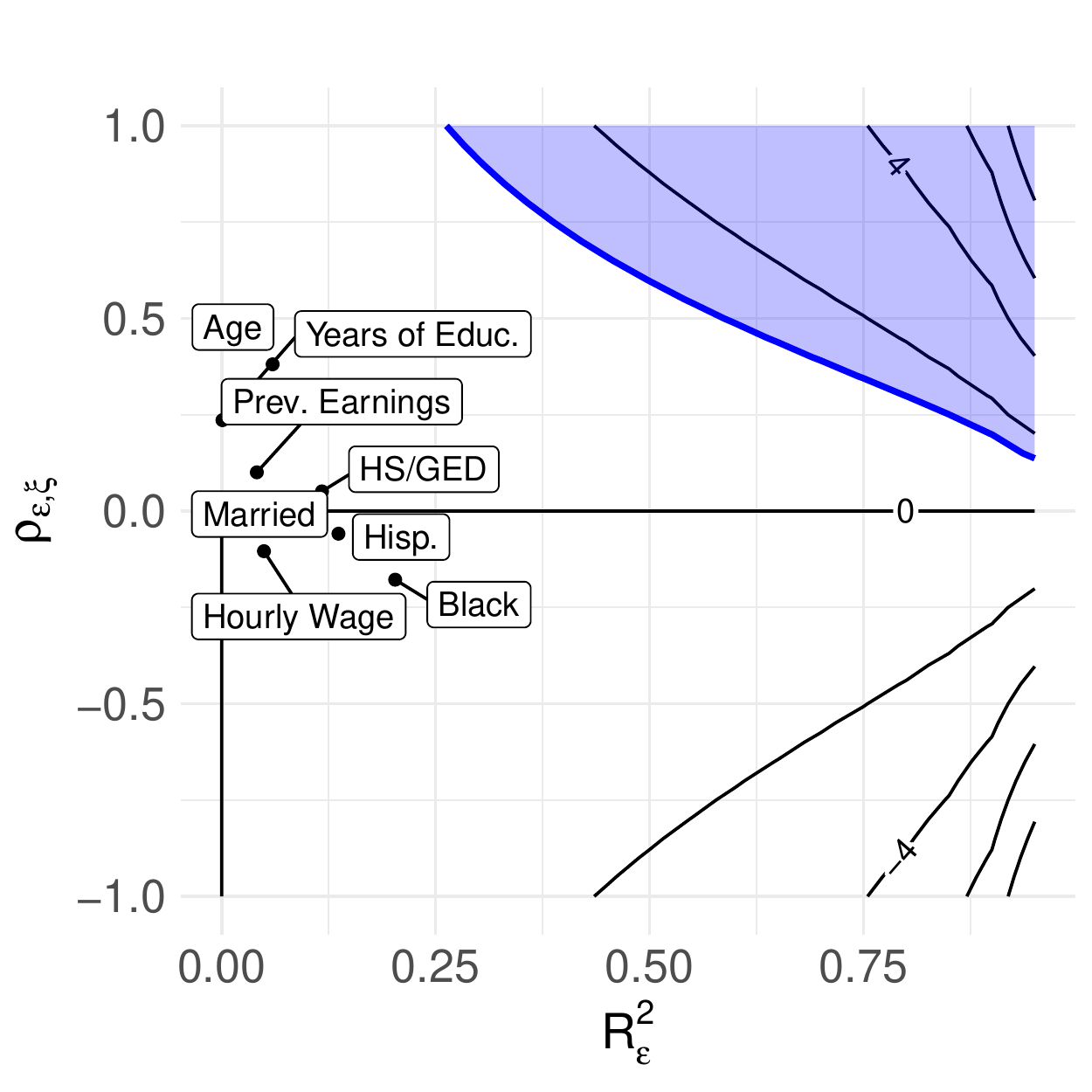}
\caption{Bias Contour Plot for Omaha, Nebraska, using an Augmented Weighted Estimator. Akin to Figure \ref{fig:jtpa_contour}, the shaded blue region represents the killer confounder region, for which a confounder will result in a directional change of the point estimate. We also plot the formal benchmarking results. We see that the points are even further away from the killer confounder region than in the weighted estimator setting.} 
\label{fig:jtpa_contour_aug} 
\end{figure}

\subsection{Extreme Scenario Analysis} 
For the extreme scenario analysis, we examine the potential bias that may occur if the correlation term is equal to the maximum possible value of $\sqrt{1-\cor_\Sc(w_i, \tau_i)}$. Then, we evaluate the $R_\varepsilon^2$ value that corresponds to this maximum correlation term, when $|\cor_\Sc(w^*_i, \tau_i)| = 1$ or $0.5$. In general, we expect this to be an extremely conservative estimate for the maximum amount of bias incurred by an omitted confounder. 

\begin{table}[!ht]
\centering 
\begin{tabular}{lcccccc} \hline
& Estimate & $\rho_{max}$ & $R^2_{max}$ & Est. Bias & $R^{2(0.5)}_{max}$ & Est. Bias \\ \hline
Weighted & 1.36 & 0.997  & 0.994 & 34.796 & 0.786 &  4.838 \\
Augmented & 1.31 & 0.998 & 0.997 & 45.4872 & 0.778 & 4.667 \\ \hline 
\end{tabular} 
\end{table} 

The general plausibility of an omitted confounder with the degree of explanatory power and imbalance seems relatively low. In particular, comparing $\rho_{max}$ and $R^2_{max}$ with the calibrated parameters shows that the omitted confounder would have to be significantly stronger than any of the observed covariates for the extreme scenario to occur. 

In cases when researchers do not feel that the benchmarked parameters are representative of the potential confounders, it can be difficult to justify the plausibility or implausibility of such an extreme $\rho_{\varepsilon, \tau}$ (or $\rho_{\varepsilon, \xi}$) term. An alternative approach is for researchers to vary different $\cor_\Sc(w^*_i, \tau_i)$ (or $\cor_\Sc(w_i^*, \xi_i)$) values, which can be easier to assess the plausibility of, because they can directly  compare the posited $\cor_\Sc(w^*_i, \tau_i)$ (or $\cor_\Sc(w_i^*, \xi_i)$) with the observed correlation values calculated using the estimated weights. $\cor_\Sc(w^*_i, \tau_i)$ represents the maximum amount of variation that the selection weights can explain in the treatment effect heterogeneity. For example, if researchers assume that the (true) selection weights are highly correlated with the treatment effect heterogeneity, then $\cor_\Sc(w^*_i, \tau_i)$ should be close to 1. 

To visually represent this, we generate plots where the $x$-axis represents the $R^2$ value, and the $y$-axis represents the adjusted point estimate. We fix $\cor_\Sc(w^*_i, \tau_i)$ and $\cor_\Sc(w_i^*, \xi_i)$ to a set of values: \{-0.9, -0.5, 0.25, 0.25, 0.5, 0.9\}. The estimated correlation value between the estimated weights and the individual-level treatment effect is 0.07, while the estimated correlation value between the estimated weights and the idiosyncratic treatment effect is 0.11. Thus, even for the case that $|\cor_\Sc(w^*_i, \tau_i)|$ or $|\rho(w_i^*, \xi_i)|$ to equal 0.25 would imply that additionally balancing on an omitted confounder would result in a significantly higher amount of variation explained. We see that for both the weighted and augmented weighted estimators, it is only when the correlation term switches signs that the point estimate is at risk of being zero, or negative. In other words, additionally balancing on the omitted confounder would have to alter the direction of the correlation between the weights and $\tau_i$ (or $\xi_i$) for the point estimate to become negative. 

\begin{figure}[!ht]
\centering
 \textbf{Extreme Scenario Analysis Plots for NE} \\ \vspace{2mm} 
\includegraphics[width=0.45\textwidth]{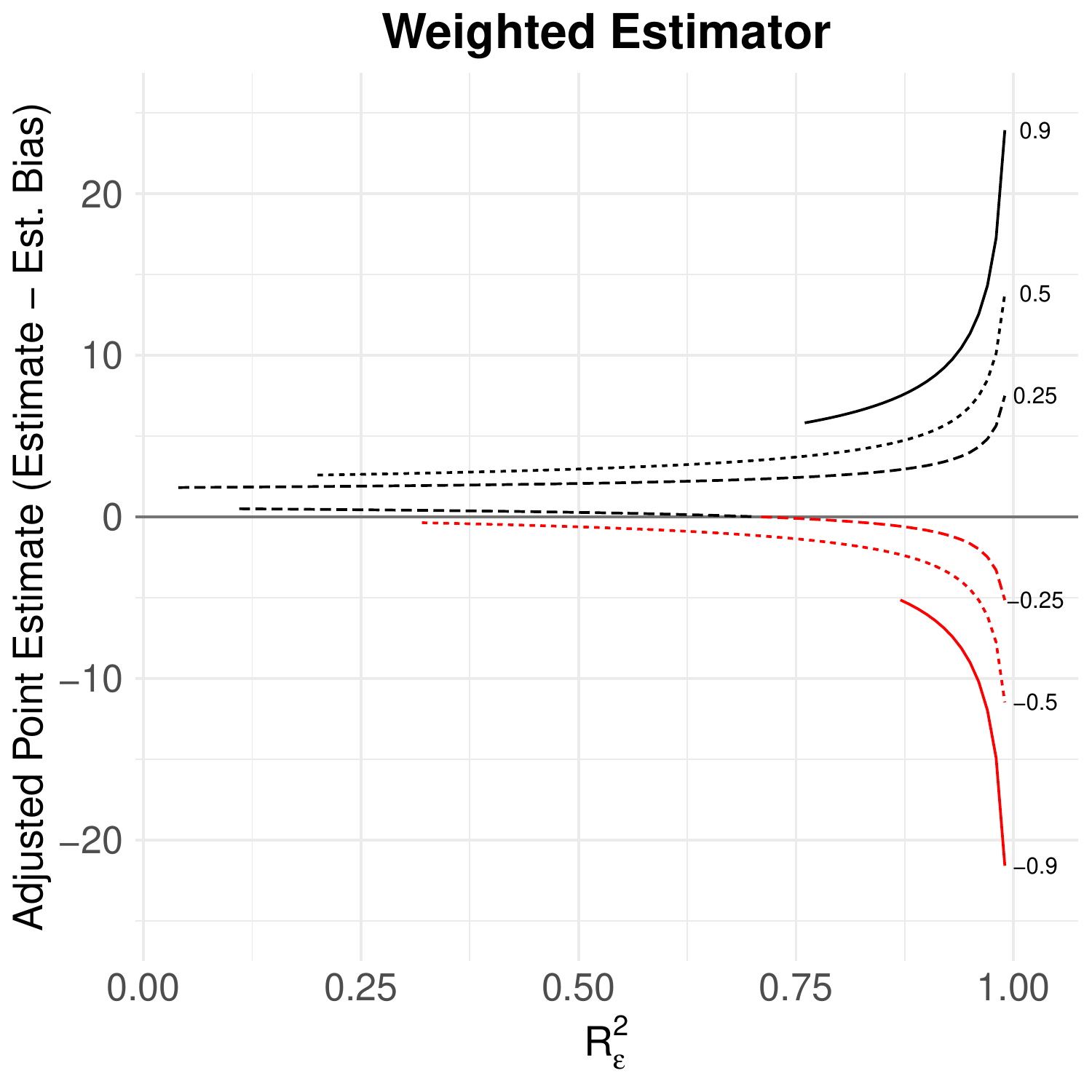}
\includegraphics[width=0.45\textwidth]{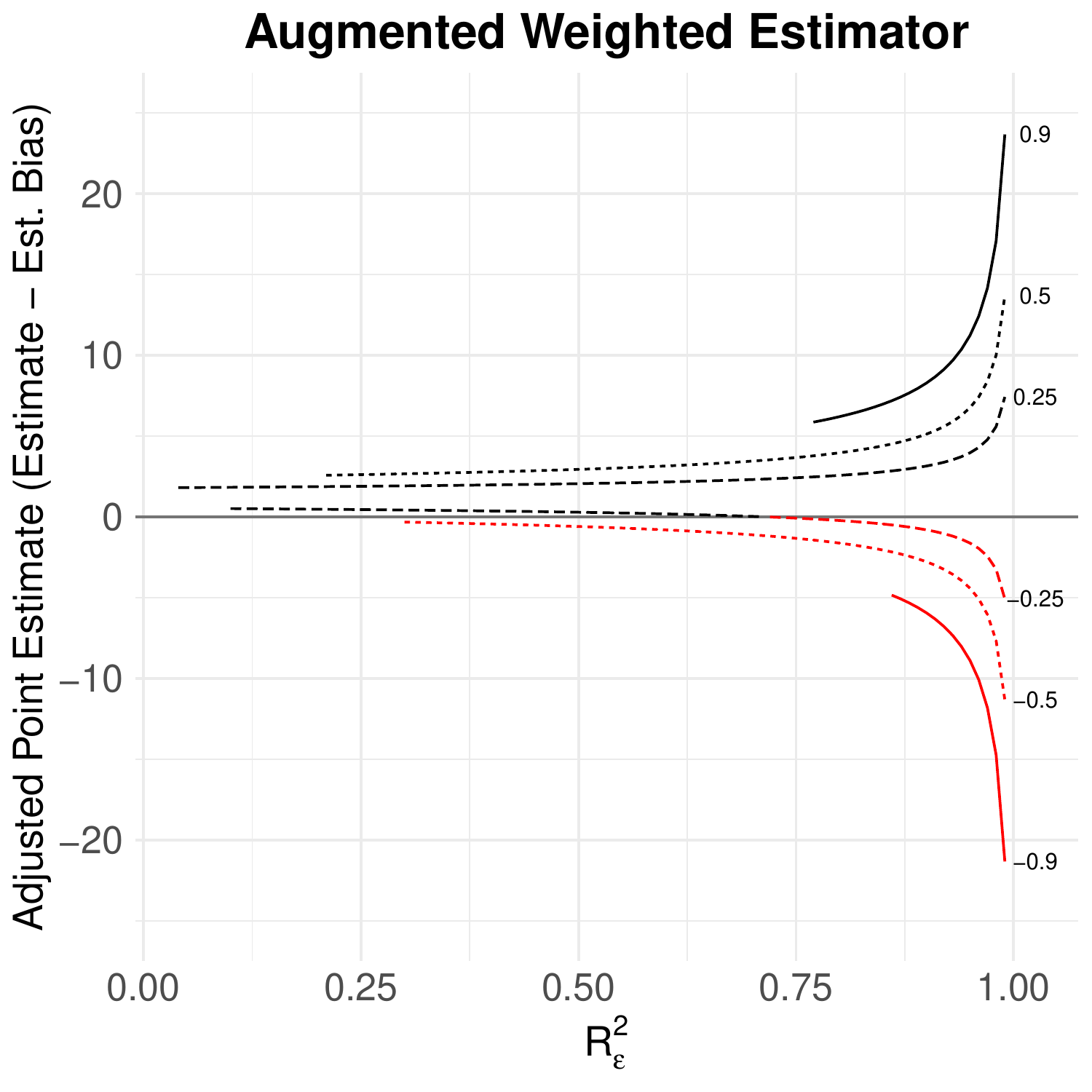}
\caption{We vary different values of $\cor_\Sc(w^*_i, \tau_i)$ (and $\cor(w_i^*, \xi_i)$). We set the $x$-axis to be different $R_\varepsilon^2$ values, and the $y$-axis to be the adjusted point estimate (i.e., the point estimate minus the estimated bias). The lines marked by red represent results that would alter the interpretation of the point estimate.} 
\end{figure} 

\clearpage
\subsection{Different Experimental Site: Corpus Christi, Texas} 
While the experimental site of Omaha, Nebraska was relatively robust to confounders, we now run the same sensitivity analysis across an experimental site with more sensitivity to confounders. The experimental site chosen is Corpus Christi, Texas. The site is similarly sized to Omaha, Nebraska, and has 524 individuals, while the population consists of 5,701 individuals. Consistent with before, the same set of pre-treatment covariates are used to estimate selection weights with entropy balancing (see Table \ref{tbl:cov_bal_cc} for covariate balance summary). The point estimate for the weighted estimator is $0.73$.

\paragraph{Summary of Sensitivity Statistics} 
We estimate an upper bound of 10.04 for $\sigma^2_\tau$. A summary of the point estimates and robustness values is provided in the following table: 
\begin{table}[!ht]
\centering 
\begin{tabular}{lccc} \toprule 
& Unweighted & Weighted & $RV_{q=1}$\\ \midrule
Impact of JTPA access on earnings & -0.21 & 0.73 & 0.14 \\  \bottomrule 
\multicolumn{3}{l}{\small $\hat \sigma^2_{\tau,max} = 10.04$; $\widehat{\cor}_\Sc(w_i, \tau_i) = 0.20$ (Extreme scenario bound: 0.98)}
\end{tabular}
\end{table} 

The weighted estimator has an associated robustness value of 0.14. This implies that the error in the weights for omitting a confounder must explain 14\% of the variation in the individual-level treatment effect, and 14\% of the variation in the true weights, in order for the resulting bias to be large enough to reduce the treatment effect to zero. Recall that for the site of Omaha, Nebraska, the robustness value was estimated to be around 0.41. As such, for the site of Corpus Christi, Texas, the omitted confounder has to explain much less relative variation in both the treatment effect heterogeneity and the selection process to reduce the treatment effect to zero. We conclude that in comparison to Omaha, Nebraska, there is less robustness in the site of Corpus Christi, Texas. 

\paragraph{Formal Benchmarking Results.} 
We now perform formal benchmarking, and calculate the MRCS and $\{k_{\sigma}^{min}, k_\rho^{min}\}$. We see that omitting a confounder with similar confounding strength to the covariates of whether or not the individual is black or whether or not the individual has a high school diploma will result in the largest amount of bias ($0.48$). Furthermore, the greater sensitivity to a potential killer confounder is reflected in the MRCS values. Recall for the site of Omaha, Nebraska, the smallest MRCS values were around 3-4, indicating that an omitted confounder would have to be 3-4 times stronger than the strongest observed covariate in order to be a killer confounder. In contrast, for Corpus Christi, Texas, we see from the MRCS values that an omitted confounder need only be 1-2 times stronger than the strongest covariates in order tor result in a killer confounder.
\begin{table}[ht]
\centering
\begin{tabular}{lcccccc}
  \toprule
Covariate & $R^2_\varepsilon$ & $\rho_{\varepsilon, \tau}$ & Est. Bias & MRCS & $k_\sigma^{min}$ & $k_\rho^{min}$ \\ 
  \midrule
Prev. Earnings & 0.07 & 0.06 & 0.08 & 9.31 & 2.01 & 6.31 \\ 
  Age & 0.00 & -0.35 & -0.03 & -22.45 & 377.41 & -1.07 \\ 
  Married & 0.00 & -0.48 & -0.03 & -21.27 & 637.56 & -0.78 \\ 
  Hourly Wage & 0.01 & -0.04 & -0.02 & -42.97 & 14.18 & -10.63 \\ 
  Black & 0.01 & 0.94 & 0.48 & 1.52 & 12.55 & 0.40 \\ 
  Hisp. & 0.43 & 0.06 & 0.27 & 2.74 & 0.33 & 5.82 \\ 
  HS/GED & 0.10 & 0.30 & 0.48 & 1.52 & 1.44 & 1.24 \\ 
  Years of Educ. & 0.03 & 0.22 & 0.17 & 4.41 & 5.60 & 1.75 \\ 
   \bottomrule 
   \multicolumn{5}{l}{\small Point Estimate: 0.73; $\hat \sigma^2_{\tau, max} = 10.04$; $RV_{q=1} = 0.14$}
\end{tabular}
\caption{Formal benchmarking results for the site of Corpus Christi, Texas. The greater degree of relative sensitivity in this site, compared to the site of Omaha, Nebraska, is reflected in the formal benchmarking results, in both the smaller MRCS values, and the smaller $k_\sigma^{min}$ and $k_\rho^{min}$ values.} 
\end{table}

\paragraph{Bias Contour Plot.} We generate the bias contour plot for Corpus Christi, Texas and plot the killer confounder region. See Figure \ref{fig:contour_plot_CC}. We see that consistent with our numerical results from before, the killer confounder region spans a larger portion of the plot in contrast to the contour plot for Omaha, Nebraska. Furthermore, we visually confirm the results from our benchmarking exercise, in which we see that for the weighted estimator, a confounder with similar confounding strength as indicators for HS/GED or whether or not the individual is black or Hispanic will result in a bias large enough to result in the point estimate being brought down close to zero. 

\begin{figure}[!ht]
\centering
\includegraphics[width=0.49\textwidth]{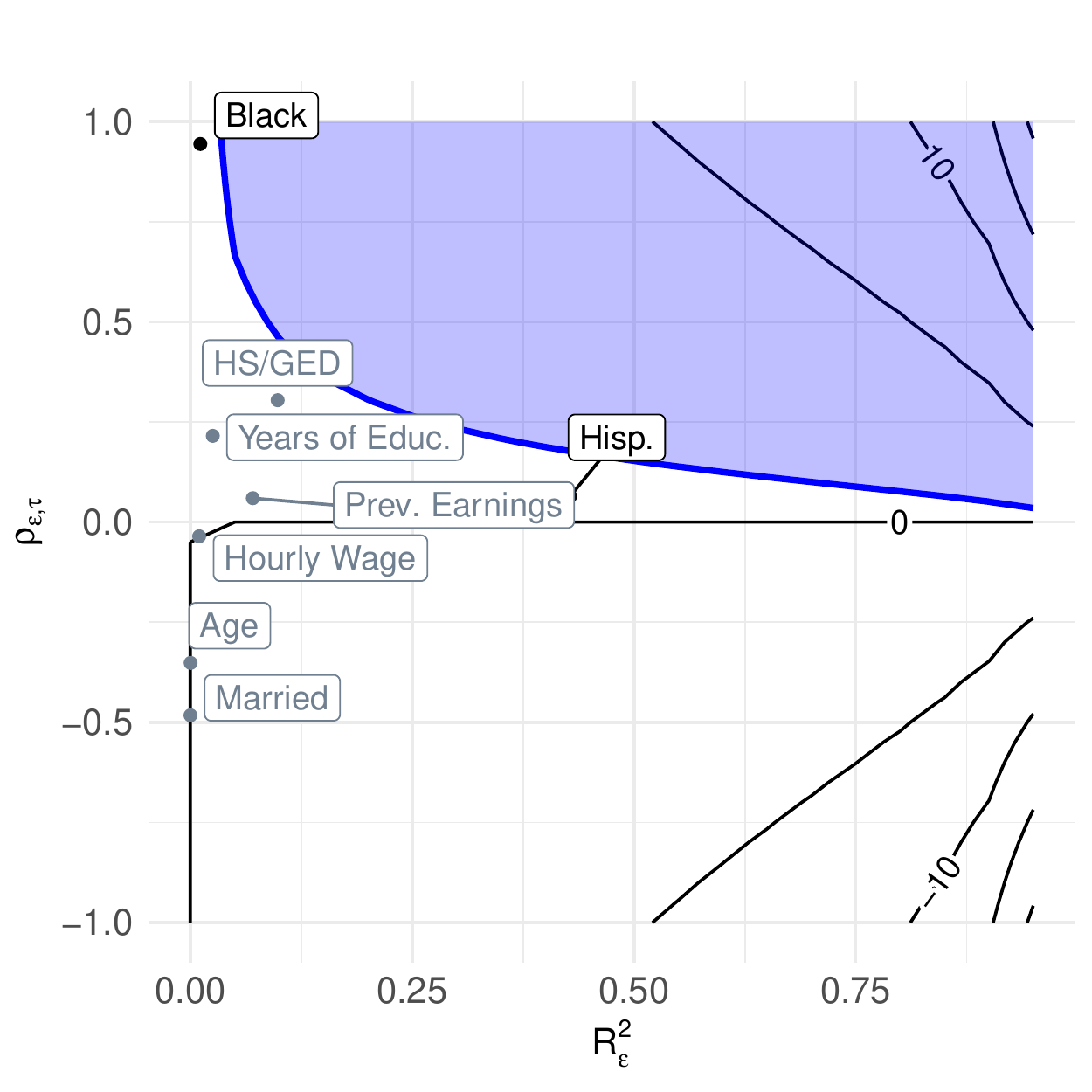}
\caption{Bias contour plot for Corpus Christi, Texas. We see that relative to the contour plot for Omaha, Nebraska, the killer confounder region is much larger. Furthermore, the formally benchmarked points, which correspond to the parameter values associated with a confounder with similar confounding strength to the observed covariate, are closer to the killer confounder region. This indicates a greater degree of sensitivity to a potential killer confounder.} 
\label{fig:contour_plot_CC}
\end{figure} 

\paragraph{Comparison with Benchmark PATE.} We conclude that the weighted estimator has greater sensitivity to potential confounders. We can compare the point estimate with the benchmark PATE, which is 1.37. As such, in alignment with what we expect from the sensitivity analysis results, there is more error in recovering the PATE for the site of Corpus Christi, Texas than in the experimental site of Omaha, Nebraska. 

\clearpage 
\subsection{Supplementary Tables} 
We provide the covariate balance tables for both Omaha, Nebraska (Table \ref{tbl:cov_bal}) and Corpus Christi, Texas (Table \ref{tbl:cov_bal_cc}). 

\begin{table}[ht]
\centering
\textbf{Summary of Covariate Balance in Omaha, Nebraska} \\ \vspace{2mm} 
\begin{tabular}{lccccc}
  \hline
  & \multicolumn{2}{c}{Experimental Sample} & \multicolumn{1}{c}{Population} & \multicolumn{2}{c}{Std. Difference}\\
  \cline{2-6}
Covariates & Unweighted & Weighted & Average & Unweighted & Weighted \\ 
   \hline
Previous Earnings & 21.60 & 25.18 & 25.18 & -0.11 & 0.00 \\ 
  Age & 31.68 & 33.53 & 33.53 & -0.19 & 0.00 \\ 
  Married & 0.10 & 0.21 & 0.21 & -0.28 & 0.00 \\ 
  Hourly Wage & 4.77 & 4.68 & 4.68 & 0.03 & 0.00 \\ 
  Black & 0.51 & 0.24 & 0.24 & 0.65 & 0.00 \\ 
  Hispanic & 0.04 & 0.13 & 0.13 & -0.27 & 0.00 \\ 
  HS/GED & 0.84 & 0.73 & 0.73 & 0.26 & 0.00 \\ 
  Years of Education & 11.52 & 11.29 & 11.29 & 0.12 & 0.00 \\ 
   \hline
\end{tabular}
\caption{A summary of the covariate balance in the experimental site of Omaha, Nebrasks, comparing the weighted and unweighted standardized difference across the pre-treatment covariates. We see that the covariates with largest degree of imbalance (prior to weighting) are racial indicators for black and Hispanic individuals, marital status, and high school education.} 
\label{tbl:cov_bal} 
\end{table} 

\begin{table}[ht]
\centering
\textbf{Summary of Covariate Balance in Corpus Christi, Texas} \\ \vspace{2mm} 
\begin{tabular}{lccccc}
  \hline
  & \multicolumn{2}{c}{Experimental Sample} & \multicolumn{1}{c}{Population} & \multicolumn{2}{c}{Difference$^*$}\\
  \cline{2-6}
Covariates & Unweighted & Weighted & Average & Unweighted & Weighted \\ 
  \hline
Previous Earnings & 18.55 & 25.40 & 25.40 & -0.22 & -0.00 \\ 
  Age & 32.14 & 33.45 & 33.45 & -0.13 & -0.00 \\ 
  Married & 0.22 & 0.20 & 0.20 & 0.05 & 0.00 \\ 
  Hourly Wage & 4.79 & 4.68 & 4.68 & 0.05 & 0.00 \\ 
  Black & 0.10 & 0.28 & 0.28 & -0.40 & -0.00 \\ 
  Hispanic & 0.69 & 0.07 & 0.07 & 2.54 & 0.00 \\ 
  HS/GED & 0.73 & 0.74 & 0.74 & -0.01 & -0.00 \\ 
  Years of Education & 10.81 & 11.36 & 11.36 & -0.30 & -0.00 \\ 
  \hline
\end{tabular}
\caption{A summary of covariate balance in the experimental site of Corpus Christi, Texas, comparing the weighted and unweighted standardized difference across the pre-treatment covariates. We see that the experimental site has relatively more Hispanic individuals and fewer black individuals, in comparison to the rest of the population.} 
\label{tbl:cov_bal_cc}
\end{table}

\subsection{Replication Code}
Replication code for the JTPA application can be found at the following link: \\
\texttt{https://github.com/melodyyhuang/senseweight}. 
\endgroup 
\end{document}